# OSIRIS-REx Contamination Control Strategy and Implementation

***Space Science Reviews*, accepted 24 October 2017**

J.P. Dworkin[1], L.A. Adelman[1,2], T. Ajluni[1,2], A.V. Andronikov[3], J.C. Aponte[1,4], A.E. Bartels[1], E. Beshore[5,6], E.B. Bierhaus[7], J.R. Brucato[8], B.H. Bryan[7], A.S. Burton[9], M.P. Callahan[10], S.L. Castro-Wallace[9], B.C. Clark[11], S.J. Clemett[9,12], H.C. Connolly Jr.[13], W.E. Cutlip[1], S.M. Daly[14], V.E. Elliott[1], J.E. Elsila[1], H.L. Enos[5], D.F. Everett[1], I.A. Franchi[15], D.P. Glavin[1], H.V. Graham[1,16], J.E. Hendershot[1,17], J.W. Harris[6,7], S.L. Hill[12,14], A.R. Hildebrand[18], G.O. Jayne[1,2], R.W. Jenkens Jr.[1], K.S. Johnson[7], J.S. Kirsch[12,14], D.S. Lauretta[5], A.S. Lewis[1], J.J. Loiacono[1], C.C. Lorentson[1], J.R. Marshall[19], M.G. Martin[1,4,6], L.L. Matthias[14,20], H.L. McLain[1,4], S.R. Messenger[9], R.G. Mink[1], J.L. Moore[7], K. Nakamura-Messenger[9], J.A. Nuth III[1], C.V. Owens[14], C.L. Parish[7], B.D. Perkins[14], M.S. Pryzby[1,21], C.A. Reigle[7], K.Righter[10], B. Rizk[5], J.F. Russell[7], S.A. Sandford[22], J.P. Schepis[1], J. Songer[7], M.F. Sovinski[1], S.E. Stahl[9,23], K. Thomas-Keprta[9,12], J.M. Vellinga[6,7], M.S. Walker[1]

[1] *NASA Goddard Space Flight Center, Greenbelt, MD, USA*
(jason.p.dworkin@nasa.gov)
[2] *Arctic Slope Research Corporation, Beltsville, MD USA*
[3] *Czech Geological Survey, Prague, Czech Republic*
[4] *Catholic University of America, Washington, DC, USA*
[5] *Lunar and Planetary Laboratory, University of Arizona, Tucson, AZ, USA*
[6] *Retired*
[7] *Lockheed Martin Space Systems, Littleton, CO, USA*
[8] *INAF Astrophysical Observatory of Arcetri, Florence, Italy*
[9] *NASA Johnson Space Center, Houston, TX, USA*
[10] *Boise State University, Boise, ID, USA*
[11] *Space Science Institute, Boulder, CO, USA*
[12] *Jacobs Technology, Tullahoma, TN, USA*
[13] *Rowan University, Glassboro, NJ, USA*
[14] *NASA Kennedy Space Center, Titusville, FL, USA*
[15] *The Open University, Milton Keynes, UK*
[16] *University of Maryland, College Park, MD, USA*
[17] *Ball Aerospace, Boulder, CO, USA*
[18] *University of Calgary, Calgary, AB, Canada*
[19] *SETI Institute, Mountain View, CA, USA*
[20] *Analex, Titusville, FL, USA*
[21] *ATA Aerospace, Albuquerque, NM, USA*
[22] *NASA Ames Research Center, Moffett Field, CA, USA*
[23] *JES Tech., Houston, TX, USA*

**Abstract** OSIRIS-REx will return pristine samples of carbonaceous asteroid Bennu. This article describes how pristine was defined based on expectations of Bennu and on a realistic understanding of what is achievable with a constrained schedule and budget, and how that definition flowed to requirements and implementation. To return a pristine sample, the OSIRIS-REx spacecraft sampling hardware was maintained at level 100 A/2 and <180 ng/cm$^2$ of amino acids and hydrazine on the sampler head through precision cleaning, control of materials, and vigilance. Contamination is further characterized via witness material exposed to the spacecraft assembly and testing environment as well as in space. This characterization provided knowledge of the expected background and will be used in conjunction with archived spacecraft components for comparison with the samples when they are delivered to Earth for analysis. Most of all, the cleanliness of the OSIRIS-REx spacecraft was achieved through communication among scientists, engineers, managers, and technicians.

# Acronyms

| | |
|---|---|
| AAM3 | Asteroid Approach Maneuver 3 |
| ACS | attitude control system |
| ARC | NASA Ames Research Center |
| ATLO | Assembly, Test, and Launch Operations |
| ATP | adenosine triphosphate |
| CCWG | Contamination Control Working Group |
| CDR | Critical Design Review |
| CEWG | Contamination Engineering Working Group |
| CFD | computational fluid dynamics |
| DAC | DSMC Analysis Code |
| DART™ | Direct Analysis in Real Time |
| DNA | Deoxyribonucleic acid |
| DSMC | Direct Simulation Monte Carlo |
| DTGS | deuterated glycine trimer |
| EA | combustion elemental analyzer |
| EACA | ε-amino-*n*-caproic acid |
| EDL | Earth descent and landing |
| EDU | engineering design unit |
| EDX | energy dispersive X-ray spectroscopy |
| EMP | electron microprobe |
| FTIR | Fourier Transform infrared spectroscopy |
| GC | gas chromatography |
| GC-IRMS | GC combustion isotope ratio MS |
| GSFC | NASA Goddard Space Flight Center |
| ICP-MS | inductively coupled plasma MS |
| IEST | Institute of Environmental Sciences and Technology |
| IRMS | isotope ratio mass spectrometer |
| ISO | International Organization for Standardization |
| JSC | NASA Johnson Space Center |
| KSC | NASA Kennedy Space Center |
| LA-ICPMS | laser ablation inductively coupled plasma MS |
| LC | liquid chromatography |
| LM | Lockheed Martin Space Systems |
| LPF | Large Payload Fairing |
| MAVEN | Mars Atmosphere and Volatile EvolutioN Mission |
| MRD | mission requirement document |
| MS | mass spectrometry |
| NASA | National Aeronautics and Space Administration |
| NRC | National Research Council |
| NVR | nonvolatile residue |
| OCAMS | OSIRIS-REx Camera Suite |
| OCSSG | Mars Organic Contamination Science Steering Group |
| OLA | OSIRIS-REx Laser Altimeter |
| OSIRIS-REx | Origins, Spectral Interpretation, Resource Identification, and Security–Regolith Explorer |
| OTES | OSIRIS-REx Thermal Emission Spectrometer |
| OTU | operational taxonomic unit |
| OVIRS | OSIRIS-REx Visible and Infrared Spectrometer |
| PDR | Preliminary Design Review |
| PFTE | polytetrafluoroethylene |
| PHSF | KSC Payload Hazardous Servicing Facility |
| PI | Principal Investigator |
| ppb | parts per billion |
| ppm | parts per million |
| REXIS | Regolith X-ray Imaging Spectrometer |
| SEM | scanning electron microscopy |
| SRC | sample return capsule |
| TAG | touch-and-go |
| TAGSAM | touch-and-go sample acquisition mechanism |
| TEM | transmission electron microscopy |
| ToF-SIMS | time-of-flight secondary ion MS |
| UHP | ultrahigh purity |
| ULA | United Launch Alliance |
| UV | ultraviolet |
| VC-HS | visibly clean-highly sensitive |
| VIF | Atlas V Vehicle Integration Facility |
| XANES | X-ray absorption near edge structure |
| $\mu$-$L^2$MS | microprobe two-step laser desorption/laser ionization MS |

## Table of Contents

# 1 Introduction

The OSIRIS-REx mission (Origins, Spectral Interpretation, Resource Identification, and Security Regolith Explorer) is the third mission selected under NASA's New Frontiers Program. The mission was approved for initial competitive development (Phase A) on December 29, 2009. The contamination control strategy for OSIRIS-REx evolved from the Organic Contamination Science Steering Group (OCSSG) approach developed for Flagship missions to Mars (Mahaffy et al. 2004) to one tailored and implementable in a cost-capped NASA program to a primitive asteroid. This manuscript describes the lessons and results in the seven years of implementation and development through launch on September 8, 2016.

The primary objective of the mission is to return and analyze at least 60 g of "pristine" (see below) carbonaceous asteroid regolith (Lauretta et al. 2017). The OSIRIS-REx team selected the B-type near-Earth asteroid (101955) Bennu due to its accessibility and spectral similarity to CI and CM carbonaceous chondrites (Clark et al. 2011). Carbonaceous chondrite meteorites are hypothesized to be fragments of carbonaceous asteroids. These are the among the oldest and most primitive solids in the solar system and contain up to 3% carbon, and can include parts per million (ppm) or lower abundances of soluble organic compounds. Meteorite studies suggest that these types of asteroids may have contributed a wide range of organic compounds such as amino acids to the Earth, possibly supporting the emergence of life (e.g., Burton et al. 2012). The spacecraft will rendezvous with Bennu in 2018, then spend over a year characterizing the asteroid before executing a touch-and-go (TAG) maneuver to collect a sample of regolith, which will be returned to Earth for worldwide study on September 24, 2023. The analysis of pristine asteroid regolith samples from a well-characterized geological context will provide key constraints in the history of asteroid Bennu. This encompasses the epoch before it was accreted, through when it may have been geologically active and part of a larger body, to its dynamical orbital evolution from the main belt to Earth-crossing. The team will apply what they learn from the history of Bennu through sample analysis to the potential history of other asteroids.

The OSIRIS-REx spacecraft will collect surface regolith via a touch-and-go sample acquisition mechanism (TAGSAM) that fluidizes loose particles with high-pressure, high-purity $N_2$/He gas (Bierhaus et al. 2017). The $N_2$/He gas carries the samples into a cylindrical sample container, enclosed by biaxially oriented polyethylene terephthalate (e.g., Mylar®) flaps; 5% He is added for leak checking prior to launch. The gas escapes through a metal mesh that serves as the outer wall of the cylinder, and entrained particles, up to 2.5 cm for roughly equidimensional particles, or >2.5 cm in the longest dimension for oblong particles, are trapped. Contact pads of stainless steel loops also collect small particles for investigation of the properties of the space-exposed asteroid surface.

The value of these samples could be reduced by the addition of terrestrial contamination, which can directly obscure results and undermine the confidence of measurements and conclusions. For these reasons, the control of the access of contamination to the sample is key.

## 1.1 Defining Pristine

The driving Mission Level 1 requirement is to "return and analyze a sample of *pristine* carbonaceous asteroid regolith in an amount sufficient to study the nature, history, and distribution of its constituent minerals and organic material." The team designed this Mission Level 1 requirement to capture the importance of contamination by elevating it to the highest level of mission requirements, with enough flexibility to allow Mission Level 2 and 3 requirements to focus on the implementation. In the strictest sense, the "pristine" state is violated by any alteration of the physical, chemical, textural, or other state that compromises sample integrity. Alteration includes changing inherent states, losing sample components, or adding extraneous components. These could include changes in bulk chemistry/mineralogy, trace components, stable isotopic ratios, volatiles (ices and organics), crystallinity and phase state, remnant magnetism, grain-size distribution, grain/clast integrity, texture/structure/layering, and chemical/electronic activation state. This overly broad definition of contamination is beyond the scope of the science requirements of OSIRIS-REx.

Some level of contamination and alteration of the sample is probable. Decisions and actions which impact sample cleanliness can occur at any time in the lifecycle of spacecraft fabrication, operations, and sample curation. Mitigation, therefore, needs to be carefully planned from mission conception. Thus, it is important to strategize about what levels of contamination and alteration of the sample to accept to ensure the success of the mission. Overly aggressive requirements, which do not directly serve the investigations, can drive mission architecture. These driving cases can result in non-value-added cost growth which, if unchecked, can lead to reduction in the scope (descope) of the contamination requirements or even project cancelation.

Instead, the aim of the team was to develop a set of realistic contamination requirements as well as a number of planned descope options to allow a graceful relaxation in case of technical or cost avoidance needs. Maintaining schedule for a planetary mission is paramount; a schedule slip that consumes the launch period creates a delay until the Earth and target orbits next align. Such a delay comes at significant economic and political cost. OSIRIS-REx was required to launch within a 39-day planetary launch period or delay a full year. A one-year delay would cause the mission to consume all available cost reserves and was not a programmatically viable option.

A recommendation from NASA's Stardust mission (Sandford et al. 2010) was that a mission needs to define what is meant by "clean" (a.k.a. "pristine" for OSIRIS-REx) from the very beginning (Table 1). *OSIRIS-REx defines pristine to mean that no foreign material is introduced to the sample in an amount that hampers the ability to analyze the chemistry and mineralogy of the sample*. Specific contaminant abundances are set to a level necessary to achieve the NRC (National Research Council) recommended "±30 percent precision and accuracy" (National Research Council 2007) on measurements. The team will carry out a wide range of sensitive and high-spatial-resolution chemical and mineralogical studies of the sample. Accordingly, contamination control must simultaneously preserve, to the extent necessary, the original organic and inorganic compositions of the sample from collection through curation. Achieving this in the New Frontiers-dictated cost-controlled environment is a significant challenge. Fundamental to the OSIRIS-REx mission's approach to contamination control is the belief that judicious knowledge of the nature of low levels of contaminants can effectively mitigate their impact to science analysis.

[Insert Table 1 here]

# 2 Organization

While OSIRIS-REx launched in 2016, initial planning for contamination control and assessment began in 2009 and matured via weekly input from scientists, engineers, and managers from across the Project. During mission Phase A and B the Contamination Control Working Group (CCWG) was chaired by the OSIRIS-REx Project Scientist and met weekly to define and refine the contamination requirements presented at the mission's Preliminary Design Review (PDR). CCWG shared membership with the Curation, Sample Return Capsule (SRC), and Sample Analysis Working Groups. The Sample Analysis Working Group was tasked with the implementation of the contamination knowledge requirements. Once the control and knowledge requirements were set, the CCWG was dissolved and the Contamination Engineering Working Group (CEWG) was formed. CEWG was chaired by the Lead Project Contamination Engineer (NASA Goddard Space Flight Center; GSFC) with routine participation by contamination engineers from Lockheed Martin Space Systems (LM), NASA Kennedy Space Center (KSC), United Launch Alliance (ULA), and the Project Scientist representing the scientists. Other engineers, curators, and managers were invited depending on the topic, but these meetings served to focus the implementation decisions and socialize the key participants who would be working together for years, and very intensely during launch operations.

Since returning a pristine sample is a Mission Level 1 requirement, contamination control plans were regularly analyzed during mission lifecycle reviews by a panel of external experts. To explore the plans and design in greater depth the team held an all-day contamination control and knowledge peer review in November 2013, shortly after the NASA MAVEN (Mars Atmosphere and Volatile EvolutioN Mission) launch. The timing of this review also captured the recent experience of MAVEN and was approximately midway between the mission PDR and mission Critical Design Review (CDR). It provided the ability to study the details of the contamination control and knowledge planning before flight hardware construction began. The review covered all aspects of contamination control and knowledge, from the requirements through the flight hardware implementation to the operations at the launch site and recovery site including curation of the science sample and the knowledge samples. The team brought in experts and walked through the plans, ensuring that nothing major had been missed.

The authority to implement these requirements derives from sample cleanliness being a Mission Level 1 requirement and the fact that the OSIRIS-REx Project Scientist was tasked with spearheading the contamination effort. This meant that contamination science was reported directly to the Principal Investigator via two members of the Science Executive Council (Figure 1). Simplicity and cost control derive from the presence of a graceful descope plan for cost growth avoidance. It is crucial for the success of this process that the same people who wrote the mission concept and requirements are the ones who implement the cleaning and analysis.

[Insert Figure 1 here]

# 3 Contamination Control

Based on our mission definition of pristine, the team derived Mission Level 2 requirements for contamination control that were (1) traceable to an independent document or analysis, (2) achievable within the project budget and schedule, and (3) rapidly verifiable without impacting the overall mission schedule, particularly during Assembly, Test, and Launch Operations (ATLO). ATLO is the phase of a mission which typically involves the most number of people and organizations, is the most expensive, and is the most time-critical. This time criticality is even more pronounced for a mission with a limited launch period, such as OSIRIS-REx.

Since verification is performed on a surface, the allowable contamination level in the sample was converted to a surface area requirement. Sample scientists generally refer to contamination in samples as mass ratios (e.g., parts per billion, ppb, or ng contaminant per g of sample). Therefore, the derived surface contamination requirement is based on (1) the expected sample mass to be collected, (2) the area of the spacecraft surfaces that may contaminate the sample, and (3) an assumption of how efficiently surface contaminants are transferred to the sample. OSIRIS-REx will return a minimum of 60 g of asteroid material, so a reasonably conservative value for the contamination requirement is based on this sample mass being contaminated by contact with TAGSAM interior surfaces (1920 $cm^2$). The most likely contamination risks arise from contact with the TAGSAM head itself, the gas employed during the collection, storage conditions in curation, and sample handling and processing. By comparison, the risk of contaminants reaching the sample by outgassing or surface creep from other spacecraft components is low, but nonzero. Therefore, for spacecraft construction the team focused on controlling and monitoring contamination on those surfaces closest to the sample storage capsule, and assumed 100% transfer of contamination. While this is a worst-case scenario, it provides sufficient margin for ensuring the pristine nature of the collected sample. This approach also provides a way to prioritize controlled surfaces; starting with the most distant, covered, small, and unlikely sources of contamination, the risk, and thus the attention, grows as proximity or line-of-sight to the sample increases.

Careful consideration was given to what contaminants would be monitored and controlled for during ATLO. An extremely broad range of scientific investigations will be carried out on the returned sample, and dozens of minerals and thousands of molecular species are of interest. It is clearly impractical to control for the full range of target materials. Table 2 lists several potential organic limits based on various guidelines that were considered and discussed in more detail below.

Initially, the team looked to the Mars Organic Contaminants Science Steering Group (OCSSG) (Mahaffy et al. 2004) for organic contamination requirements (Table 2) and functional contamination control performance (Table 3). The OCSSG molecular requirements would not have been quickly verifiable on TAGSAM surfaces during construction and ATLO because the number of species and required sensitivity of most of the tests are outside of what is possible with routine analyses and rapid turnaround times. Delays in verification would lead to delays in ATLO procedures, with concomitant cost and schedule overruns, and threaten missing the launch period. The cleanroom performance specified by the OCSSG was divided into three categories (Table 3): level 1 (general surfaces of the martian spacecraft carrying organics detectors), level 2 (general martian sample handling and processing facility surfaces), and level 3 (specific sample handling elements coming in direct contact with martian samples). The specifications of level 1 are readily achievable by the OSIRIS-REx ATLO facilities at LM and most aerospace cleanrooms. Level 2 can be achieved, although using direct verification on hardware rather than indirect witness plates is impossible due to contamination imparted by the measurement. Finally, the team was unaware of any industrial facility that can meet and verify level 3 cleanliness, including those used to construct martian probes. The shortfalls of level 3 requirements mean that a replacement, scientifically valid requirement needed to be found. Given that carbonaceous chondrites are orders of magnitude richer in organic compounds than martian meteorites, Bennu is also expected to be similarly richer in organic compounds than Mars. Mars-based requirements are more stringent than needed for OSIRIS-REx.

[Insert Table 2 here]

[Insert Table 3 here]

As described in the previous section, the OSIRIS-REx definition of pristine included "Quantitation of the amount of organic carbon present to ±30 percent precision and accuracy over a range of 0.1 ppm to 1 percent" (National Research Council 2007). This rationale was converted into a requirement (see NRC-derived entry in Table 2). However, there were two serious concerns with this approach. First, the detection methods (e.g., Fourier transform infrared spectroscopy, FTIR) for prescreening followed by liquid chromatography-mass spectrometry (LCMS), gas chromatography-mass spectrometry (GCMS), and Direct Analysis in Real Time-mass spectrometry (DART™-MS) (Loftin 2009)) each are inherently restricted to a specific range of compounds. But since these methods are similar to the types of measurements commonly carried out in meteorite studies, this is acceptable. Second, analysis costs and potential schedule delays for such open-ended compound searches present an unacceptable risk to OSIRIS-REx.

The team investigated (for example, see "worst-case" (soluble organic-depleted CI meteorite, e.g. Yamato 980115; Burton et al. 2014) and "reasonable" (soluble organic-containing CM meteorite, e.g. Murchison; Glavin et al. 2006) meteorite entries in Table 2) a series of other benchmarks for organic contamination limits, including CI and CM meteorites as Bennu analogs (Clark et al. 2011). However, the range of plausible meteorite organic abundances in Bennu analogs varies by orders of magnitude, and the uncertainty that a new meteorite discovery could change the requirements was a threat to the stability of the requirements and thus to cost. The complexity of a replacement requirement depends on the definition of "pristine." One definition would be to examine a meteorite that has been explicitly identified as "pristine" in the literature (e.g., the Antarctic CR2 carbonaceous chondrite Graves Nunatak (GRA) 95229 (Pizzarello et al. 2008)). However, any claim of a "pristine" meteorite is subjective and neither sufficiently documented nor universally accepted. Furthermore, demonstrating the contamination in a complex, difficult to characterize system is challenging and open-ended. Finally, as future studies of any sample used as a contamination archetype, particularly one as complex and heterogeneous as a meteorite are performed, the requirements derived from the sample could change. Such requirement changes during the development of a mission levies an unacceptable risk to cost and schedule. However, the bulk organic carbon abundance across carbonaceous chondrites is far less variable than particular soluble compounds; with a total carbon abundance of 1-3% in CI and CM meteorites. The team determined that 30 ppm contamination of 2% carbon bulk carbon measurement affords better than 30% precision to the planned analyses. Yet, the soluble organics, especially those relevant to astrobiology, are more sensitive. Instead, amino acids were used as a target species.

The rationale was threefold. First, amino acids are among the most pervasive compounds in the biosphere (e.g., Friedel and Scheller 2002). Second, modern detection methods are extremely sensitive (femtomole; (Glavin et al. 2006)). Third, amino acid data already exists on Stardust aluminum foil samples (Elsila et al. 2009). Stardust in many ways is an intellectual predecessor to OSIRIS-REx and was also constructed, integrated, and tested at the LM Waterton Plant. While a different alloy of aluminum (1100 in the Stardust collector versus 6061 in TAGSAM) (Tsou et al. 2003) was used, Stardust aluminum foils witnessed similar ATLO procedures, the deep space environment, return to Earth in a SRC, and curation at NASA Johnson Space Center (JSC) as the OSIRIS-REx TAGSAM head will experience. Elsila et al. (2009) studied several Stardust foils and determined that the most contaminated sample was foil C2092S,0. The low amino acid abundances of Stardust material and the confirmation of the cometary nature of the amino acid glycine by $^{13}C$ isotope analyses provide confidence that useful measurements can be made in the presence of the nylon-derived contamination observed on Stardust foil C2092S,0. Since amino acids have never been explicitly monitored and controlled as part of contamination control for a NASA mission, the risk of imposing a novel requirement was best understood by setting the contamination of this Stardust foil sample as the upper allowed limit of OSIRIS-REx contamination. Table 5 shows the data on this foil and a total amino acid contamination level of 186 ng/cm$^2$, which was rounded down to 180 ng/cm$^2$. Of particular note in the Stardust samples is the high relative abundance of ε-amino-*n*-caproic acid (EACA), the hydrolysis product of nylon 6, which was used in recovery, curation, and distribution of Stardust materials (Sandford et al. 2010). Amino acid–based polymers, such as nylon and latex, were prohibited on OSIRIS-REx. That action alone would have eliminated 185 ng/cm$^2$ from Stardust foil C2092S,0, leaving <1–2 ng/cm$^2$ of amino acid contamination.

Like with total carbon, the use of meteorite analogs for inorganic contamination limits is more straightforward because bulk elemental abundance varies less than organics across carbonaceous chondrites. Sufficient limits based on 10% of chondritic abundances were designated (Table 4). For further simplicity, a restricted set of indicator elements was selected as proxies for contamination monitoring that represented distinct and critical areas of scientific study. These elements are all measurable by scanning electron microscopy energy dispersive X-ray spectroscopy (SEM/EDX). To make these requirements useful across the team, they were converted to the language of contamination engineering, based on films and particles described in IEST-STD-CC1246D (IEST 2002) and assuming a collection of 60 g of sample (the minimum expected) inside the 1916 cm$^2$ surface area TAGSAM head.

IEST-STD-CC1246D defines particulate contamination per 0.1 $m^2$ ($N$) in terms of levels ($L$) for particles size $x$ according to the equation $\log N = -0.926(\log^2 x - \log^2 L)$. The results of this equation are then binned by particle size. For example, level 100 has a maximum of 1780 5µm, 264 15µm, 78.4 25 µm, 10.7 50 µm, and 1 100 µm particles per 0.1 $m^2$. Films or nonvolatile residue (NVR) are simply defined relative to "A" which is 1000 ng/$cm^2$ of contamination, so A/2 is 500 ng/$cm^2$.

Under the anticipated conditions of ATLO, inorganics are expected to be mostly particulates, but organics should still dominate the particulate population. Thus, the inorganic elements in Table 4 are expected to be a small component of the total. With these assumptions, a theoretical worst case with pure elemental particles is generally met by an achievable level 100 particulate requirement and a NVR level of A/2 for the sensitive areas of the OSIRIS-REx flight system. Since there are still pathological conditions that would violate the intent behind the requirement while still meeting level 100 A/2 standard (e.g., 100 ng/$cm^2$ of tin particles is below level 100 but exceeds the total science contamination limit for tin), the team requested the ability to check for the unexpected.

[Insert Table 4 here]

The team adopted an organic contamination requirement based on the 100 A/2 limit for carbon and the other elements in Table 4, with the addition of amino acids as a test for organic contamination. The goal was to minimize amino acid contamination as much as practical, since the control of amino acids was novel. This approach included measuring the amino acid, particle, and NVR contamination on proxy witness plates throughout ATLO. This effort is needed because the act of measuring flight hardware via established methods (washes, wipes, tape lifts) is likely to contaminate the hardware. Furthermore, since cleaning is impossible after launch, and verification of contamination after launch would be impossible until the sample acquisition hardware was returned to Earth, the team set all contamination control limits to conservative levels at time of launch.

OSIRIS-REx uses high-purity hydrazine monopropellant thrusters. Hydrazine is a strong base and powerful reducing agent, largely due to the adjacent nitrogen lone pair electrons making it an alpha effect nucleophile. The team thought it prudent to recognize the potential reactivity of hydrazine on the sample and limit the exposure of unreacted hydrazine on the sample.

Different subdisciplines use different terminology, so definitions had to be standardized for the different types of contamination witnesses (Table 6). Level 2 requirements as described in Table 7 for both contamination control and contamination knowledge were established. These requirements are more stringent than those already imposed by NASA for OSIRIS-REx to meet the requirements of Planetary Protection, Category II outbound and Category V unrestricted Earth return (NASA 2011). The flow of requirements and documentation is shown in Figure 2.

Contamination control requirements are defined to provide a simpler and more generic test. But in anticipation of "unknown unknowns" the Sample Analysis Working Group was responsible for observing and cataloging the unexpected via contamination knowledge investigations (see section 5). Contamination knowledge analyses are open-ended, but only on limited samples and with no impact to schedule. This effort provided the information needed to maximize the scientific benefit of the returned sample, without potentially halting ATLO for months.

[Insert Table 5 here]

[Insert Table 6 here]

[Insert Table 7 here]

[Insert Figure 2 here]

## 3.1 Amino Acid Transfer Efficiency

Once the team identified amino acids as a critical analyte, they performed a simple test to evaluate the transfer efficiency of dry amino acids. This test allowed for the determination of the probability of adhering amino acids (and presumably other charged species) entering the sample. To simulate regolith, the team used silica fume due to its high surface areas and copper-clad steel balls to grind the silica fume into the TAGSAM aluminum surfaces; 75% of the simulated regolith by mass was steel balls.

Two 60-g-total identical mixtures of silica fume and copper-clad steel balls were cleaned by heating at 700°C in air for 24 hours in a muffle furnace. A TAGSAM engineering design unit (EDU) with a film of known contamination (1.0 mg of D-isovaline dissolved in 1:1 water:methanol) applied to the interior surfaces of TAGSAM was filled with regolith simulant and openings sealed with Kapton tape. The doped TAGSAM, one sample of regolith simulant, and a custom-made vibration fixture plate were taken to a Ling B335 Shaker/SAI120 Amplifier in Building 7 at GSFC (Figure 3) and shaken for 1 minute at 20 Hz with a 2-cm vertical displacement and maximum 5*g* acceleration.

[Insert Figure 3 here]

After vibration was complete, the mount and unit was removed and disassembled. The sample and blank were separated into balls and fume for analysis. The samples were extracted in 100°C water for 24 hours, split with half hydrolyzed in 150°C HCl vapor for 3 hours, both halves were separately desalted, derivatized with *o*-phthaldialdehyde/N-acetyl-L-cysteine (OPA/NAC), and analyzed via a Waters® ACQUITY™ ultraprecision liquid chromatograph and fluorescence detector couplet to a Waters® LCT Premier™ time-of-flight mass spectrometer LCMS according to Glavin et al. (2006; 2010). The results indicate that (1) the worst-case transfer efficiency of an aliphatic amino acid from TAGSAM to this simulant is 0.03 ng/g from 1 mg coating the interior of TAGSAM (an efficiency of 0.5 ppm); and (2) amino acids from an essentially uncleaned TAGSAM surface appear at only 0.22 ng/g of regolith. This abundance of total amino acid contamination is actually *below* even the 1 ng/g level of amino acids specified by OCSSG.

These results differ from previously reported lysine transfer tests performed in relation to Mars sample handling requirements (0.1% from aluminum to sand without agitation) (Mahaffy et al. 2004). We suggest the difference may be that the sand could have contained far more moisture which would greatly aid in transfer. It is reasonable that dry transfer, such as expected on airless Bennu regolith is more relevant to OSIRIS-REx.

### 3.2 Amino Acid Cleaning

Since amino acid requirements for ATLO are novel, the team performed a number of tests to determine the effectiveness of precision cleaning techniques on the removal of amino acids, as well as the potential for amino acid contamination derived from the solvents and gloves used at LM and GSFC.

Common steel screws were used as the substrate to test LM precision cleaning protocols relative to a procedural blank (a glass vial cleaned by heating to 500°C in air for >8 hours with no screw). The glass vials were borosilicate conical screw cap test tubes with a piece of aluminum foil used to prevent the polytetrafluoroethylene (PFTE, e.g., Teflon®) lined cap from touching the vial. The vials were shipped to LM for use, where samples were placed in the glass vials to be returned to GSFC for analysis. The identical procedure was performed for all amino acid contamination monitoring plates from LM. The plates collected at KSC were wrapped in 500°C cleaned foil and sent to GSFC for subdivision and analysis.

The samples were as follows: a screw removed from the parts box without any cleaning "uncleaned" placed in a sample vial in the LM cleanroom, a screw which was taken to the cleanroom and dirtied by being exposed to human breath sufficient to provide some condensation, an identical dirtied screw which was cleaned by sonicating in Brulin 815 GD™ detergent, rinsed with water, and then precision cleaned using a pinpoint spray of polished water (135 ± 5°F and 45 ± 5 psi), a screw which had been heat sealed in a nylon bag, and an identical packaged screw which was subsequently cleaned as above. Three types of gloves and six types of bags were analyzed after exposure to 5 mL room temperature water for 24 hours; and two types of 2-propanol were analyzed (Table 8).

Each sample was analyzed via LCMS with the AccQ•Tag™ protocol (Boogers et al. 2008) on a Waters® ACQUITY™ and LCT Premier™ time-of-flight mass spectrometer equipped with an electrospray ionization source (positive ion mode), mass resolution setting of 5,000 m/Δm. We elected to use this protocol over the OPA/NAC method and chromatography Glavin et al. (2006; 2010) since the derivatization product is stable enough to allow for unattended sequential analysis, AccQ•Tag™ does not require desalting because it is not susceptible to multivalent cation interference, and chiral separation was not required—combined, this resulted in more rapid analyses to meet the 1-week requirement for ATLO amino acid analyses. Sample was introduced via a Waters® ACQUITY UPLC® with fluorescence detector. For LC analysis a 250-μL syringe, 50-μL loop, and-30 μL needle were used. The total injection volume was 1 μL. A set of nine calibrators of proteinogenic amino acids (0.25 to 250 μM) was prepared in water and analyzed. A linear least-square model was fit to each analyte. Both mass and fluorescence traces were quantitated. The blank sample was used to subtract procedural and laboratory background; trace levels of glycine were observed in the blank. Sample transfers were performed in an International Organization for Standardization (ISO) 5 laminar flow bench. The identical analytical procedure was used on authentic contamination monitoring plates. Each amino acid was individually quantified. This analytical method was used on all amino acid contamination control analyses, with the more involved Glavin et al. (2006; 2010) method reserved for contamination knowledge analyses (below).

The cleaning method tested was determined to be effective in removing amino acid contamination. The cleaning appears more efficient at removing bound amino acids than free. This is reasonable because they are most likely present in particulates (e.g., skin flakes).

[Insert Table 8 here]

The team also estimated the amount of amino acid loss during spacecraft thermal-vacuum testing. Thermal-vacuum testing was performed after assembly when component cleaning is no longer possible, but could serve to further decrease the contamination acquired during earlier assembly and test operations. The team simulated LM thermal-vacuum conditions at GSFC with a laboratory manifold. The experiment was simple, yet sufficient for the purpose and time available.

Each sample for this experiment was created by adding an amino acid solution (392 µL of 1.5 µM each of 16 biological amino acids) to a 12-mm outer diameter (10-mm inner diameter, fill height of 5 mm) amber vial, which was then dried at <30°C under reduced pressure (~1 torr). If this solution dried evenly over the interior of the vial and the vial was a perfect cylinder, then the amino acid film would have been nearly identical to the total amino acid abundance in the "dirtied" screw in Table 8. However, in that experiment only 27% of the amino acids were free, as opposed to bound in peptides or cells.

Each vial was then placed individually in a quartz finger and held at 100°C under vacuum (~$1 \times 10^{-5}$ torr) in a tube furnace. Six vials were used in total, each heated for a different time period (unheated, 1 hour, 7 hours, 24 hours, 48 hours, and 120 hours). After heating, each sample was analyzed via the AccQ•Tag™ method.

Analysis of the free amino acids showed a decrease in concentration over time. The analysis showed a reduction of approximately 50% of each free amino acid after 24 hours of vacuum heating. Due to the plausible concentrations used and the small volume permitted in the experimental setup, the signal-to-noise ratio for a given peak was insufficient to allow the accurate quantitation of rates. Regardless, half-lives are in the range of hours, not minutes nor days. A trade between the effectiveness of the precision cleaning and the cost and schedule for thermal-vacuum cleaning resulted in the flight TAGSAM head being heated to 95±5°C for 24 hours at $\leq 1 \times 10^{-5}$ torr.

### 3.3 Spacecraft Requirements and Implementation

The OSIRIS-REx spacecraft was processed in an ISO 7 cleanroom at LM, with some tests performed in an ISO 8 cleanroom. These environments were monitored with contamination monitoring and contamination knowledge plates both prior to and during occupation by OSIRIS-REx hardware and personnel. Some of these were shared cleanrooms, so the LM contamination engineers required knowledge and the ability to control the activities and materials used by the other programs.

Hardware verification samples were also collected at key times. To minimize contamination, sensitive surfaces were bagged in PFTE whenever possible (Figure 4), and a qualification TAGSAM head was used instead of the flight head for most of the ATLO. The identical, but clean, flight TAGSAM head was integrated just prior to a final SRC fit-check and final stowage in the launch container. The launch container was maintained under a near continuous positive pressure purge. This procedure provided for the minimal environmental exposure of the sampling hardware. Furthermore, when sampling hardware was exposed, only the minimum number of personnel required to perform the work were allowed in the room. All personnel in the facility were gowned in nylon-free cleanroom suits with the nose and mouth covered. Gloves were taped to the gown and wiped with Fisher Optima 2-propanol. Double gloves were used when working with critical hardware. Makeup, perfume, and cologne were prohibited; tobacco users were required to rinse their mouth with water 30 minutes before entering the cleanroom. Sensitive surfaces were cleaned to 50 A/2 to meet the 100 A/2 at launch requirement. Exterior surface of the spacecraft was maintained at 500 A/2 and internal surfaces at visibly clean–highly sensitive (VC-HS) levels. VC-HS level is defined by NASA-SN-C-0005 as "The absence of all particulate and nonparticulate matter visible to the normal unaided (except corrected vision) eye…[when viewed with] ≥100 foot candles [of light at a distance of] 6 to 18 inches…[from] exposed and accessible surfaces…Particulate is identified as matter of miniature size with observable length, width, and thickness. Nonparticulate is film matter without definite dimension." (NASA, 1998). Details of the spacecraft contamination control implementation are in the Mission Contamination Control Plan (see Supplemental Material S1).

[Insert Figure 4 here]

The launch vehicle fairing was cleaned to VC-HS levels under ultraviolet (UV) illumination. This effort was necessary to further minimize particulate contamination on the TAGSAM and system, since all the instruments were uncovered and pointed up at launch. The fairing interior environment was additionally sampled with a 930 $cm^2$ (1x1 ft) aluminum foil contamination-monitoring plate. This plate was assembled from a KSC-supplied clean ASTM E1235-12 NVR plate as the substrate since these NVR plates are routinely used to monitor Atlas V fairings. This substrate, which was wrapped with the same 500°C heat-cleaned aluminum foil used for other amino acid contamination monitoring plates, served as a clean backing for the amino acid collection surface; a second smaller clean aluminum foil was attached to the lower foil with Kapton tape (Figure 5). This ensured that the amino acid monitoring surface did not contact the NVR plate (since both sides of the amino acid monitoring foil are extracted for analysis), that the geometry did not require any changes to the existing fairing mounting hardware, and that there were no risks of foreign object debris generated by the plate. The amino acid monitoring plate was held vertically on a bracket inside the fairing between encapsulation on August 24, 2016, and final fairing closeout on September 6, 2016. After any parts of the foil touching tape were torn off and discarded during preparation in an ISO 5 laminar flow bench, approximately 10% of the foil was measured for amino acid abundances, 15% for other contamination knowledge analyses, and the remainder archived at JSC.

[Insert Figure 5 here]

### 3.4 Contamination Control Results

Unexpected events are possible during spacecraft processing. For contamination control, the two events with the most significant impacts were: first, an unexpected SRC outgassing event that took place during spacecraft thermal-vacuum testing, and second, that more mechanical testing than anticipated was required. The SRC outgassing event was caused by higher than modeled temperatures on the SRC due to reflections. It was fully mitigated with an additional higher temperature vacuum bakeout of the backshell and spot cleaning of the spacecraft. The additional mechanical testing meant that the SRC and TAGSAM head were actuated more than anticipated. This allowed for more particulates (primarily SRC heatshield material) to collect on hardware verification samples. These impacts lead the team to believe that there might be a violation of the level 100 particulate requirement. However, due to adherence to protocols and cleaning for amino acid mitigation, the NVR values were substantially below their requirements, which proved to balance the contamination budget.

The intent of the 100 A/2 requirement is to meet the elemental abundances in Table 4. The contamination knowledge plates were routinely analyzed for particulate elemental distribution at JSC via SEM/EDX spectroscopy. A JEOL 7600 field emission SEM at 15 kV in backscatter mode and EDX using a Noran microanalysis detection system with acquisition times ranging from 20 to 100 s per ≥0.05µm particle was used. Using contamination knowledge plates, the team confirmed that the particles on the contamination control plates were below the levels of concern for the critical inorganic elements and that the majority of the material (as expected) is carbon based. Assuming the worst-case assumption that the particles are graphite, the total carbon contamination was determined to be below 534 ng/cm$^2$ of carbon (Table 9).

[Insert Table 9 here]

Though amino acids had never been regulated for contamination control, the performance was far below requirements without the use of nonstandard or "heroic" cleaning procedures (Table 9). Analyses were performed at GSFC using the identical analytical procedure as with the amino acid cleaning test. All analyses were conducted and written reports delivered to the contamination engineering within one week of receipt. The dominant amino acid detected was glycine, as expected. In addition to exceptional performance on the sensitive hardware, all ATLO facilities performed very well. Figure 6 shows the sum of amino acids collected on contamination monitoring plates in the LM cleanrooms, KSC cleanroom, and Atlas V fairing. Depending on the activity, new plates were exposed days before the old plates were collected, so Figure 6 overestimates the exposure time by 6%. The team confirmed that the amino acid contamination was linearly correlated with exposure time by comparing a contamination monitoring plate deployed for three months concurrently with three one-month plates.

[Insert Figure 6 here]

## 3.5 Hydrazine

Hydrazine is known to react with organics via a Wolff-Kishner reduction, and reactions based on semicarbazide formation (e.g., Kolb et al. 1994) are also possible. The team conducted simple tests of the reactivity of various organic compounds exposed at room temperature for five minutes with anhydrous hydrazine at vapor pressures ranging from $9 \times 10^{-4}$ to 15 torr. The exposed species included 2 mmol each of methanol, ethanol, isopropanol, and acetone; 80 µmol 1-butanol; 1 and 50 µmol pyruvic acid; and a solid film made of a mixture of 0.2 µmol of each of the following amino acids: aspartic acid, glutamic acid, serine, glycine, D,L-alanine, β-alanine, D,L-α-, D,L-β-, γ-aminobutyric acid, α-isobutyric acid, D,L-isovaline, D,L-valine, D,L-isoleucine, and D,L-leucine. Amino acids were dissolved in room temperature polished water and analyzed by LCMS according to Glavin et al. (2006). Other species were measured by headspace injection in a Thermo Scientific™ Trace DSQ™ GCMS (with cryo-oven) with a Restek Rtx®-35 amine column (30 m, 0.25 mm internal diameter, 0.5 µm $d_f$) at 1 mL/min He constant flow from 30°C for 3 minutes ramping at 10°C/minute to 250°C for 5 minutes and a split injector set to 200°C at 10 mL/min.

Following exposure to hydrazine, the acetone was lost, presumably reduced to propane (which was not observed under the GC conditions), and the pyruvic acid was reduced to propionic acid in all experiments within the 5 minutes required to collect and analyze the sample. Since most Wolff-Kishner reductions are performed in the presence of a strong base under reflux conditions for hours, the reactions observed were faster than anticipated under ambient temperatures and low pressure. As expected the alcohols were unaffected. Though structures can be drawn to cyclize or dimerize the amino acids, no loss of amino acids or appearance of new peaks was observed even when amino acids were dissolved in liquid anhydrous hydrazine at room temperature. On the basis of these tests, the team decided that it is sufficient to design the spacecraft to cant the thrusters away from the sampling site and determined that the collection process with this thruster design will deposit <180 ng/cm$^2$ hydrazine on the TAGSAM surface. This hydrazine will rapidly evaporate from bare metal at sampling temperatures but traces might be adsorbed by minerals or react with free carbonyls.

While the science team for NASA's Phoenix mission to Mars was interested in understanding thruster plume products (Plemmons et al. 2008), OSIRIS-REx is the first mission to impose a maximum hydrazine flux as a scientific requirement, and as such there was no existing precedent (model-based, testing-based, or otherwise) to aid in defining the appropriate limit. In the absence of historical knowledge, the team used analogy to the amino acid limit of 180 ng/cm$^2$ on the TAGSAM head. To minimize contamination from all sources, the TAGSAM head remains in the launch container until just prior to Asteroid Approach Maneuver 3 (AAM3), at which point the launch-container cover is ejected and the head is removed from the container to its "parked" position just outside the launch cover. In this configuration, there is expected to be no measurable amount of hydrazine deposited on the head. The two other primary configurations of the head are sampling configuration (Figure 7), and sample-mass measurement (Figure 8). All spacecraft motion and articulation in the sample-mass measurement configuration is done via reaction wheels, and so thruster plume impingement (and therefore hydrazine deposition) is not a factor.

Hence, the only times when the spacecraft thrusters could deposit hydrazine onto the TAGSAM head are when the head is in the sampling configuration (Figure 7). This occurs during initial deployment and checkout, baseline sample-mass measurements, the TAG rehearsals, and the TAG event(s). The thruster firings that occur during these times are the checkpoint burn, the matchpoint burn, and the backaway burn. These cases were modeled to determine the amount of hydrazine deposited on the TAGSAM head. The quantity of hydrazine that may reach and react with the regolith is a function of the plume dynamics, the fraction of unreacted hydrazine in the plume, and the vapor pressure of hydrazine in vacuum on the warm TAGSAM surface. Different TAGSAM components are predicted to be 25°C to 55°C during TAG with a maximum temperature requirement of 75°C, all well above the condensation temperature of hydrazine under vacuum, from –93°C to –133°C with a maximum around –108°C (Weijun et al. 2008).

[Insert Figure 7 here]

[Insert Figure 8 here]

Limited data exist on the amount of unreacted hydrazine in a thruster plume. Most contamination-focused plume impingement analyses assume steady-state consumption of 100%, leaving no unreacted hydrazine in the plume. Testing done in support of the Phoenix Mars mission (Plemmons et al. 2008) suggests the amount of unreacted hydrazine is < 0.05%, and likely < 0.01%. However, the Phoenix thruster type tested was different than OSIRIS-REx attitude control system (ACS) thrusters, and the measurement was conducted over a steady-state burn and did not include (or at least did not isolate) initial less efficient transient period at burn start-up nor operate in the pulsed mode employed by OSIRIS-REx. Testing an OSIRIS-REx ACS thruster under the relevant conditions proved to be cost prohibitive. Instead, the team took a worst-case value of 0.05% unreacted hydrazine from the upper limit for the Phoenix tests.

The primary tools used to determine the flux of hydrazine on the TAGSAM head are the ANSYS Fluent computational fluid dynamics (CFD) solver and DAC (Direct Simulation Monte Carlo (DSMC) Analysis Code). The CFD tool (Figure 9) is used for the volume in which the gas density is sufficiently high that continuum solutions are accurate descriptions of the thruster plume dynamics. These solutions formulate boundary surfaces, which are the initial conditions for the DSMC code, which then simulates the dynamics of individual particles.

[Insert Figure 9 here]

Analysis of DSMC results revealed two subcases for the TAG geometry: one in which the TAGSAM head is in free space and the other when it is near the surface of Bennu (Figure 10). The near-surface case is distinct because plume interactions with the surface result in density contours that are different from those when the spacecraft is far from the asteroid (i.e., in “free space”). In particular, the presence of the surface creates a recirculation that increases the amount of thruster plume flux on the head, including the unreacted hydrazine.

The first is the portion of the burn that occurs when the head is on or near the surface. “Near the surface” is conservatively defined as ≤7 m range between the thrusters and the surface. Since the thrusters are nominally 3 m from the surface at the time of TAG, a 7 m range threshold for this condition means that there is a measurable enhancement of plumes on the head for an additional 4 m as the spacecraft backs away from the asteroid. The second is the portion of the burn that occurs beyond 7 m, at which point the plume geometry is equivalent to firing thrusters in free space.

[Insert Figure 10 here]

Using the calculated hydrazine flux on the head and the planned mission thruster profile during TAG, the team derived the total hydrazine fluence on the TAGSAM head. The analysis also utilized the following reasonable assumptions: condensation temperature of hydrazine in vacuum is –108°C; the TAGSAM head will be warmer than –108°C for the TAG maneuvers; for any deposited hydrazine on clean head (prior to first TAG), all hydrazine will leave the TAGSAM surface because of TAGSAM surface temperatures (Chirivella 1975; Carré and Hall 1983); and after first TAG, the team assumed 100% sticking coefficient. The last assumption implies all hydrazine deposited on the head stays on the head and is available to contaminate the sample. This assumption derived from the possibility that the TAGSAM head may be coated in a thin layer of potentially reactive dust after the first TAG. The results of these assumptions applied to the DSMC code (Table 10) demonstrate that under nominal conditions (one TAG), if the worst-case assumptions hold (0.05% unreacted hydrazine and 100% sticking coefficient) OSIRIS-REx will collect 120 ng/cm$^2$ hydrazine. However, if subsequent TAGs are required, but the TAGSAM head becomes covered with dust from Bennu, this hydrazine requirement will need to be waived in favor of collecting a sample under these off-nominal conditions. If a second TAG is required on a dirty head, <400 ng/cm$^2$ hydrazine could be accreted, and <650 ng/cm$^2$ hydrazine for a third TAG. These are conservative values based on the above assumptions; actual values are likely to be lower.

[Insert Table 10 here]

# 4 Materials Restrictions

To help meet the contamination control requirements there were a number of materials that were prohibited or restricted for areas adjacent to the sampling head in addition to the high-outgassing materials typically prohibited on spacecraft (https://outgassing.nasa.gov/). Areas with no plausible path to the sample were not subjected to these added restrictions. For example, the OSIRIS-REx Thermal Emission Spectrometer (OTES) (Christensen et al. 2017) detector is a deuterated glycine trimer (DTGS)—a potentially very concerning contaminant in both amino acid and isotopic measurements. But the DTGS is essential for OTES operation and has no reasonable path to the sample from deep within the instrument. Conversely, there was a risk that the Regolith X-ray Imaging Spectrometer (REXIS) cover release mechanism Frangibolt® could be powered on long enough not only to break the titanium bolt (~1 minute of heating and ~150°C) to release the cover but also to unnecessarily continue heating the unit (~2 minutes of heating and over 350°C). Experiments in an instrumented vacuum chamber showed that the extra heating decomposes the outer polymer coating to an oily mixture of silicones, hydrocarbons, and esters. The mitigation was the addition of additional software controls and the addition of a separation switch into the mechanical design.

Principal compounds that decompose to amino acids or contain biological impurities were prohibited (Table 11). Nylon and other polyamides and latex are amino acid polymers and were prohibited. Nomex® and Kevlar® also degrade to amino acids, though with structures unexpected in Bennu samples. Regardless, the use of Nomex® was limited to technician's suits during hazardous operations. Natural rubber was prohibited to avoid the protein contamination. To reduce the risk of mercury vapor exposure, all fluorescent lights were required to be encapsulated in a secondary shield to prevent release of mercury in case of breakage.

[Insert Table 11 here]

Table 11 also gives an illustrative list of compounds that, although long, is not comprehensive. A list of all prohibited chemicals is impossible to compile because it is dependent on the location and application, and often requires too much knowledge of organic chemistry by nonspecialists to decipher. Instead materials engineers and scientists reviewed materials lists for compounds of concern using their knowledge of chemistry to approve or recommend alternatives (see below).

It turned out that the most difficult material restriction was nylon. Nylons are very common in cleanrooms, spacecraft, and launch vehicles. The prevalence of nylon (bags, ties, tethers, wipes, casters, thermocouples, etc.) was not anticipated. Moreover, communicating the banning of nylon with all mission partners proved more difficult than expected. The difficulty is partly due to the prevalence of nylon, the lack of nylon labeling on many products, and occasional confusion over polyamides and polyimides (the latter of which are not a contamination concern). Nylon is spread via contact transfer, and this becomes efficient when wet, so it was better to vacuum nylon that could not be removed than to wipe it with solvents. Afterward it could be covered, for example, with Kapton tape. This protocol was even applied to journalists on the August 20, 2016, Media Day (Clark 2016a). The overall effort to mitigate nylon contamination was demonstrated to be very effective, as nylon monomers were near or below detection limits in amino acid analyses of witness plates.

To the extent possible, the team attempted to minimize the diversity of organic polymers (e.g., silicones, lubricants, adhesives) in sensitive areas of the spacecraft. Such polymers are necessary for spacecraft construction, but minimizing chemical *diversity* of the contaminating species reduces the complexity of the contamination and therefore simplifies identification and interpretation of contaminants. This required excellent communication within the team, particularly among the scientists, contamination engineers, and materials engineers. The minimization of diversity was also aided by the archiving requirement: to supply a sample of each material to be archived at JSC, should a scientist need to analyze a suspected contaminant in parallel with samples from Bennu. Other restrictions were simpler to implement: the use of fluorescent lamps is rapidly declining, and those present were already encapsulated; and natural rubber is uncommon. Including them is important, however; it prevents missed restrictions by not second-guessing the facilities and provides an easy accomplishment for those laboring to meet the more difficult requirements.

In spite of these restrictions, some materials came as a surprise and created the need for late changes to materials and procedures. The late discovery of "surprise materials" was due to insufficient communication across engineering disciplines and scientists having limited understanding of the materials used in spacecraft construction. For example, one process required diamond abrasive, while another used a coating that included amorphous silica. Since both nanodiamonds and amorphous silica are of scientific interest in primitive asteroids, the diamond-abraded surface was cleaned and verified diamond-free at JSC via FTIR, and the silica-containing material was removed. The diversity of materials and processes in spacecraft construction and testing is enormous. It is vitally important to specify all materials of concern with spacecraft partners even if the scientists on the team have no expectation that they are used in engineering applications. Engaging the full set of engineers and technicians on the rationale behind the contamination requirements and empowering them to speak up when they see that a process poses an avoidable risk can reduce the use of high-heritage but undesirable procedures.

Another example occurred when there was a change in the materials in the SRC avionics deck. This decision created a situation where the fasteners were of like metals and would gall. The galling problem was not discovered until it was too late to implement a mechanical solution. Instead a film of Braycote 601EF lubricant was used. Though Braycote 601EF is used elsewhere on the spacecraft, this is a location where it could creep to the sample. Since the surfaces are in the cold and dark interior of the SRC, it is expected not to photo-degrade as seen by Rosetta, for example (Schläppi et al. 2010). In addition, the mission caries a residual risk of Braycote contamination, to remind the team of this event when the sample is analyzed starting in 2023. The flight witness plates, discussed below, will be studied to determine the impact of this lubricant. Nevertheless, this contamination risk could have been avoided with more cross-communication between engineering disciplines.

## 4.1 Materials Testing

Conversely, excellent communication between the subdiscipline engineers led to the chemical investigation of products whose chemical makeups were unclear and/or proprietary. In one case, Sonotech® Soundsafe® ultrasonic couplant was to be used during the testing of Frangibolts®. GCMS and LCMS analyses at GSFC showed myriad organic compounds with varying degrees of concern, principally hydantoin and various amines. Upon review of the required properties, pure glycerol was substituted with excellent results.

In another case, selection of a scientifically acceptable adhesive was required for the exterior of OSIRIS-REx Visible and Infrared Spectrometer (OVIRS). Engineers suggested Bondline™ 6460, but manufacturer's literature indicated that it contained polyoxypropylenediamine. Analysis at GSFC (Figure 11) was conducted via LTQ Orbitrap XL hybrid mass spectrometer equipped with DART™ source (He gas, 350°C, positive ion mode), with a mass resolution of 60,000 and lock mass enabled (on a polysiloxane compound found in air background). Results indicated the presence of at least the trimer through heptamers of polyoxypropylenediamines along with other compounds. Subsequent LCMS analysis of unhydrolyzed OPA/NAC derivatized methanol solution of Bondline™ 6460 also determined that the polyoxypropylenediamines are of mixed chirality (e.g., all 14 diastereomers of tripolyoxypropylenediamine were likely observed in Figure 11a). This could complicate, or at least cast doubt on, enantiomeric analyses in the returned sample. Fortunately, EPO-TEK® 353ND is an able replacement and appears to primarily uses 2-ethyl-4-methylimidazole instead. Though imidazoles are of interest, they are achiral. A sample of the EPO-TEK® 353ND as used was archived at JSC in the event that it presents a concern in the returned sample. This along with other materials and contamination control reports, was shared with the contamination knowledge scientists and placed on the internal science team website for review.

[Insert Figure 11 here]

# 5 Contamination Knowledge

The contamination control efforts described are based on reasonable assumptions of the composition of contaminants and provide no information on the contamination after launch. While the adopted 100A/2 contamination control limit has the advantage of being verifiable without the need for complex measurements that could pose schedule risk during ATLO, little is learned about the nature of the contaminants. A separate and parallel contamination knowledge effort was necessary to ensure that sample measurements are well understood and accurately corrected for background and are not compromised by unexpected composition of the contamination. Thus, in addition to samples collected during ATLO for particles, films, and amino acids, contamination knowledge witness plates were regularly deployed throughout the course of ATLO in the vicinity of TAGSAM and spacecraft assembly operations (Figure 12).

Similarly, an array of sapphire and Al witness plates are flown on the spacecraft and exposed before, during, and after sampling. These plates are then returned along with the samples to understand the contamination acquired during flight.

Contamination knowledge was also employed to investigate anomalies. For example, the REXIS detector assembly mount with detector flexible printed circuits was inadvertently contaminated by a defective heating element during component-level thermal-vacuum testing. The contamination knowledge scientists were enlisted to analyze several samples and controls within in a few days of the event. SEM/EDX was used to determine that the contamination was composed of numerous elements (e.g., Na, Mg, S, K, Ca, Cr, Fe, Ni, Cu, Zn, Cd, Sn, Ba, Pb, Bi) including several from Table 4. The Principal Investigator (PI) used this information to decide that this contamination posed an unacceptable risk to sample science. Since this was irreparable damage, the backup detector had to be used on REXIS instead (Masterson et al. 2017).

[Insert Figure 12 here]

To help determine the sources of collected contaminants, selected sample return capsule materials, purge filters, and gloves used in the ATLO facilities have been archived and will be distributed for analysis in parallel with samples of Bennu, as requested. Finally, samples of the spacecraft monopropellant (high-purity anhydrous aniline-free hydrazine), gas used for sample collection, and cleanroom air samples were collected and analyzed for trace volatile organics before and after launch.

## 5.1 Contamination Knowledge Plates

During spacecraft assembly, the curators and other science team members worked with the mission engineers and ATLO personnel to archive materials from the spacecraft, and to monitor cleanliness levels in the LM and KSC cleanrooms through deployment of Si wafer and Al foil witness plates (Figure 12). To minimize particle loss during shipping, a pair of plates was hand-carried to JSC. The collection of archived items and witness plates are stored in a dedicated stainless steel nitrogen-purged cabinet in Class ISO 7 cleanroom at JSC.

Each contamination knowledge plate exposed four Si and four Al surfaces; three of each (75%) were archived to be inspected later in parallel with the returned asteroid samples; the remainder was analyzed to provide relatively prompt information on the contamination environment of the spacecraft assembly facility. Thousands of particles were examined by SEM for size, texture, and bulk elemental abundances. This work served the long-term need of assessing the contamination background that will be important for interpreting returned sample measurements. But these studies were also carried out within 1–2 weeks of delivery to JSC so that unexpected contaminants that could pose unacceptable science risks could be identified in time to mitigate the issue. This approach also protected the ATLO schedule from delays associated with the scientific investigations of contamination. This reporting structure, however, also allowed the Principal Investigator the ability to promptly review contamination knowledge data to determine if an interruption in ATLO was warranted (Figure 13). In addition, all reports were shared with the contamination engineers and placed on the internal science team website for any member of the team to review.

[Insert Figure 13 here]

The design of the contamination knowledge program allowed the analytical arsenal of the OSIRIS-REx scientists to be engaged to study samples as necessary. Given the complexity, time required, and cost of some analyses, they were not to be used unless a previous test indicated a need (Figure 14). Due to the general high cleanliness of the samples, the most arduous techniques were not employed. However, all collected particles and 75% of the Al foils remain available for much more detailed analyses if necessary.

[Insert Figure 14 here]

### 5.2 Contamination Knowledge Plate Results

Each contamination knowledge plate was designed for easy subdivision for analysis by SEM/EDX on silicon wafers and organic analysis on aluminum foils. A detailed description of the results is outside the scope of this manuscript. However, some representative findings are below.

Contamination knowledge plate #4 was exposed in the OSIRIS-REx cleanroom June 12, 2015, to July 14, 2015, at LM. During this time OTES and the OVIRS were installed, and a number of power subsystems were tested on the spacecraft. SEM examination of one Si wafer from contamination knowledge plate #4 identified ~40 particles and particle groups (excluding Si particles from the mount) 1.5–32 µm in size when measured along the longest dimension (Figure 15). Most particles are carbonaceous material and metal/metal oxides that could be attributed to aluminum and stainless steel. One Pb-bearing brass 7 × 16 µm particle was identified. One siliceous mineral particle contained K. Three fiber-like particles were observed: one was C-rich, and two were Al-rich. Other than the Pb particle, these particle counts and compositions were acceptable, and there were generally no unexpected elements to invalidate the assumptions used to derive the level 100 particulate requirement, and the contamination level of the indicator elements (Table 4) were not violated. However, the Pb-bearing particle was of concern since Pb is a key element of scientific interest. Analysis of the Pb-bearing particle indicated a texture and elemental composition consistent with leaded-brass (Pb being a common additive to brass to improve machinability). After a review of drawings and discussions with the LM contamination engineers it was discovered that a brass set-screw was used adjacent to the contamination knowledge plate. This screw was removed. Since it was relatively far from the spacecraft, the team has confidence that no Pb-bearing brass particles found their way to the spacecraft, let alone into the sample-collection hardware.

[Insert Figure 15 here]

In parallel, one aluminum foil from contamination knowledge plate #4 was analyzed for organic compounds. The contamination monitoring amino acid analysis showed this to be the dirtiest single exposure during ATLO (Figure 6). It was an early and busy period in the cleanroom (which was also shared with the NASA InSight spacecraft at the time), and the highest 1-month quantity of total amino acids were detected (9.8 ng/cm$^2$). It is unclear if the source of this higher level of contamination was the level of activity, the time it took personnel to learn the new procedures, or another source. Contamination knowledge amino acid analysis agreed qualitatively with contamination control analysis and confirmed that the glutamic acid detected was exclusively the L-enantiomer dominant in biology, using the derivatization and workup of Glavin et al. (2006; 2010) as previously described. (Note, however, the LCMS analysis was performed on a different Waters® ACQUITY™ coupled to a Waters® Xevo™ quadrupole-time of flight mass spectrometer, as the Waters® LCT Premier™ time of flight mass spectrometer was occupied with AccQ•Tag™ analyses.) A sample of foil which was not water extracted was also analyzed by pyrolysis GCMS (CDS Analytical Pyroprobe 5200 fed into a Thermo Scientific™ Trace gas chromatograph coupled with a Thermo Scientific™ DSQII™ quadrupole mass spectrometer) using a Restek RT-Q-Bond®, 30-meter, 0.25-mm internal diameter, 8-µm $d_f$ column to allow for the analysis of small volatile compounds. GC flow rate was 1.5 mL/min in the constant flow mode. The temperature program was 50°C–250°C at 10°C/min with a 20-min final hold time. The quadrupole mass analyzer was scanned from 20 to 500 m/z. A procedural blank foil was analyzed before each sample. A number of small organics were observed (methanol, acetaldehyde, 1-butene, propenal, acetone, cyclopentane, 1-hexene, benzene, and 1-heptene). All of these species were also seen in the blank, but at lower abundances, and no compounds not also detected in the blank were observed. It is therefore concluded that these highly volatile compounds were more representative of the laboratory environment where the analyses were made than ATLO exposure. DART™-MS analysis of the extract was indistinguishable from a procedural blank consistent with a very clean sample.

### 5.3 Microbial DNA Analysis Results

Evaluations of cleanrooms have revealed that, while they are generally low in microbial number, there is substantial diversity, often with unique extremophiles represented (Mahnert et al. 2015). The team performed a single spot check of one cleanroom; a more thorough study is planned for the future from archived contamination knowledge plates. To identify potentially contaminating microorganisms the team assessed via 16S and ITS metagenomic sequencing a sample of the Al foil from contamination knowledge plate #8 exposed in an ISO 8 cleanroom during vibration testing. While in the ISO 8 cleanroom, the instruments and sensitive hardware were bagged, and the spacecraft was spot-cleaned after testing, so the spacecraft should have a lower level of contamination than experienced by the contamination knowledge plate exposed to the room.

To gauge microbial diversity, DNA was extracted from a polyester swab (Puritan) used to collect a surface sample from the knowledge plate and molecular biology grade water in which the plate had been submerged with continuous vortexing for 5 minutes following swab collection. DNA extraction was carried out via a combination of custom and kit methodologies. Custom extraction involved processing the swab tip with a Mini-Beadbeater (BioSpec Products) and subsequent DNA collection and cleanup with the QIAamp BiOstic Bacteremia DNA Kit (Qiagen). DNA was extracted from water with the DNeasy PowerWater Kit (Qiagen). An identical swab tip and aliquot of molecular grade water were also processed in parallel with accompanying reagent and standard negative controls. DNA concentration was determined with a Qubit fluorimeter (ThermoFisher Scientific™). The extracts were amplified with 16S primers for bacteria and archaea (515F-806R and 27Fmod-519Rmod) and ITS primers for fungi (ITS1F-ITS2R) with barcodes attached to the forward primer. Prior to library preparation, the amplified products were pooled and purified using Agencourt® AMPure® XP beads (Beckman Coulter®). Illumina® library preparation and sequencing with the MiSeq platform followed the manufacturer's recommended protocols (Illumina®). Operational taxonomic units (OTUs) were generated from the resulting paired-end sequence data after it was joined, and barcodes, ambiguous base calls, and sequences <150 bp were removed. The OTUs were further defined by clustering at 3% divergence threshold. UCIHIME was used to remove chimeras (Edgar et al. 2011). Taxonomic classifications were generated using BLASTn against curated databases resulting from GreenGenes (http://greengenes.lbl.gov/cgi-bin/nph-index.cgi), RDPII (http://rdp.cme.msu.edu/), and NCBI (https://www.ncbi.nlm.nih.gov/). Sequences identified in the control samples were subtracted from the knowledge plate samples, ensuring that the microorganisms identified were unique to the knowledge plate. A summary of the results is shown in Table 12.

[Insert Table 12 here]

While a complete microbial census of the cleanroom was not carried out, 16S rRNA gene signatures from knowledge plate #8 revealed a pattern of microbial diversity consistent with full-scale assessments (La Duc et al. 2012). The majority of OTUs belong to microorganisms that are human-associated or common in the environment. However, sampling the small surface area of the knowledge plate did reveal the presence of organisms with increased capabilities of survival under extreme conditions (e.g., *H. werneckii* and *N. amylolyticus*). As the stringent cleanliness standards governing cleanrooms often selects for these types of microbes (Mahnert et al. 2015), it is these characteristics of persistence that are of utmost concern to planetary protection officials (e.g., Smith et al. 2017), but of less importance for contamination control. As such, a comprehensive microbial evaluation of the remaining knowledge plates is planned, as it may be useful for future missions with stricter planetary protection requirements than OSIRIS-REx.

## 5.4 Gas Analysis Results

Contamination knowledge analysis of TAGSAM gas, purge, and air samples (Table 13) was done by the JSC Toxicology and Environmental Chemistry group with the same type of canisters (Figure 16) and the same target analytes as was done for Stardust (Sandford et al. 2010). GCMS measurements of one hundred target volatile organics typically showed only low levels of acetaldehyde (~0.03 ppm), acetone (0.04 ppm), and more 2-propanol (~0.1 ppm) in air – 2-propanol was used as part of the cleaning process of the test hardware, and as a wipe for gloves used in the cleanrooms. For most of the collected gas samples, all other target molecules were below ~1 ppm detection limits.

The gas to be loaded into the TAGSAM bottles, which was collected for analysis directly from the manufacturer, showed no compounds above detection limits (~0.01 ppm). Analysis of the gas when collected through the flight-loading manifold found 0.02 ppm of acetone and 0.6 ppm of 2-propanol. After an engineering model TAGSAM bottle was loaded with collection gas, then heated at 40-45°C for 24 hours, the gas showed 0.01 ppm of acetone and 1.1 ppm of 2-propanol. These trace contaminants likely arose from the cleaning of the bottle and manifold.

[Insert Table 13 here]

[Insert Figure 16 here]

A potential contamination risk was identified in the way the high-pressure 95:5 $N_2$:He gas is injected into the asteroid surface during the TAG event. The pressurized TAGSAM collection gas is released by firing a NASA standard initiator (NSI) pyrovalve. The gas is directed into the asteroid surface through a short 316L stainless steel convoluted tube that connects the gas bottle to the TAGSAM head. The firing of a pyrovalve produces particulate debris and combustion byproducts that may be entrained in the gas flow. However, there are very few published reports on the nature of pyrovalve "blowby," and it is very likely that the nature and abundance of blowby materials is highly dependent upon the particular pyrovalve used, associated plumbing, and the composition of pyrotechnic initiator/booster charge and its decomposition products (Bement, 1997). One combustion modeling study identified over 40 chemical compounds produced during the pyrotechnic detonation (Woods et al., 2008). In addition, the high-pressure gas mobilizes particles from the pyrotechnic initiator, valve material and plumbing interior surfaces (Groethe et al., 2008). We selected a pyrovalve, however, which mitigates the blowby of combustion products by formation of a metal-metal seal during the pyrotechnic event. This impulsive sealing event causes fractures which were found to release some particulates into the gas stream.

To assess the potential contamination of the pyro device, the team carried out a test firing of a NSI connected to a high-purity TAGSAM gas bottle to collect and characterize particles and volatiles entrained in the gas for contamination knowledge. The NSI and gas bottle were flight spares, and the associated plumbing system was composed of stainless steel but was not a flight-like configuration for this test. The gas was directed through a PFTE filter and collected in a 6-L canister that had been provided by the JSC Toxicology and Environmental Chemistry group. The canister preparation was similar to those used for the other gas analyses, but was larger to partially accommodate the high-pressure gas from the TAGSAM bottle (the total gas loaded was less than half the actual TAGSAM bottle pressure). The collected gas was analyzed for the same suite of 100 species targeted for all other OSIRIS-REx gas analyses. As with the contamination knowledge plates, the PFTE filter was inspected by optical microscopy and SEM/EDX to determine the compositions of the impacting particles.

The collected gas showed no combustion byproducts but higher levels of the same compounds attributed to pre-cleaning of the gas manifold: 0.06 ppm acetone, 3.3 ppm 2-propanol, and a trace (~0.01 ppm) acetaldehyde. Again, these are the same species found from cleaning procedures and unlikely from blowby.

Visual inspection of the PFTE filter showed much more particulate debris than expected. The filter was penetrated by a number of large (hundreds of μm) impactors, including two recovered metal grains larger than 1 mm. Numerous 1- to 10-μm-sized particles were also found embedded in the filter surface, but they comprise a small fraction of the particulate mass. Examination by SEM/EDX showed that the largest metal fragment and a subsampling of the small particles on the PFTE filter are composed of stainless steel (~99% by number).

Far more particulate was generated in the NSI firing than anticipated. However, the particles are of uniform composition, the particles from this test and an identical NSI are available for study from the materials archive, and the absence of pyro gas contamination should simplify the task of background correction during returned sample analysis. The TAGSAM blowby test showed the importance of not making assumptions about the nature of contaminants and the valuable role that sample analysis can play in guiding mission operations and design.

## 5.5 Monopropellant Analysis Results

To better understand the potential impact of any impurities in hydrazine on the sample, the team collected flight samples of aniline-free ultrahigh-purity flight hydrazine monopropellant during spacecraft fuel loading (Figure 17). Since one complete thruster was archived, it will be possible to recreate the monopropellant as flown and perform any needed experiments with the spare thruster and effectively identical fuel if future scientific results require better knowledge of potential contamination caused by the monopropellant.

Two 125-mL samples were collected in cleaned glass bottles with PFTE caps cleaned to IEST (Institute of Environmental Sciences and Technology) level 25 A. Level 25 A is the cleanliness level used for standard NASA Kennedy Space Center (KSC) propellant analyses. One set of samples collected was dried in PFTE beakers; these are archived at JSC for future inorganic analysis. The second sample of liquid hydrazine was sent for organic and stable isotopic analysis at GSFC. In addition, an identically cleaned bottle was sent and filled with Millipore water to serve as an organic blank.

[Insert Figure 17 here]

The stable nitrogen and hydrogen isotopic compositions of hydrazine propellant were measured to be $\delta^{15}N_{AIR}$ = +4.7 ± 1.5‰ and $\delta D_{VSMOW}$ = +154 ± 23‰. The $\delta^{15}N$ analysis was carried out using a Costech ECS 4010 combustion elemental analyzer (EA) connected through a Thermo Finnigan™ Conflo III interface to a Thermo Finnigan™ MAT 253 isotope ratio mass spectrometer (IRMS). Seven tin cups were individually loaded with ~0.2 µL of hydrazine and were then sealed and introduced to the EA-IRMS for analysis through the zero-blank autosampler of the EA. Three cups containing solid L-alanine of known isotopic composition ($\delta^{15}N_{AIR}$ = –5.56‰, Iso-Analytical) were also analyzed in order to calibrate the measured isotopic values. For δD analyses, 0.5 µL of a 1:50 hydrazine:1,4-dioxane solution were injected into a Thermo Scientific™ Trace™ GC whose output is split, with approximately 10% directed into a Thermo Scientific™ DSQII™ electron-impact quadrupole MS that provides mass and structural information for each eluting peak. The remaining ~90% passes through a Thermo Finnigan™ GC-TC interface, where amino acids are quantitatively pyrolyzed to hydrogen gas, and into the Thermo Finnigan™ MAT 253 IRMS for δD analyses. High-purity $H_2$ ($\delta D_{VSMOW}$ = +75.2‰, Oztech) was introduced through the dual inlet of the IRMS and used as a reference gas, while a solution of biphenyl of known isotopic composition ($\delta D_{VSMOW}$ = –41.2‰; Indiana

University) was injected through the GC-IRMS and used for isotopic calibration. The GC was outfitted with a Restek 30-m Rxi®-5mx column, and a flow rate of 0.5 mL/min was used. The following oven program was used for hydrazine analysis: splitless injector held at 220°C, initial oven temperature of 45°C held for 6 minutes, ramped at 5°C/min to 65°C, ramped at 30°C/min to 300°C, held for 3 minutes.

GCMS analysis showed only hydrazine with all trace species consistent with column bleed or column stationary phase. Amino acid analyses of lyophilized hydrazine showed 0.01 ng/g β-alanine and 0.45 ng/g γ-amino-*n*-butyric acid. These are very close to blank levels and may derive from contamination sublimating into the hydrazine during workup. Diluted hydrazine was infused into a Thermo Scientific™ LTQ Orbitrap XL™ hybrid mass spectrometer using positive ion mode electrospray, but no masses, other than hydrazine, were observed which were not also present in a base extraction of the infusion capillary tube.

# 6 Materials Archive

Hardware and process coupons of materials that have plausible access to the sample or were of contamination concern due to the materials involved were archived at JSC in six dedicated nitrogen-purged cabinet desiccator boxes housed in an ISO 7 cleanroom. Prior to the start of this archive, the team monitored the background cleanliness levels in the archiving cabinet. They deployed witness plates in an empty desiccator purged with curation grade nitrogen over a one-year period. A total of eight witness plates (10 $cm^2$ aluminum foils) were initially deployed and collected at the following exposure times: 1, 2, 5, 14, 28, 60, 120, and 365 days. The samples (along with a blank) were sent after collection to GSFC for analysis. Analyses as previously described were conducted on the foils: LCMS (Glavin et al. 2008); derivatization GCMS (Mawhinney et al. 1986); pyrolysis GCMS; DART™-MS; as well as ATP luminosity analysis for cell counts (PallChek™ Rapid Microbiological System). The LCMS data showed that with the exception of the Day 1 sample (which was contaminated during handling and workup) all other samples had 0.05–0.8 $ng/cm^2$. The lowest and highest abundances were in the 365 day and 120 day exposures, respectively. So total abundance is not an accumulation of material over time. The GCMS analysis showed a steady increase of volatile compound buildup on the witness plates over the course of the experiment, but all were at very small amounts, usually equal to or less than the amount of volatiles found in the analytical laboratory background. The DART™-MS analyses of the acid-hydrolyzed extracts were all similar in appearance over the time course study. Finally, the luminosity analysis of a 5 $cm^2$ foil found that the samples were below the limit of detection for the luminometer (1 fmol ATP or ~1000 "typical" bacterial cells). While luminometer analyses are not intended to be definitive assessments of bioburden, it is interesting that the luminometer results suggest <200 cells/$cm^2$ in all samples. If all the amino acids were derived from cells, then ~5000 cells/$cm^2$ would have been expected (Neidhardt et al. 1990); thus the source is unlikely to be dominated by viable cells.

The results of this year-long monitoring showed that the curation cabinet was very clean and that buildup of volatile organic compounds was at levels at or below background or blank in the analytical labs, and thus ready to receive samples.

Archiving began in February 2014, with the reception of the first item in the collection—lubricant used on the OTES rotary actuator. As SRC and TAGSAM were built from March 2014 until their availability for ATLO and integration in summer of 2015, items were obtained and sent to JSC for archiving. Additionally, as various instruments were assembled and readied for integration, the instrument providers identified and packaged coupons to send to JSC for the archive. Finally, as instruments and subassemblies of the spacecraft were tested and integrated, coupons and items were continuously sent to JSC through integration at KSC, finishing ~90 days after launch with the final archived items being related to launch operations. In total, 395 items were received for the materials archive.

As previously described, through the ATLO process (from March 2015 until late August 2016) Si wafer and Al foil witness plates were deployed in the various cleanrooms in LM and KSC. These plates each contained four Si wafers and four Al foils, with one of each type per plate analyzed and the remainder being archived; 128 individual witness plates (64 Si wafers and 64 Al foils) were collected in total.

Key summary information for each archived item is presented in an online catalog that will be accessible via https://curator.jsc.nasa.gov/ prior to sample return. Each catalog entry for coupons and hardware lists the material, its location on the spacecraft (e.g., SRC, TAGSAM, spacecraft, instrument, or launch operations), a description of that item (including weight or dimensions), the company that made the item and its webpage or other contact information, the archiving location, archiver, archive date, part number, and photo or drawing of the location on the spacecraft. The materials can be grouped into several general categories including metals (stainless steel, aluminum, titanium alloys, BeCu alloy, and the brass set screw discussed previously), epoxies, paints, polymers, lubricants, sapphire, and a lengthy list of miscellaneous and support materials such as gloves, tape, and bags. A sample from the cold plate during thermal-vacuum testing was also archived; this provides a worst-case spacecraft-wide average of NVR. Materials involved in potential contamination events (e.g., the brass screw revealed by contamination knowledge plate 4), were also archived. A brief summary of materials presented by spacecraft component or location of origin and material type is presented in Figure 18. The list of materials in the catalog is in the Online Supplemental Material 2.

[Insert Figure 18 here]

# 7 Flight System

Once a spacecraft leaves Earth, additional cleaning and testing is impossible. However, the team needs a method to gain contamination knowledge of the state of the sampling system and provide contamination knowledge for the sample's return to Earth in 2023. Thus, contamination control and knowledge also extended to aspects of the design of the OSIRIS-REx flight system. Naturally, these design specifications could not be allowed to increase the risk to the spacecraft nor cause harm to the sample. Due to cost and complexity concerns, it was decided not to include active contamination monitoring (e.g., a spacecraft mass spectrometer, pressure gauge, or quartz crystal microbalance).

## 7.1 Flight Witness Plates

The most cost-effective method of contamination monitoring is a laboratory analysis of returned blanks or control samples in the form of witness plates. The first decision is the composition of the witnesses, which is a compromise between science and engineering. Based on the recommendation of Sandford et al. (2010), the team required two different materials. One should be similar to the sampling system to serve as a good proxy to the surfaces that could collect contamination, and the other should be chemically similar, but distinct from the sample. Electrical conductivity of one set of plates is desirable to facilitate electron-beam analyses.

The team considered aluminum, gold, titanium nitride, and silicon for the conductive material. The sampler is primarily 6061 aluminum alloy, so high-purity aluminum was used instead of an alloy to simplify mass spectral analyses. Aluminum is monoisotopic, which results in less interference in some analyses. Yet pure aluminum rapidly forms a dielectric oxide surface coating which is very hard to clean of particulates; this made the precision cleaning of these components more laborious. However, since our sampling system is likely to gouge the sampler and make aluminum debris, aluminum flakes are expected, well understood, and easily compensated for in analyses.

For the other witness, the team discussed zeolites, Tenax or related resins, and other adsorbents. Anything particulate was rejected for foreign object debris concerns and the risk of it contaminating the sample. Based on meteorite studies, the sample is expected to contain ferromagnesian silicates, so the scientists opted to use quartz, as it is unlikely to be in the sample. However, there was an engineering concern that a quartz plate could shatter and damage mechanisms. Instead, the team chose to use sapphire ($Al_2O_3$), as flown successfully on NASA's Stardust mission. This is an example where spacecraft safety tipped the balance between engineering and science recommendations.

Each plate is a monolith, and the thickness of each plate is unique (similar to what was done on NASA's Genesis mission (Burnett 2013), meaning that plate identity can be verified by measurement if there is a mix-up or breakage. Sandford et al. (2010) also advised that the witnesses should be prepared in a way that they can be easily subsampled. However, pre-scoring the witnesses was an engineering concern due to possible breakage in space. Instead, the exposed surface of the sapphire was diamond abraded (and cleaned and verified as diamond-free by FTIR). This provides a unique signature of the exposed surface, allowing the witness to be shattered for distribution and allowing the curators to select exterior pieces, which record exposure. This rough surface also provides a modest increase in surface area, but prevents reflectance spectroscopy. It was later determined that the witness plates on the TAGSAM head create a glint into SamCam of the OSIRIS-REx Camera Suite (OCAMS), so both the sapphire and aluminum witness plates had to be abraded to mitigate this problem. Based on the sensitivity requirements of the techniques expected to be used for sample analysis, and reserving 75% to archive, the team needs a minimum of 10 $cm^2$ surface area on each of the two types of witnesses.

The simplest case would be to fly a single pair of passive witness plates. However, witness plates cannot establish the direction of molecular flow. This means that it would be impossible to determine if a compound found on a witness plate that was exposed to both the sample and the spacecraft is extraterrestrial, contamination, or both. For a witness plate to be scientifically useful, it must have the same history as the sample collector, with the exception of the presence of sample. This means that the witnesses must be physically close to the sample, but cannot be contaminated by the sample. For OSIRIS-REx, the sample is exposed to two different environments: the TAGSAM head prior to collection, which is exposed to the inside of the launch container; and the spacecraft and the TAGSAM head during and after collection, which is exposed to the spacecraft and the inside of the SRC canister.

If asteroid material outgases onto a witness plate or sheds dust onto it, it may be impossible to determine if the analyte on the witness plate is sample or contamination. This renders the witness plate useless. Thus, it is necessary to protect the witness plate from the sample with a physical barrier to preserve the reconstruction of the contamination history of the sample by comparing witness plates. Ideally, this can be accomplished with a witness plate on TAGSAM that is exposed until the TAG event (α), a witness plate on TAGSAM that is exposed at the moment of the TAG event (β), and a witness plate in the SRC that is not exposed until the SRC is opened in proximity to the sample (γ). Thus, contaminant would then be materials found in $\alpha - \beta + \gamma$.

However, for the purpose of simplicity and cost effectiveness, it is important that any required movements of witness plate covers are leveraged off of other spacecraft actions. The lack of dedicated witness plate motors means that the exposure of witnesses is dependent upon other spacecraft actions; this left a gap in the exposures, so an additional witness plate is continuously exposed to span the gap in time. The arrangement and exposure sequence of the flight witness plates is shown in Figure 19. The witness plates close via spring actuation and are not hermetically sealed, but the material captured on the witness is preserved via a tortuous path seal. The timeline on the bottom panel is schematic and not to scale, but the gap in exposure of b to c is driven by the ejection of the launch cover prior to TAG, and the gap in exposure between 2 and 3 is driven by the verification of TAGSAM stowage in the SRC. More detailed timing is described in Williams et al. (2017) and Beirhaus et al. (2017).

[Insert Figure 19 here]

It is likely that sample dust will be shaken loose from the TAGSAM head inside the SRC canister. Thus, it is possible that particles could be ground into a witness plate. A screen is placed over half of the witness plates that will have potential exposure to regolith (witnesses c, 2, 3 shown in Figure 19) to minimize the amount of regolith allowed to touch the witness surface, while still permitting volatiles to encounter the witness. A 400-mesh (37-µm) screen is permeable enough to allow contamination gases to pass, but should prevent significant quantities of regolith from being ground into the witness plates during Earth descent and landing (EDL) when there is maximum mechanical stress on the sample in the SRC.

### 7.2 Sample Return Canister Air Filter

The SRC is not hermetically sealed, but allows gas to exit during launch and enter during EDL. All this gas flow is directed through the SRC Sample Canister Air Filter, which prevents gas and particulate contaminants from entering the sample canister.

In addition, the filter could also capture asteroid-derived volatiles evolving from the sample after SRC closure. Any outgassing that occurs from the collected samples in TAGSAM after it is stowed in the sample canister could result in the deposition of escaping volatiles on the inside of the sample canister or on the avionics deck or in the filter. Any areas in the enclosed sample canister/TAGSAM/avionics-deck volume that are on average cooler could serve as cold traps that concentrate these volatiles. After reentry and recovery of the capsule and extraction of the sample canister at the Utah Test and Training Range, a $N_2$ gas purge of the canister will be started through the canister septum. The resulting flow of air will exit through the canister air filter, and this will encourage any volatiles located in the canister into the filter. Thus, if the TAGSAM contains volatiles that can outgas from the collected samples and cold trap within the canister, it is in the air filter where there is the best chance of detecting them.

The Sample Canister Filter used on the OSIRIS-REx Sample Canister is a nearly identical copy of the filter used on the Stardust spacecraft and consists of a structure containing alternating layers of filtrette material and active absorbing materials. The location of the filter is indicated in Figure 19, and Figure 20 shows a schematic of the filter's cross section. Since the OSIRIS-REx filter is nearly identical to the filter used by Stardust, performance testing the OSIRIS-REx filter was modeled on the procedures used for testing the Stardust filter (Tsou et al. 2003) to test the efficiency of the filter at capturing various organic gases, water vapor, and particulates. Three filters identical to the flight unit were tested for their organic, moisture, and particulate performance.

[Insert Figure 20 here]

### 7.3 SRC Filter Efficiency for Organics

The filter's ability to capture a variety of contaminant gases was tested. Tests were made using a specially designed apparatus that allowed controlled gas flow through the filter and that allowed the filter to be degassed in a vacuum prior to the test, as would be the case for the flight filter, at NASA Ames Research Center (ARC). A 2-liter gas bulb containing 1230 mbar $N_2$, 7 mbar ethanol, 7 mbar acetone, 7 mbar hexane, 7 mbar benzene, and 2 mbar CO was prepared and mounted on the inlet side of the filter (SRC exterior). A second evacuated 2-liter receiving bulb was placed on the opposite side of the filter so that it could capture any gases passing through the filter. The filter was then pumped on for an extended period of time. As expected, the filter initially pumped down at a nearly exponential rate, but slowed as the filter degassed. It took several days for the filter to approach the ambient pressure of $\sim 1 \times 10^{-5}$ mbar of the vacuum system (Figure 21).

[Insert Figure 21 here]

Once the filter had been largely degassed, the test gases were flowed through the filter using a flow rate and duration like that expected for the reentry of the SRC (vacuum to 1 atm in 10 minutes). The original test gas bulb and receiver bulb were then removed from the apparatus so the composition of the gases in each could be measured and compared. Each bulb was used to deposit sample gas onto a CsI window cooled to 10 K in the vacuum chamber of a FTIR equipped cryo-vacuum system. The infrared spectrum of the resulting mixed-molecular ice was then obtained, and the positions and strengths of any absorption bands detected were measured (Allamandola 1984; Hudgins et al. 1994). The measured band strengths of individual molecular species in the samples were then used to compare the filtered and unfiltered gases to determine how efficiently the filter stopped individual molecular components of the original gas mixture.

Figure 22 and Table 14 show the results of this test. In most respects the results of this filtering test are very similar to those seen for the Stardust test filters (Tsou et al. 2003). In virtually all cases where absorption bands could be detected in both the unfiltered and filtered samples, the filter stopped ~99+% of the ethanol, hexane, acetone, and benzene. CO is filtered with lower efficiency, but was included in the test as a calibration tracer and is not considered to be an issue as a sample contaminant. Nonetheless, it is interesting to note that the OSIRIS-REx filter stopped the CO with better efficiency than the Stardust filters (passing 16% versus >70%).

[Insert Figure 22 here]

[Insert Table 14 here]

Since the flight filter is expected to be heated by conduction from the SRC heatshield (heat soak) following the completion of reentry, the filter's ability to trap and hold contaminant gases when heated was tested. After the filter was used for the gas trapping test, it was briefly pumped to $<10^{-3}$ mbar followed by testing at the above conditions, first at 50°C then at 70°C at similar timescales to the heat soak of the returned SRC (peak temperatures at 20 and 30 minutes, respectively). These temperatures are the expected and maximum temperature of the SRC interior. Figure 23 shows the infrared spectra of the condensed gases that escaped the tested filter during the two heat soak tests.

[Insert Figure 23 here]

The dominant absorption features are due to $H_2O$ and $CO_2$, molecules that were not components of the original test gas. These species indicate that the heat soak liberated $H_2O$ and $CO_2$ that were present in the filter prior to the original trapping test. Since the SRC (and filter) were subjected to spacecraft thermal-vacuum testing, higher temperatures may be needed to liberate these atmospheric gases in the flight filter compared to this test.

## 7.4 SRC Filter Efficiency for Water Vapor

As in Tsou et al. (2003), another flight-design air filter was tested for humidity-trapping efficiency at JSC. The filter received a bakeout and dry air purge before each test to ensure the test started with a dry filter. Room-temperature air samples with both 90% and 40% humidity were tested at a 1.5-L/min flow rate. The flow rate was calibrated against Dry-Cal DC Lite primary flow meter and ranged from 1.48 to 1.52 over five readings within a 30-second period. Humid air was supplied and hydrated via a bubbler (Figure 24), which was then sent through a Vaisala model HMT333 humidity sensor to measure inlet humidity, then sent through the filter in the test housing, and then through a flow controller and into an outlet humidity sensor.

[Insert Figure 24 here]

The 90% test showed an initial output humidity which stayed very low, indicating the filter has high efficiency over the first ~20 minutes of testing. For the first 20 L (the approximate volume of the Sample Canister) of humid air that passed through the filter, the outlet humidity stayed below 10% (Figure 25, Table 15). The total duration of the test was 225 minutes, during which humidity on the outlet side of the filter reached a maximum of 73%. The 40% humidity test was carried out using exactly the same procedures. Nearly 22 L of air passed through the filter before the efficiency dropped below 90%. It took ~40 minutes for the relative humidity of the output air to increase above 10%, so the filter's ability to keep outlet relative humidity <10% was once again achieved well after the ~20-liter capacity of the SRC was attained. These results are comparable to or better than those for the Stardust filters (Tsou et al. 2003).

[Insert Figure 25 here]

[Insert Table 15 here]

## 7.5 SRC Filter Efficiency for Particulates

A separate OSIRIS-REx filter was tested for its ability to trap particulates. This test used a P-Trak Model 8525 (TSI Inc.) ultrafine particle counter with a probe that provided a 0.6-L/min flow rate to measure particles in the size range of 0.02 to 1 µm. Initially, measurements of unfiltered air were made at JSC in three different environments: (a) the interior of Building 229, where particle counts were ~3,000 counts/cm$^3$; (b) the exterior of building 229, where particle counts were ~25,500 counts/cm$^3$; and (c) with a candle smoke source with particle counts of 100,000 to 300,000 counts/cm$^3$. After each of these measurements, a new set of measurements was made of air run through the OSIRIS-REx test filter. Filtered readings for a, b, and c were 4 to 23 counts/cm$^3$ (over a 25-minute period of time), 20 to 46 counts/cm$^3$ over a 5-minute period of time, and 4 to 12 counts/cm$^3$ over a 10-minute period of time. The results are shown in Table 16.

[Insert Table 16 here]

The trapping efficiency of Stardust filters was measured for particles in 0.3- to 0.5-µm range to be 99.9% or better, and the efficiency was greater than this for larger particles. Our results extend to smaller particle sizes and are comparable to or better than those for the Stardust filters. In particular, the Stardust results on cigarette smoke, which is dominated by small particles (<1 µm), show 99.9% efficiency compared to the OSIRIS-REx results on comparably sized candle smoke (<<1 µm) that show 99.996% efficiency. Again, these results compare favorably with results from Stardust test filters (Tsou et al. 2003).

## 7.6 Analysis of the Returned Air Filter

The team will further improve our knowledge of the degree of sample contamination by analyzing the SRC air filter after the SRC is recovered. It will be important to analyze each of the layers of the filter independently, since gradients in detected molecules and particulates within the vertical structure of the filter will provide information concerning whether any filter-trapped materials were leaving the canister (escaping asteroidal materials) or entering the canister (external contaminants).

Furthermore, as with Stardust (Sandford et al. 2010), the team will collect gas, soil, and related samples from the SRC recovery site and purge the SRC with $N_2$ upon recovery. Samples found in the SRC can then be compared with these materials to ascertain whether or not they are contaminants associated with recovery of the SRC.

# 8 Launch Vehicle

Perhaps the most hectic and critical periods in ATLO are the final preparations for launch and the launch itself. This period also has the most number of organizations working together; the PI and project office, LM, KSC's Launch Service Program personnel assigned to the OSIRIS-REx mission, the launch service provider (ULA) personnel, and Eastern Range personnel. Each has its own bureaucracy and culture. To better unite the team, the Principal Investigator and the Project Scientist had casual conversations and gave presentations to the technicians so they would better understand the importance of the OSIRIS-REx mission and the rationale behind the atypical contamination requirements that impacted their activities. This gave them ownership in the mission and encouraged them to rethink their process from a contamination perspective—proactively addressing any concerns that came up. It was also helpful to explain the rationale behind the contamination requirements to the stakeholders and vendors to improve compliance and encourage suggestions for solving issues.

Launch site activities start in the PHSF. One of the factors that led to the selection of the PHSF was that the air handling system for the building only serves one spacecraft. In a facility with a shared air system, an anomaly in one cleanroom can impact another. If the spacecraft generating the contamination is for a classified project, it could be difficult or impossible to obtain information about the event.

The OSIRIS-REx contamination requirements imposed changes to the spacecraft processing. Activities like mating of the spacecraft to the payload adapter and payload fairing encapsulation prior to transportation to the ULA Atlas V Vehicle Integration Facility (VIF) had never been done before. Furthermore, although the PHSF cleanroom is ISO 8, the facility needed to be able to maintain an ISO Class 7 for a short period of time for the final closeout of the SRC.

The PHSF cleanroom was tested for NVR and particulate contamination prior to spacecraft arrival, and a detailed crane inspection and facility walk-down were performed to ensure the facility was ready. Advanced preparation included collecting one month of NVR and particle fallout data. The facility's maintenance schedule included daily cleanings of the facility and garmenting to be consistent with a Class 7 cleanroom. Since transportation of the encapsulated spacecraft to the VIF required a ULA diesel truck, the exhaust was pumped away from the airlock via a positive flow snorkel. Hydrocarbons were monitored during a rehearsal and found to be sub-ppm. Schedule (and thus cost) was controlled during hazardous operations in the PHSF with a slight modification for access to the airlock for launch vehicle operations. This allowed for processing flight hardware in the airlock to prepare for spacecraft mate operations.

The use of a modified witness plate bracket (Figure 5) inside the fairing to capture additional amino acid data was new, and needed to be demonstrated to do no harm in the high airflow environment of the fairing. Since the VIF is not an environmentally controlled facility, clean enclosures were required for the four "boat tail" doors for access to the Centaur Equipment Module and spacecraft (Figure 26). Each enclosure was cleaned and certified to meet ISO Class 8 with ISO Class 6.7 air source, and only one tent door was permitted open at a time to maintain positive flow out of the fairing at all times.

[Insert Figure 26 here]

An additional two payload access doors above the spacecraft were included in the fairing. The purpose of the doors was to allow visual inspection in the event of a major anomaly. Opening the doors would degrade the cleanliness of the spacecraft, but could prevent the potential loss of the mission's launch period. Fortunately, the doors were never opened for an anomaly. When these upper doors were opened to apply sealant (which was subsequently archived) for closeout, all personnel and equipment remained >0.6 m from the opening, and the fairing airflow was set to maximum.

The Atlas V and VIF use nylon or suspected nylon components extensively. Eliminating these components was viewed as a significant risk to the launch vehicle performance. This risk was mitigated by covering the nylon parts and requiring glove changes whenever nylon was contacted. Since the amino acid contamination monitoring plate showed no evidence of nylon hydrolysis products, these measures appear to have been successful.

The last view of the spacecraft separating from the payload adaptor on the Atlas V Centaur shows numerous particles reflecting in the sunlight (Figure 27). These are likely ice and fragments from the separation of the payload launch adaptor, more sources of unavoidable particulates. The launch container and SRC filter and seals should have protected the TAGSAM and SRC interior from these. If this is ultimately found not to be the case, the contamination was captured and recorded on the flight witness plates.

[Insert Figure 27 here]

# 9 Conclusion

NASA's Viking landers were assembled at LM (then Martin Marietta); and every organic component, no matter its location, was analyzed by FTIR and MS and cataloged; and each entire lander was heat-sterilized. This level of contamination control (and planetary protection) is beyond the scope of a New Frontiers-class mission. Instead, OSIRIS-REx benefitted from previous missions' innovations, including the development of aniline-free hydrazine for Viking and the use of proxy materials as indicators of contamination. OSIRIS-REx offers lessons for future missions: the demonstration that amino acids can be controlled to such low levels in an industrial cleanroom, the use of the science team for contamination knowledge analyses, and the close interaction between engineers and scientists for contamination control.

The process of developing and implementing contamination requirements cannot start too early in the mission planning process, and must be maintained throughout implementation across the whole project. This required authority behind the requirement, in this case by making contamination a Mission Level 1 priority and having it be shepherded by project leadership.

Team communication was harder than expected—and it was already expected to be difficult from the start. Learning from our experience, future missions should allocate even more time and cost margin, with ready descopes for cost containment agreed upon. A unification of language is helpful, particularly to have the science team develop requirements in the language of the engineers and technicians responsible for implementing it. Likewise, it is important for the scientists to understand what is and is not possible and verifiable in a nonlaboratory environment. Just because a measurement can be made under ideal laboratory conditions does not mean it can be made under the schedule pressure and environment of ATLO. Close communication among the science, engineering, and management stakeholders is the best path to understand what changes are possible and reasonable. Communication lapses were evident late in the overall process, showing up as missed requirements. Most notably, rework was required since the science team writing the requirements did not include a prohibition on amorphous silicates, because the team did not know that such materials had an application in aerospace. It is also important that the people who write the requirements are involved in the implementation of the requirements. This allows for an understanding of the intent of the requirement in marginal conditions, prevents the creation of requirements that are impossible to implement, and prevents implementation that undermines the science behind the requirements.

There is a constant struggle between maintaining spacecraft and ATLO processing heritage and contamination requirements—especially novel ones. To compromise, the contamination engineers tried to be flexible, but that was sometimes interpreted at the working level as indecision or as an indication that all the requirements were not well thought out and arbitrary. The solution seems obvious—increase communication and inclusiveness across all those involved in making the mission a success, from scientists to managers to engineers to technicians, early and often. Yet communication via presentations and working groups cannot replace the timely production and approval of configuration managed documentation.

The use of contamination knowledge and the materials archive enabled considerable flexibility. High-heritage materials of contamination concern could be used, with contamination knowledge responsible for unraveling their impact. At the same time, ongoing analyses of the ATLO contamination monitoring and contamination knowledge plates both created confidence in the methods and allowed prompt reaction to anomalies and detection of trends.
Finally, the unexpected can happen (Figure 28), and having a committed, connected team with enough freedom and flexibility to act in the face of obstacles is crucial.

[Insert Figure 28 here]

# 10 Acknowledgments

This material is based upon work supported by the National Aeronautics and Space Administration under Contracts NNH09ZDA007O, NNG12FD66C, and NNM10AA11C issued through the New Frontiers Program. We wish to thank the hundreds of people and their families who labored and sacrificed to make OSIRIS-REx a reality.

## FIGURE CAPTIONS

**Fig. 1** The top (reporting in green) shows the organizational chart for contamination control (CC) and knowledge (CK). Direct lines of authority for contamination knowledge are through science and contamination control and materials and processes (M&P) are through engineering. Cross-communication (dashed lines) ensures information transfer across disciplines. The middle (document in orange) shows the written products generated by the element above. The Sample Analysis Working Group developed the contamination knowledge plan and the Curation Working Group developed the curation plan. These plans were synthesized into the archiving plan. The archiving plan was included as an appendix of the contamination engineering generated contamination control plan (Figure 2 and Supplemental Material S1) with input from contamination science (united in the CCWG). The materials engineering generated M&P plan was made with knowledge of the contamination control plan but without a direct reference. The bottom (implementation in blue) shows the different aspects of the OSIRIS-REx construction and test that used these documents. Note that the implementation exclusively relied on engineering documents.

**Fig. 2** The flow of requirement documentation from the NASA planetary protection and OSIRIS-REx Level 1 requirements to Level 2 documents (mission requirement documents (MRD) numbers in Table 7 are shown). These Level 2 documents are used by numerous Level 3 documents for the flight system, each instrument (OCAMS, OTES, OVIRS, OLA, and REXIS), launch service provider, and launch vehicle.

**Fig. 3** Setup for the transfer efficiency test with **(a)** 60g of fume and balls prior to cleaning. **(b)** TAGSAM EDU wrapped in Kapton on the shaker table.

**Fig. 4** Personnel properly gowned in the LM ISO 7 cleanroom and TAGSAM head protected by a PFTE bag.

**Fig. 5** Amino acid contamination monitoring plate adapted from a standard 1 ft. x 1 ft. ASTM E1235-12 NVR plate for the 4-m Atlas V fairing.

**Fig. 6** Total amino acid abundance on environmental monitoring plates in LM and KSC cleanrooms and Atlas V Large Payload Fairing (LPF). The blue line is exposure at the LM cleanrooms (ISO 7 and 8) (the gap is during the thermal-vacuum testing, when no monitoring

plates could be deployed). The pink dashed line indicates exposure in the KSC Payload Hazardous Servicing Facility (PHSF) cleanroom (ISO 8). The green dotted line indicates exposure inside the Atlas V fairing. Periods in ISO 8 cleanrooms show steeper slopes than periods in the ISO 7 cleanroom.**Fig. 7** Sampling configuration. This is the spacecraft configuration that introduces hydrazine contamination on the TAGSAM head (indicated).

**Fig. 8** Sample mass measurement configuration. The change in moment of inertia with the TAGSAM arm extended in two positions before and after sampling are measured while the spacecraft rotates about the indicated axis to determine the collected sample mass. No thrusters are used in this configuration, and thus these events are not contributors to hydrazine contamination.

**Fig. 9** An example output of the CFD code shows that the proximity of two thrusters (from the left) creates a nonuniform plume flowfield and may contribute to enhanced flux on the TAGSAM head that is not captured by scaling results from a single thruster. Plume color relates to plume speed, from low (blue) to high (red). The interaction between the two plumes can be seen in the slow region in the center.

**Fig. 10 (a)** Illustrates plume behavior in free space, while **(b)** illustrates plume behavior when at the asteroid surface. The interaction with the asteroid surface causes an enhancement of thruster plume deposition on the TAGSAM head relative to the free space geometry. Plume color relates to plume density, from low (blue) to high (red).

**Fig. 11 (a)** The isomers of OPA/NAC derivatized tripolyoxypropylenediamine in Bondline™ 6460 as seen in the single ion chromatogram centered at 452.2214 m/z. by LCMS. Individual isomers were not identified, but likely isomers are numbered, chromatographic conditions were not optimized. **(b)** Polyproplyenediamine trimer through heptamer plus additional larger species as seen by DART™-MS. **(c)** The comparatively simpler mix of compounds observed in EPO-TEK® 353ND as seen by DART™-MS.

**Fig. 12 (a)** Contamination knowledge plates consisted of precision cleaned silicon wafers mounted on SEM sample holders to collect particles and high-purity aluminum foils for organic NVR analysis. These were deployed in parallel with the contamination monitoring plates. Following one month of exposure, the entire unit was sealed in an aluminum housing bolted to the baseplate after exposure. **(b)** Location of contamination knowledge witness plate (in red circle) on shipping container base soon after arrival in the PHSF.

**Fig. 13** The pathway from hardware to analysis to a decision. The three types of generators of contamination control, contamination knowledge, and archiving samples are shown at the bottom. Materials follow the solid lines for their destination for analysis, line thickness schematically indicates the number of samples. Contamination control samples are sent directly to the analysts, except for hydrazine (monopropellant); all other samples are sent to curation for subdivision for archiving and distribution. Once the samples are analyzed the data (dotted lines) are sent to either sample science or contamination engineering for review. Science and engineering share results. Contamination engineering assesses the results and delivers the information to project management, who makes a recommendation to the principal investigator if a decision is required on if or how to mitigate off-nominal results. Sample science passes the contamination knowledge reports directly to the principal investigator for consideration. The hydrazine analysis was performed after launch for knowledge only.

**Fig. 14** The analytical flow of contamination knowledge plates allowed analyses by a comprehensive array of instruments and techniques available. After receipt at JSC, 75% of the samples are archived (thick line) to be available for parallel analysis with Bennu samples. The blue boxes with bold text show methods performed on each sample. The orange boxes with italic text show methods performed on a small subset of samples. The white boxes show methods that were available, but not employed. Microprobe two-step laser desorption/laser ionization mass spectrometry (MS) ($\mu$-L$^2$MS), X-ray absorption near edge structure (XANES), time-of-flight secondary ion MS (ToF-SIMS), transmission electron microscopy (TEM), electron microprobe (EMP), laser ablation inductively coupled plasma MS (LA-ICPMS), inductively coupled plasma MS (ICP-MS), ATP luminosity analysis (ATP), GC combustion isotope ratio MS (GC-IRMS).

**Fig. 15 (a)** Contamination knowledge plate #4 microscope image with locations and categories of particles analyzed by EDX indicated. **(b)** EDX spectrum of the Pb-bearing particle. **(c)** Elemental distribution by number of particles (Pb is a component of <1% of the total particles).

**Fig. 16** Gas sampling was performed using 500-mL evacuated containers. This collection is of the purge system for the truck transport of the fairing-encapsulated spacecraft from the PHSF to the launch complex.

**Fig. 17** Collection of flight hydrazine monopropellant for chemical and isotopic analysis.

**Fig. 18** The number of different items in the archive shown by origin or location on the spacecraft and by material type. Abundances are shown by number and color with darker blue indicating more samples. See Supplemental Material S2 for a complete list.

**Fig. 19** **(a)** Location of TAGSAM and **(b)** SRC flight witness plates and **(c)** the timing of their exposure. TAGSAM witness *i* is exposed continuously. TAGSAM witness *ii* is covered by spring-mounted seals when the head is removed from the launch container. The six TAGSAM witness *iii* are not visible in the image and are only exposed after TAGSAM arm separation when the TAGSAM head is seated in the SRC. SRC witness *1* is exposed continuously (but under the SRC rim and not visible in the figure). SRC witness *2* is exposed in the image but covered when the SRC is opened to accept the sample. SRC witness *3* is covered until it is exposed at the same time SRC witness *2* is covered. The SRC air filter is also indicated. Dashed arrows indicate that the witness is covered in the image.

**Fig. 20** The SRC filter consists of layers of filtrete material and activated carbon. The body is 6.35 cm outer diameter and 2.2 cm thick, the same design as used on Stardust (Tsou et al. 2003).

**Fig. 21** The dynamic pressure of the SRC filter decreased at a rate that diminished with time.

**Fig. 22** Comparison of the infrared spectra of unfiltered and filtered gases. The SRC filter trapped the vast majority of the introduced contaminant gases: ethanol (E), benzene (B), acetone (A), and hexane (H).

**Fig. 23** The FTIR spectra of the condensed gases liberated from the SRC filter by the heat soak tests.

**Fig. 24** Test stand used for the SRC filter humidity tests at the JSC Gas Laboratory for Analytical Chemistry.

**Fig. 25** SRC filter performance for the 90% (top) and 40% (bottom) humidity tests. The blue lines are the inlet humidity, orange is the outlet humidity, and black is the filter efficiency. Vertical dashed green line at 13.3 min corresponds to 20 liters of air passing through the filter.

**Fig. 26** OSIRIS-REx fairing “boat tail” doors and one clean tent are nearly ready to be removed for final closeout.

**Fig. 27** The OSIRIS-REx spacecraft separating from the Atlas V Centaur stage.

**Fig. 28** The launch pad (a) at Space Launch Complex 41 with the VIF (b) with OSIRIS-REx inside is seen through the smoke from the AMOS-6/SpaceX Falcon 9 static fire text explosion and fire at the adjacent Space Launch Complex 40 (c) on September 1, 2016. Although the VIF was only about 2 km downwind from the fire and the shared water pump between the launch sites was damaged, OSIRIS-REx was protected from contamination due to a combination of planning and swift work.  Photo credit: Top: Dworkin, Bottom: (Clark, 2016b).

## TABLES

**Table 1** Summary of contamination recommendations from Stardust and their application to OSIRIS-REx, tabulated from Sandford et al. (2010).

| Stardust Team Recommendation | OSIRIS-REx Implementation |
|---|---|
| 1. Efforts need to be made both for contamination control and contamination knowledge. | 1. These activities were integrated into the OSIRIS-REx budget and schedule. |
| 2. Contamination control and assessment requires cooperative efforts be made that involve the spacecraft manufacturers, the spacecraft operators, the mission's Science Team, and the NASA Curatorial Office. | 2. Each of these groups has an individual responsible for organizing contamination control and assessment activities, and these individuals work closely together across organizational boundaries. |
| 3. Agree on what is meant by the word "clean" and how this definition will translate into operational activities. | 3. OSIRIS-REx has defined "pristine" as described in this document. |
| 4. Document what materials are used; samples of these materials should be collected and archived. | 4. A major effort on understanding and archiving spacecraft materials was performed |
| 5. Witness plates need to be removed and examined quickly so that problems associated with unexpected or problematic contaminants can be dealt with rapidly. | 5. The OSIRIS-REx team analyzed contamination control and contamination knowledge samples monthly during ATLO. Amino acid witnesses were analyzed at GSFC with a 1-week turnaround. |
| 6. Witness coupons need to be designed so that they can easily be divided and distributed to multiple analysts. | 6. A plan has been developed for division of witness plates after Earth return since we realized that an easily divisible witness is also easily fragmented during shock events. The use of the frosted surface will aid in identification of exposed surfaces after fragmentation of the witness. |
| 7. Sample return spacecraft should carry a significant number of relevant witness coupons. | 7. Ten witness coupons are flown in the SRC and fourteen on TAGSAM. |
| 8. It is generally desirable to use more than one type of witness coupon. | 8. Two types of witness plates are flown. |
| 9. Plans must be made in advance so the NASA Curatorial Office is prepared to store and distribute not only the returned samples, but also the associated contamination control and assessment materials (witness coupons, samples of spacecraft materials, etc.). | 9. The OSIRIS-REx team secured written commitments from the NASA organizations responsible for the spacecraft materials archive, samples, and "space-exposed hardware." |
| 10. Devote additional development effort to the production of the cleanest possible aerogel. | 10. Aerogel is not used by OSIRIS-REx |

**Table 2** Each row shows an example of organic contamination control rationale which was considered for OSIRIS-REx. The first column indicates how the example is described in the text. The second column lists any organic compounds specified by the rationale. The third column indicates the allowable abundance of that species in the collected sample. The fourth column gives that abundance on the surface of the TAGSAM sampling hardware. The fifth column evaluates the viability of the rationale against external traceability, achievability within the limits of the facilities available for OSIRIS-REx, and verifiability within the schedule of ATLO*.

| Name | Species | ng/g Sample | $ng/cm^2$ TAGSAM | Viability |
|---|---|---|---|---|
| OCSSG molecular guidelines[a] | Aromatic<br>Carbonyls and hydroxyls<br>Amino acids<br>Amines or amides<br>Aliphatic hydrocarbons<br>DNA<br>Total reduced carbon | 8<br>10<br>1<br>2<br>8<br>1<br>40 | 0.25<br>0.31<br>0.031<br>0.063<br>0.25<br>0.031<br>1.3 | Traceable **Yes**<br>Achievable No<br>Verifiable No |
| 200 A/100 (OCSSG Level 2 cleaning guidelines [Table 3] | N/A | N/A | 10 | Traceable ~<br>Achievable No<br>Verifiable **Yes** |
| NRC-derived (30%)[b] | All detectable species | ≤30/ compound | ≤0.94/ compound | Traceable **Yes**<br>Achievable ND<br>Verifiable No |
| 30% Precision of Worst-case meteorite (Yamato 980115) | Total carbon[c]<br>Amino acids[d] | 57,000<br>44 | 1,800<br>1.4 | Traceable ~<br>Achievable No<br>Verifiable No |
| 30% Precision of Reasonable meteorite (Murchison) | Total carbon[e]<br>Amino acids[f] | 30,000<br>6,300 | 1,000<br>200 | Traceable ~<br>Achievable **Yes**<br>Verifiable **Yes** |
| Stardust achieved[e] | Amino acids | N/A | 180 | Traceable **Yes**<br>Achievable **Yes**<br>Verifiable **Yes** |

* N/A: Not applicable; ND: Not determined; ~ external traceability is arguable

[a] Mahaffy et al. (2004)

[b] National Research Council (2007)

[c] Chan et al. (2016)

[d] Burton et al. (2014)

[d] Average values in Kerridge (1985)

[d] Glavin et al. (2010)

[e] Elsila et al. (2009)

**Table 3** OCSSG cleaning guidelines summary (Mahaffy et al. 2004). Shaded fields are outside the capabilities of identified industrial ATLO facilities including those available for OSIRIS-REx.

| OCSSG cleanliness level | 1 | 2 | 3 |
|---|---|---|---|
| Nonvolatile residue (NVR) level[a] | A/2<br>(500 ng/cm$^2$) | A/100<br>(10 ng/cm$^2$) | AA3<br>(1 ng/cm$^2$) |
| Particulate cleanliness level[a] | 400 | 200 | 25 |
| Outgassing flux | 100 ng/cm$^2$/hr | 10 ng/cm$^2$/hr | 1 ng/cm$^2$/hr |

[a] IEST (2002)

**Table 4** Indicator elements and their control limits. Individual elements serving as representatives or indicators of suites of elements important for various scientific investigations. Each element has a carbonaceous chondrite derived contamination limit in the sample, which is then converted into a surface requirement for the sampling system in the third column. The fourth column shows the abundance of each element if it were to contaminate a surface at 100 A/2 (assuming pure spherical particles). Thus, meeting the 100 A/2 requirement meets the C and Ni limits, but contamination knowledge is needed to ensure that unexpectedly high levels of K, Sn, Nd, and Pb are not major contributors to the contamination.

| Element | Indicator | Contamination limit ($ng/cm^2$) | Level 100 (particles) + A/2 (NVR) ($ng/cm^2$) |
|---|---|---|---|
| C | Organics | 1,000 | 34 + 500 |
| K | Terrestrially abundant lithophile | 170 | 14 + 500 |
| Ni | Extraterrestrially abundant siderophile | 34,000 | 143 + 500 |
| Sn | Industrial contaminant | 0.53 | 117 + 500 |
| Nd | Lanthanide, lithophile | 1.5 | 113 + 500 |
| Pb | Chalcophile, geochronology, industrial contaminant | 180 | 182 + 500 |

**Table 5** Amino acid abundance on Stardust foil C2092S,0 (Elsila et al. 2009).

| Amino acid | Primary source | pmol/cm$^2$ | ng/cm$^2$ |
|---|---|---|---|
| Glycine | Cometary | 68 | 5.1 |
| β-Alanine | Unknown | 7 | 0.6 |
| D-Alanine | Not detected | <4 | <0.4 |
| L-Alanine | Contamination | 12 | 1.1 |
| ε-Amino-*n*-caproic acid (EACA) | Contamination | 1413 | 185.2 |
| Total | | 1,500 | 192.0 |
| Total contamination | | 1,425 | 186.3 |

**Table 6** Contamination sampling terminology used across the project.

| Term | Description |
|---|---|
| Hardware (or process) coupons | Pieces manufactured from the same raw materials, using the same equipment and processes specifically for archive at JSC or offal (excess material after machining) earmarked for archive at JSC |
| Contamination knowledge plates | Silicon and aluminum foils prepared by JSC for monthly collection and replacement for the purpose of archiving the environmental effects on the spacecraft for both particulate and NVR. 25% was analyzed, while the remainder is archived at JSC for future use |
| Contamination monitoring plates—particulate | Silicon wafers collected and analyzed by LM monthly during ATLO for determining background facility particulate fallout rates |
| Contamination monitoring plates—NVR | Aluminum foil collected and analyzed by LM monthly during ATLO for determining background facility NVR |
| Contamination monitoring plates—amino acids | Aluminum foil collected and analyzed by GSFC monthly during ATLO for determining background facility amino acid load |
| Hardware verification samples—particulate | Millipore slide for limited exposure used for verification of cleanliness levels for SRC canister interior, TAGSAM head, Launch Container by LM |
| Hardware verification samples—NVR | Aluminum plate or aluminum foil for limited exposure used for verification of cleanliness levels for SRC canister interior, TAGSAM head, Launch Container by LM |
| Hardware verification samples—amino acids | Aluminum foil for limited exposure used for verification of cleanliness levels for SRC canister interior, TAGSAM head, Launch Container by GSFC |
| Flight witness plates | Designed to collect material during flight; returned to Earth for analysis and archive in 2023 |

**Table 7** OSIRIS-REx Level 2 contamination requirements (mission requirement document, MRD).

| MRD | Requirement |
|---|---|
| 107 | OSIRIS-REx shall limit the contamination on the TAGSAM head, TAGSAM launch container interior, SRC canister interior to levels at or below those specified by IEST-STD-CC1246D level 100 A/2 until launch. |
| 108 | OSIRIS-REx shall limit total hydrazine contamination on the TAGSAM Head surface to <180 ng/cm$^2$. |
| 109 | OSIRIS-REx shall return and maintain the bulk sample exposed to total amino acid contamination <180 ng/cm$^2$ on the TAGSAM head surface. |
| 110 | OSIRIS-REx shall document the contamination acquired by the TAGSAM head during flight. |
| 111 | OSIRIS-REx shall generate and follow the project contamination control plan. |
| 406 | OSIRIS-REx shall provide a ≥1 g sample of all materials of which the TAGSAM head, TAGSAM launch container interior, SRC canister interior, or witness material are composed. |

**Table 8** Amino acid testing results, blank corrected. For these experiments the samples were split, with half analyzed without hydrolysis (Free Amino Acids) and the other half acid-hydrolyzed, according to Glavin et al. (2006), to show both free amino acids and those bound (presumably as peptides in this case). Bags were extracted in water and reported as concentration in solution.

| **Sample** | **Free amino acids ($ng/cm^2$)** | **Total amino acids ($ng/cm^2$)** |
|---|---|---|
| Uncleaned screw | 14.91 | 89.64 |
| Dirtied screw | 54.58 | 252.01 |
| Dirtied and cleaned | 0.50 | 1.97 |
| Nylon packaged | 13.69 | 101.13 |
| Nylon packaged and cleaned | 0.50 | 1.50 |
| Kimberly Clark 55082 medium purple nitrile glove used at GSFC | n.d. | 67.3 ng/g |
| Kimtech G3 medium tan nitrile gloves used at LM | n.d. | 646.2 ng/g |
| TechNitrile medium blue nitrile gloves | n.d. | 150.3 ng/g |
| Dry Kimtech glove rubbed on foil | n.d. | 0.97 |
| Fisher Optima 2-propanol wet Kimtech glove rubbed on foil | n.d. | 0.13 |
| Nylon (KNF LB106) | n.d. | 38,000 ng/g |
| PFTE (KNF LB605) | n.d. | 17 ng/g |
| Polypropylene (KNF LB3022) | n.d. | 7.8 ng/g |
| Tyvek® (Beacon Converters A7373) | n.d. | 6.3 ng/g |
| Lamellar polypropylene (KNF Kenlam 7150) | n.d. | 1.4 ng/g |
| Lamellar polyethylene (Beacon Converters FR2176) | n.d. | 0.060 ng/g |
| Aluminum (Reynolds extra heavy duty foil hot bag) | n.d. | 0.015 ng/g |
| 2-Propanol (Fisher ACS) | n.d. | <0.05 ng/g |
| 2-Propanol (Fisher Optima) | n.d. | <0.05 ng/g |

n.d. = not determined

**Table 9** Summary of all contamination control results for OSIRIS-REx sampling hardware shown in the IEST-STD-CC1246D (IEST 2002) particle levels and NVR abundances measured for the indicated components. Then for a worst-case analysis, the contamination levels were converted (→) into elemental carbon abundance to allow for a comparison to the rationale in Table 4. Particulate level was converted into C by assuming spherical graphite particles of the maximum size for each bin and NVR was assumed to be pure C. Contamination knowledge SEM/EDX analyses had demonstrated that the other elements in Table 4 were well below their threshold. The total measured amino acids for these components are also shown. Amino acids were dominated by glycine.

| **Hardware** | **Particle** | | | | **NVR** | | | | **Total C** | **Amino acids** |
|---|---|---|---|---|---|---|---|---|---|---|
| | **level** | | **$ng/cm^2$ C** | | **level** | | **$ng/cm^2$ C** | | **$ng/cm^2$** | **$ng/cm^2$** |
| Requirement | **100** | **→** | **34** | + | **A/2** | **→** | **500** | = | **534** | **180** |
| SRC | 176 | → | 323 | + | A/5.6 | → | 180 | = | **503** | **13.1** |
| Launch container | 100 | → | 34 | + | A/10 | → | 100 | = | **134** | **2.32** |
| TAGSAM head | 116 | → | 61 | + | A/16 | → | 220 | = | **281** | **0.96** |

**Table 10** Summary of calculated hydrazine impingement on the TAGSAM head.

| Mission Phase | Fluence ($ng/cm^2$) during first TAG | Fluence ($ng/cm^2$) during subsequent TAG(s)** |
|---|---|---|
| Approach | 0.48* | n/a |
| Rehearsal #1 | 0.29* | 29.35 |
| Rehearsal #2 | 0.76* | 76.04 |
| TAG | 118.30 | 171.92 |
| **Subtotal** | **<120** | **<277** |

*This hydrazine is expected to evaporate prior to collection

**Assuming the head was covered in dust from the previous attempt

n/a Not applicable. Approach does not recur after the first TAG.

**Table 11** A representative list of compounds of contamination concern. Species in **bold** were specifically targeted by OSIRIS-REx, and items in *italics* are routinely controlled by most missions, including OSIRIS-REx. Other species are of interest but not explicitly controlled.

| Biology | People | Outside | Building materials | Polymers | Inorganic | Processes | Industrial organics | Compounds |
|---|---|---|---|---|---|---|---|---|
| *Birds*<br>microbes<br>*rodents*<br>*spiders, etc.* | *Food*<br>glossy magazines and catalogs<br>*moisturizer*<br>*perfume*<br>*sun screen*<br>*sweat and other excretions*<br>*tobacco residue*<br>vitamins | *Asphalt fumes*<br>*car exhaust*<br>*herbicide*<br>*insecticide*<br>*paint fumes*<br>*pollen*<br>*smoke* | *Carpet*<br>plywood<br>varnish | Bakelite<br>Delrin<br>**latex**<br>**nylons**<br>***silicones***<br>silk | Borax<br>diamond<br>**fluorescent lamps (Hg)**<br>lanthanides<br>silanes<br>silicates | Heat-sealing plastics<br>*soldering*<br>*welding* | *Creosote*<br>*detergents and soaps*<br>*glues*<br>*grease*<br>leather<br>*oils*<br>**natural rubber**<br>*wax* | Acetaldehyde, acetic anhydride, acetone, acetonitrile, acrylic acid, acrylonitrile, **amino acids**, aniline, barbituric acids, benzene, benzoic acid, benzyl alcohol, bromobenzene, butadiene, butyl acetate, butylamines, butyric acids, carbon disulfide, carbon tetrachloride, chloroform, cresols, cyanide, cyanoacetylene, cyanogens, dichloroethane, dichloromethane, diethyl ether, dimethyformamide, dimethylsulfoxide, dioxane, ethanolamine, ethylene glycol, ethyl acetate, ethylamine, ethylenediamine, formaldehyde, formamide, furfural, glutaronitrile, glyceric acid, glycerol, glycolic acid, glyoxal, hydantoins, hydroxy acids, imidazole, indole, malic acid, malonic acid, malononitrile, methylamine, methyl bromide, naphthalene and other polycyclic aromatic hydrocarbons, nitropropane, nitrobenzene, orotic acids, oxalic acid, paraformaldehyde, phenol, phthalic acid, piperazine, propylamines, purines, pyrimidines, pyruvic acid, pyruvonitrile, quinolines, strong acids, strong bases, styrene, succinic acid, succinic anhydride, tetrahydrofuran, toluene, urea, valeric acids, xylenes |

**Table 12** Microbes identified by DNA sequencing a sample of contamination knowledge plate #8, which was exposed during OSIRIS-REx vibration testing at LM.

| Fungi | Bacteria | Archaea |
|---|---|---|
| *Clavispora/Candida intermedia* | *Brevibacterium paucivorans* | *Natronococcus amylolyticus* |
| *Fusarium cerealis* | *Eubacterium* species | |
| *Hortaea werneckii* | *Lactobacillus fermentum* | |
| *Malassezia restricta* | *Pseudomonas alcaligenes* | |
| *Penicillium citrinum* | *Reyranella soli* | |
| *Penicillium crustosum* | *Sphingobium* species | |
| *Phoma* species | | |

**Table 13** Gas samples collected and analyzed.

| Sample Type | Location |
|---|---|
| Air | LM ISO 7 Cleanroom |
| Air | LM ISO 7 Supplemental cooling cart |
| Air | LM ISO 8 Cleanroom |
| Air | PHSF high bay |
| Air | Faring air supply used during Faring transport |
| Air | VIF Faring air supply |
| Air | VIF Centaur common equipment module air supply |
| Nitrogen/Helium | Source gas for TAGSAM bottles |
| Nitrogen/Helium | Source gas for TAGSAM bottles through bottle loading manifold |
| Nitrogen/Helium | Pressurized TAGSAM qualification bottle after 24 hours at 40-45°C |
| Nitrogen/Helium | TAGSAM gas during NSI actuation |
| Nitrogen Purge | LM purge for spacecraft |
| Nitrogen Purge | PHSF facility purge for spacecraft |
| Nitrogen Purge | Spacecraft in fairing transport purge |
| Nitrogen Purge | VIF T-0 purge system |
| Nitrogen Purge | Launch pad T-0 purge system |

**Table 14** Gas trapping efficiencies of the SRC filter.

| Compound | Integration range ($cm^{-1}$) | Filter efficiency (%) |
|---|---|---|
| Ethanol | 3534–3345 | 99.37 |
| Benzene (4 bands) | 3110–2998 | >94 |
| Hexane+ethanol+acetone | 2999–2845 | >98.5 |
| CO | 2157–2129 | 84 |
| Benzene | 1988–1960 | >90 |
| Benzene | 1843–1813 | >99.2 |
| Acetone | 1731–1698 | 99.05 |
| Benzene | 1488–1477 | 98 |
| Mostly ethanol | 1477–1407 | |
| | | 99.4 |
| Hexane+ethanol+acetone | 1397–1348 | 99.27 |
| Hexane+ethanol+acetone | 1266–1216 | 99.4 |
| Ethanol+acetone+benzene | 1106–1024 | 98.7 |
| Acetone+ethanol | 898–873 | >98 |
| Hexane | 739–722 | >95 |
| Benzene | 701–676 | >97.1 |

Note—both spectra show bands near 2350 and 663 $cm^{-1}$ due to $CO_2$ and in the 3730–3570 $cm^{-1}$ 1640–1590 $cm^{-1}$ regions due to $H_2O$. These gases were not part of the original gas mixture and are likely contaminants associated with gases already trapped in the filter due to prior exposure to atmosphere, and the cryo-vacuum system used to condense the gases for IR measurement.

**Table 15** Humidity performance versus volume of gas through the SRC filter.

| | 90% Humidity mass (g) | 40% Humidity mass (g) |
|---|---|---|
| Initial | 1196.10 | 1196.12 |
| Final | 1198.01 | 1196.72 |
| Change | 1.91 | 0.60 |

| Efficiency | 90% Humidity total volume (L) | 40% Humidity total volume (L) |
|---|---|---|
| 100% | 6.025 | 6.025 |
| 95–99.99% | 13.273 | 11.678 |
| 90–94.99% | 19.598 | 22.658 |
| 85–89.99% | 23.543 | 31.130 |
| 80–84.99% | 26.269 | 38.352 |
| 75–79.99% | 28.440 | 43.881 |
| 70–74.99% | 30.361 | 47.942 |
| 65–69.99% | 32.149 | 51.255 |
| 60–64.99% | 33.913 | 54.051 |
| 55–59.99% | 35.793 | 56.552 |
| 50–54.99% | 37.955 | 58.954 |
| 45–49.99% | 40.721 | 61.326 |

**Table 16** Particulate trapping efficiency results of the SRC filter.

| Unfiltered (counts/$cm^3$) | Filtered (counts/$cm^3$) | Efficiency (%) |
|---|---|---|
| 3,000 | 11 | 99.63 |
| 25,500 | 33 | 99.87 |
| 200,000 | 8 | 99.996 |

## Online Supplemental Materials 1

[Insert “OnlineSupplementalMaterials1.pdf”]

## Online Supplemental Materials 2

Summary table of materials archived for contamination knowledge analysis. Details on each component and how to request samples will be available at https://curator.jsc.nasa.gov/ prior to OSIRIS-REx sample return.

[Insert “OnlineSupplementalMaterials2.pdf”]

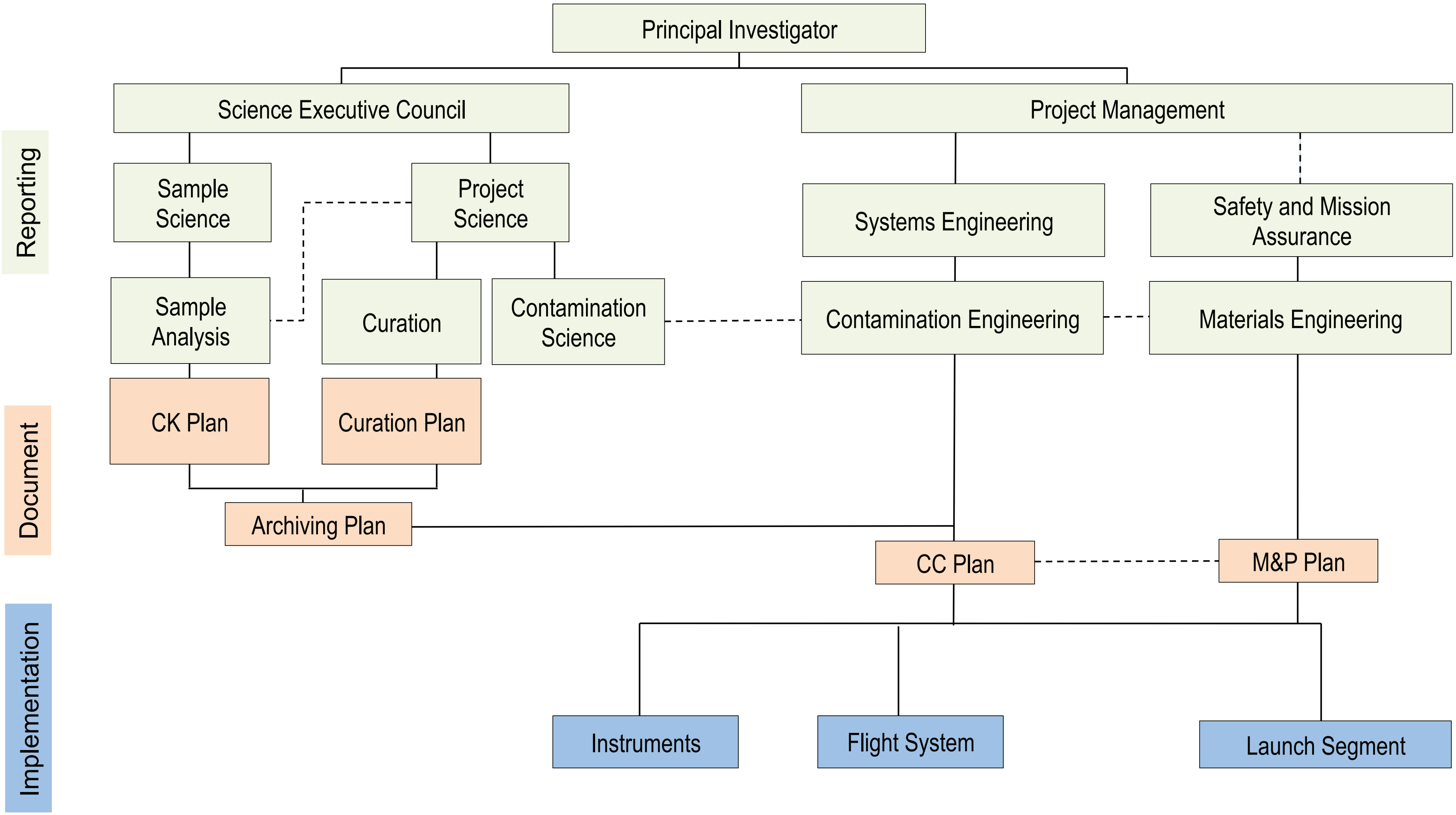
Principal Investigator
Science Executive Council
Project Management
Reporting
Sample Science
Project Science
Systems Engineering
Safety and Mission Assurance
Sample Analysis
Curation
Contamination Science
Contamination Engineering
Materials Engineering
Document
CK Plan
Curation Plan
Archiving Plan
CC Plan
M&P Plan
Implementation
Instruments
Flight System
Launch Segment

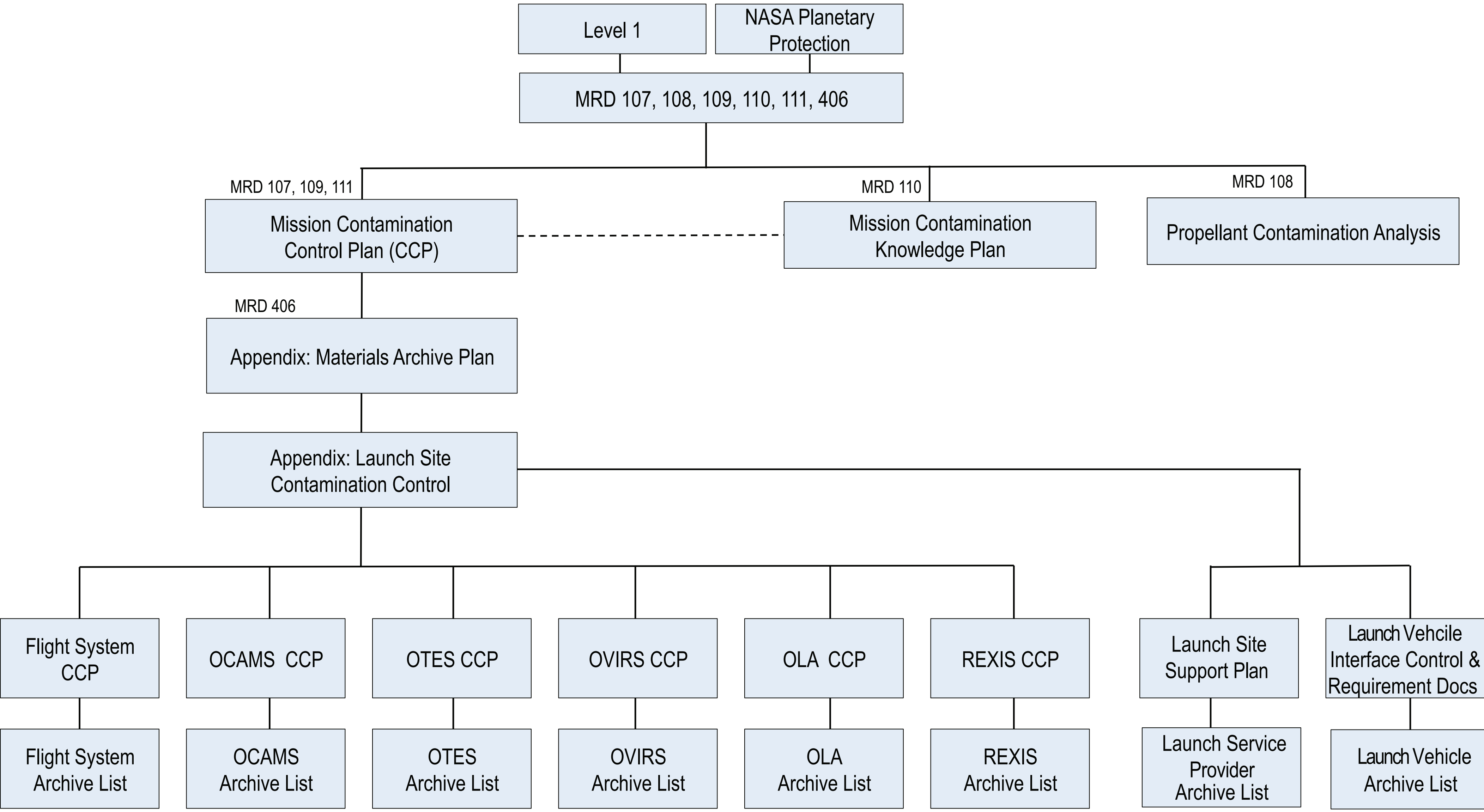
Level 1
NASA Planetary Protection
MRD 107, 108, 109, 110, 111, 406
MRD 107, 109, 111
Mission Contamination Control Plan (CCP)
MRD 110
Mission Contamination Knowledge Plan
MRD 108
Propellant Contamination Analysis
MRD 406
Appendix: Materials Archive Plan
Appendix: Launch Site Contamination Control
Flight System CCP
OCAMS CCP
OTES CCP
OVIRS CCP
OLA CCP
REXIS CCP
Launch Site Support Plan
Launch Vehicle Interface Control & Requirement Docs
Flight System Archive List
OCAMS Archive List
OTES Archive List
OVIRS Archive List
OLA Archive List
REXIS Archive List
Launch Service Provider Archive List
Launch Vehicle Archive List

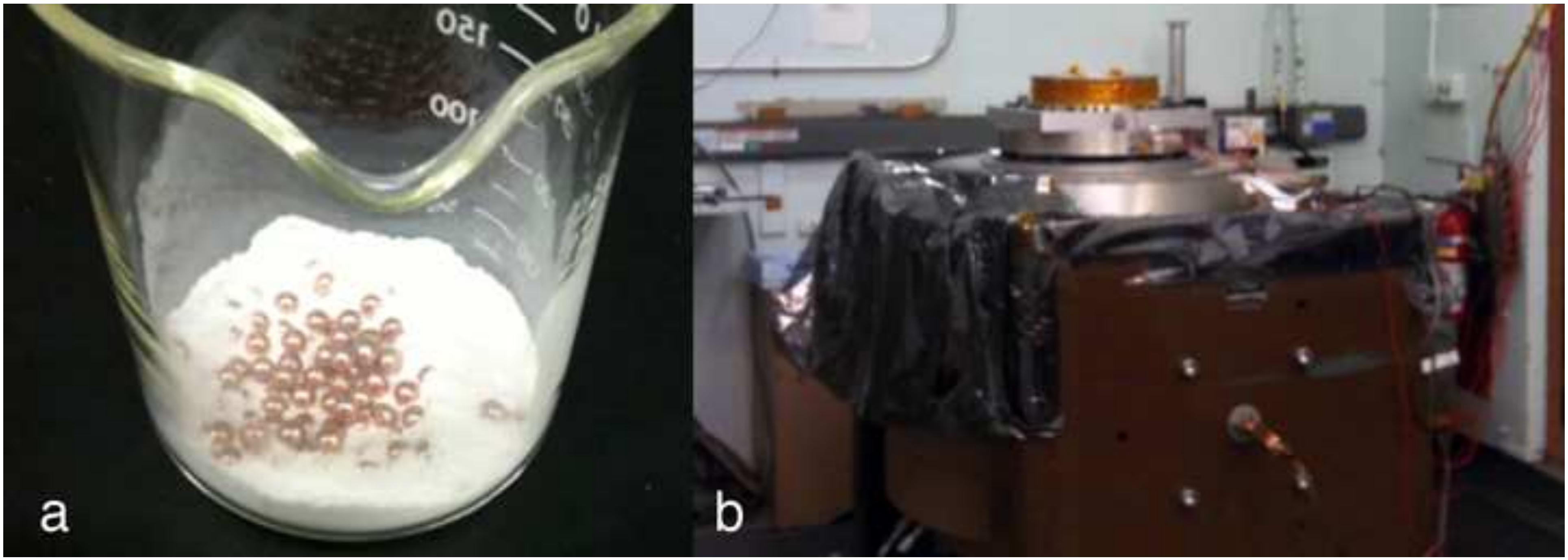
a
b

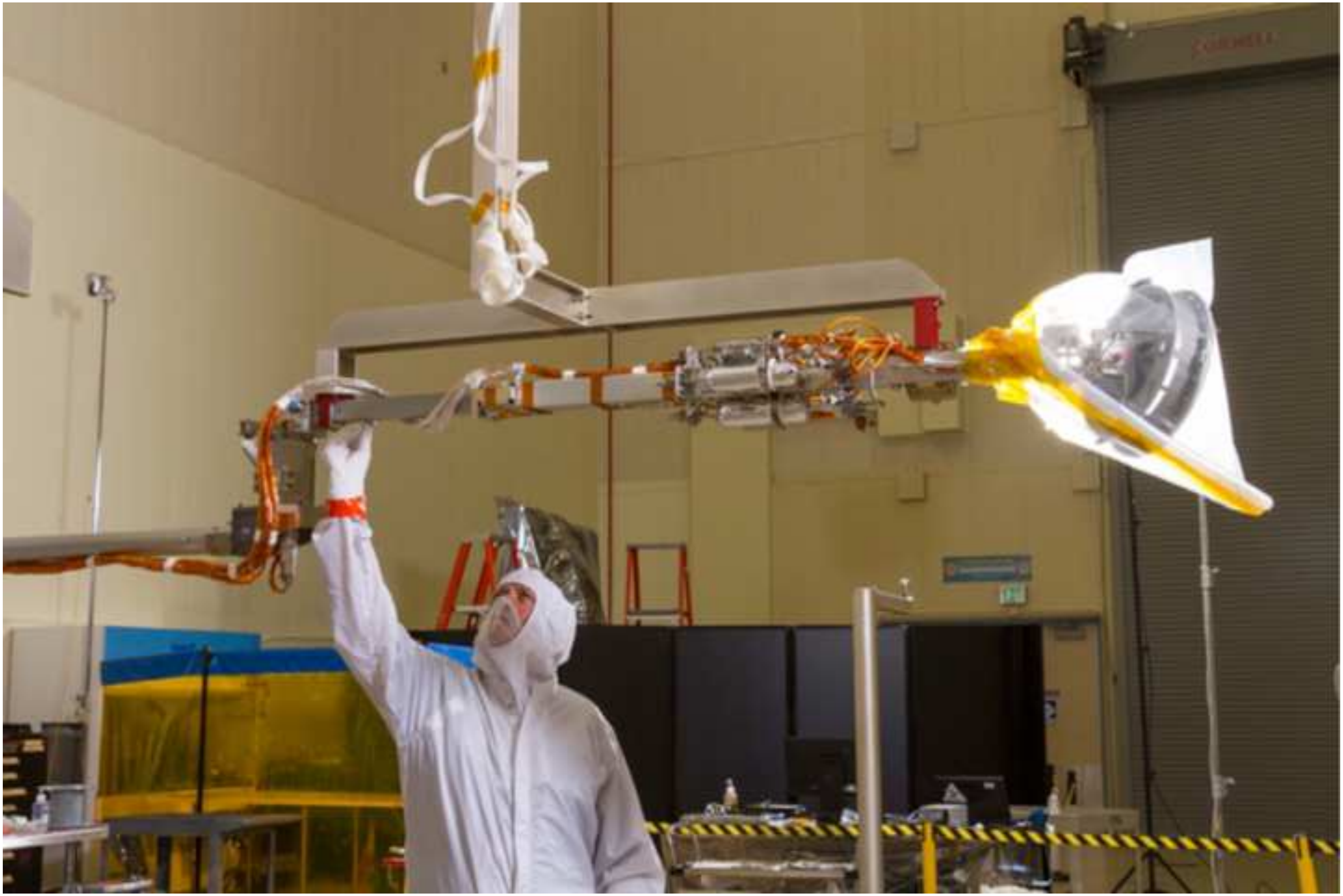

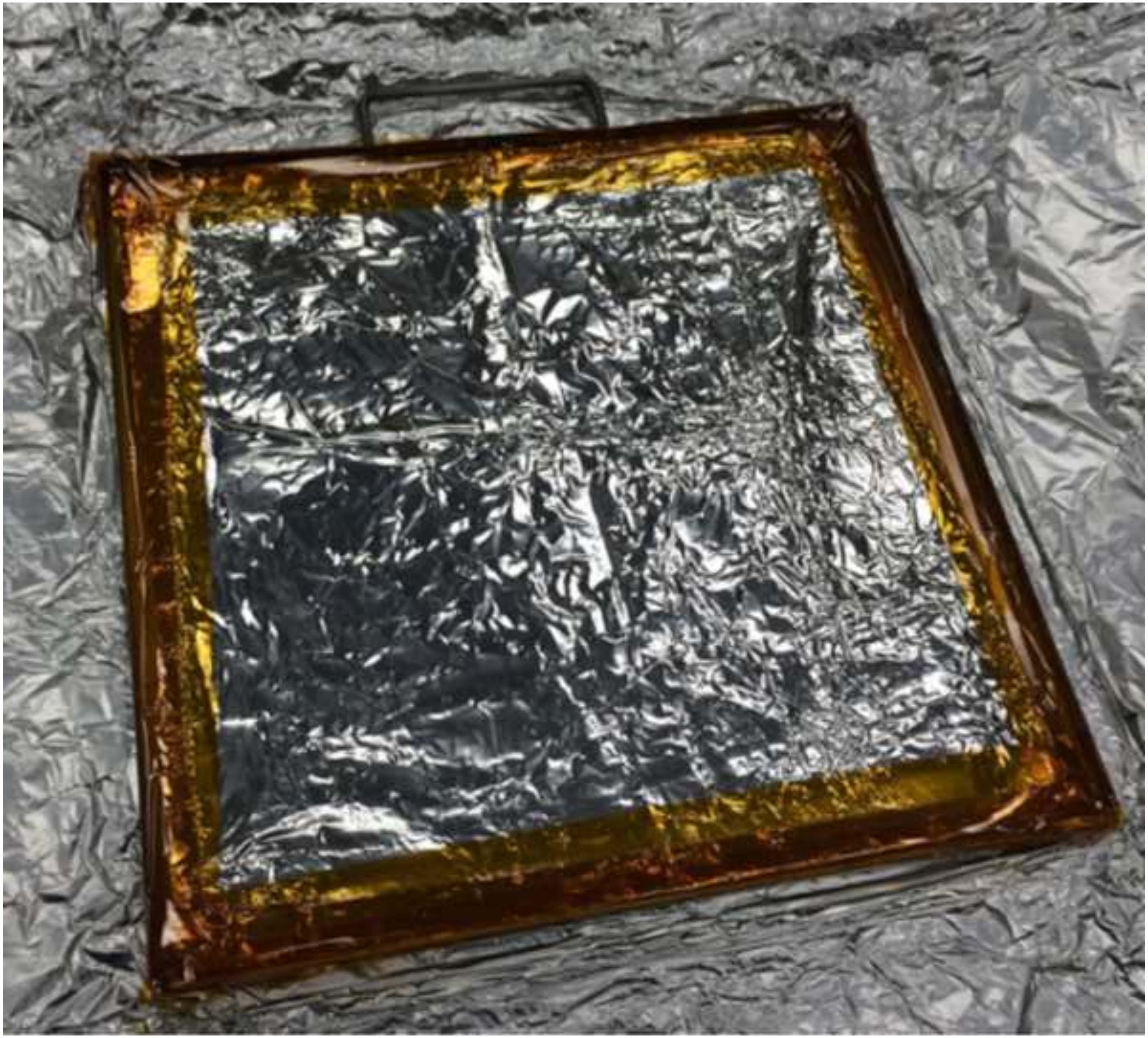

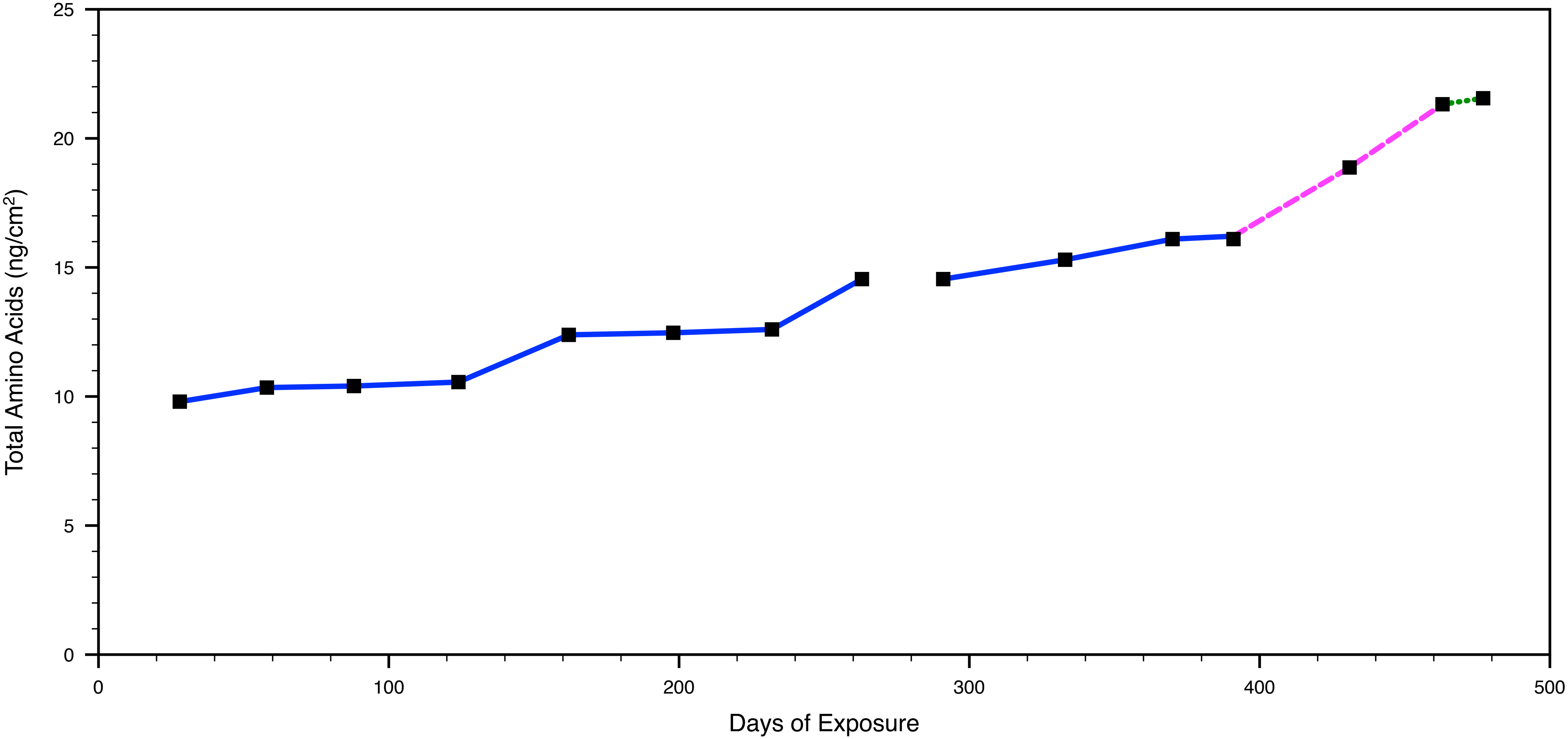
25
20
15
10
5
0
Total Amino Acids (ng/cm$^2$)
0
100
200
300
400
500
Days of Exposure

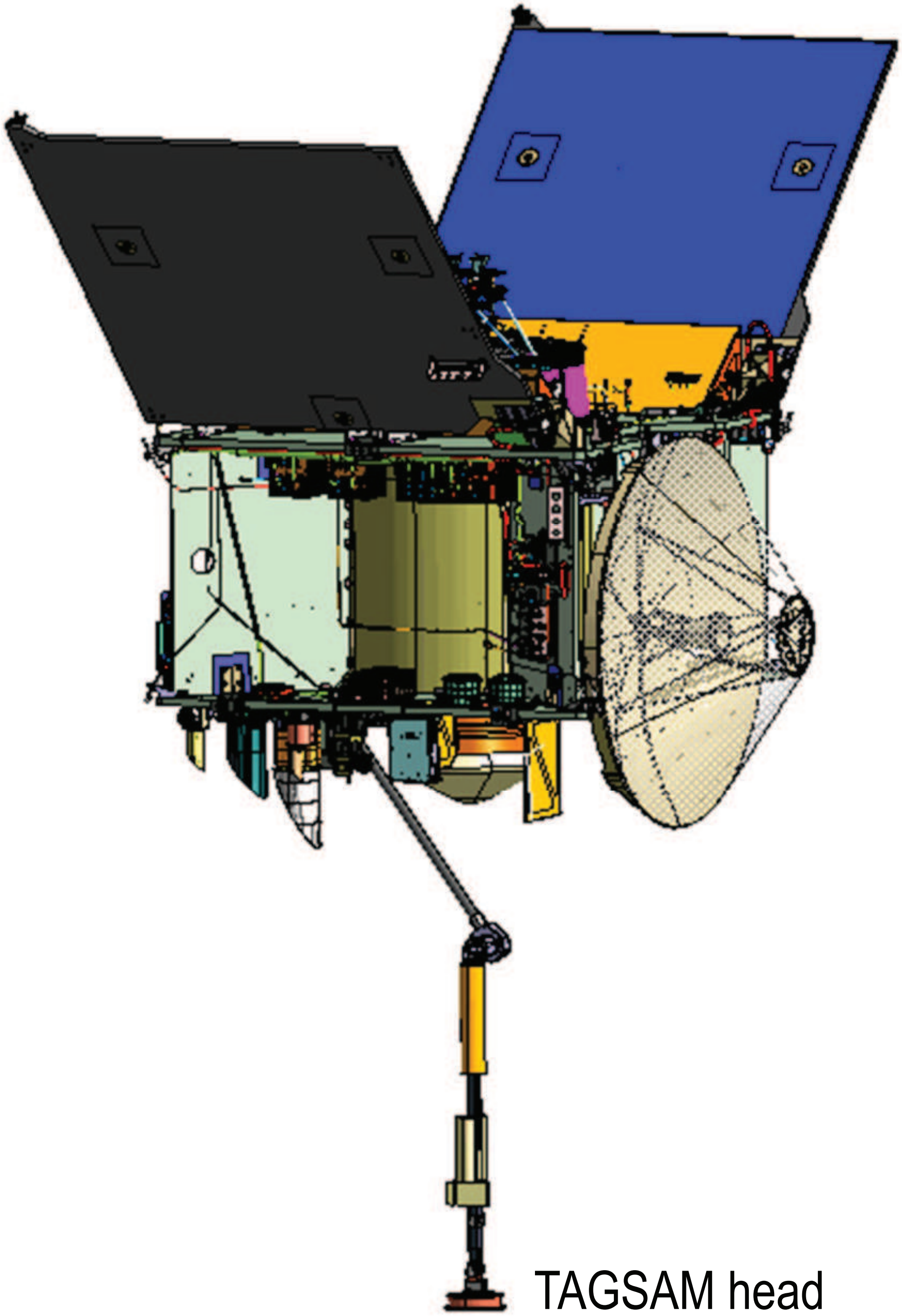
TAGSAM head

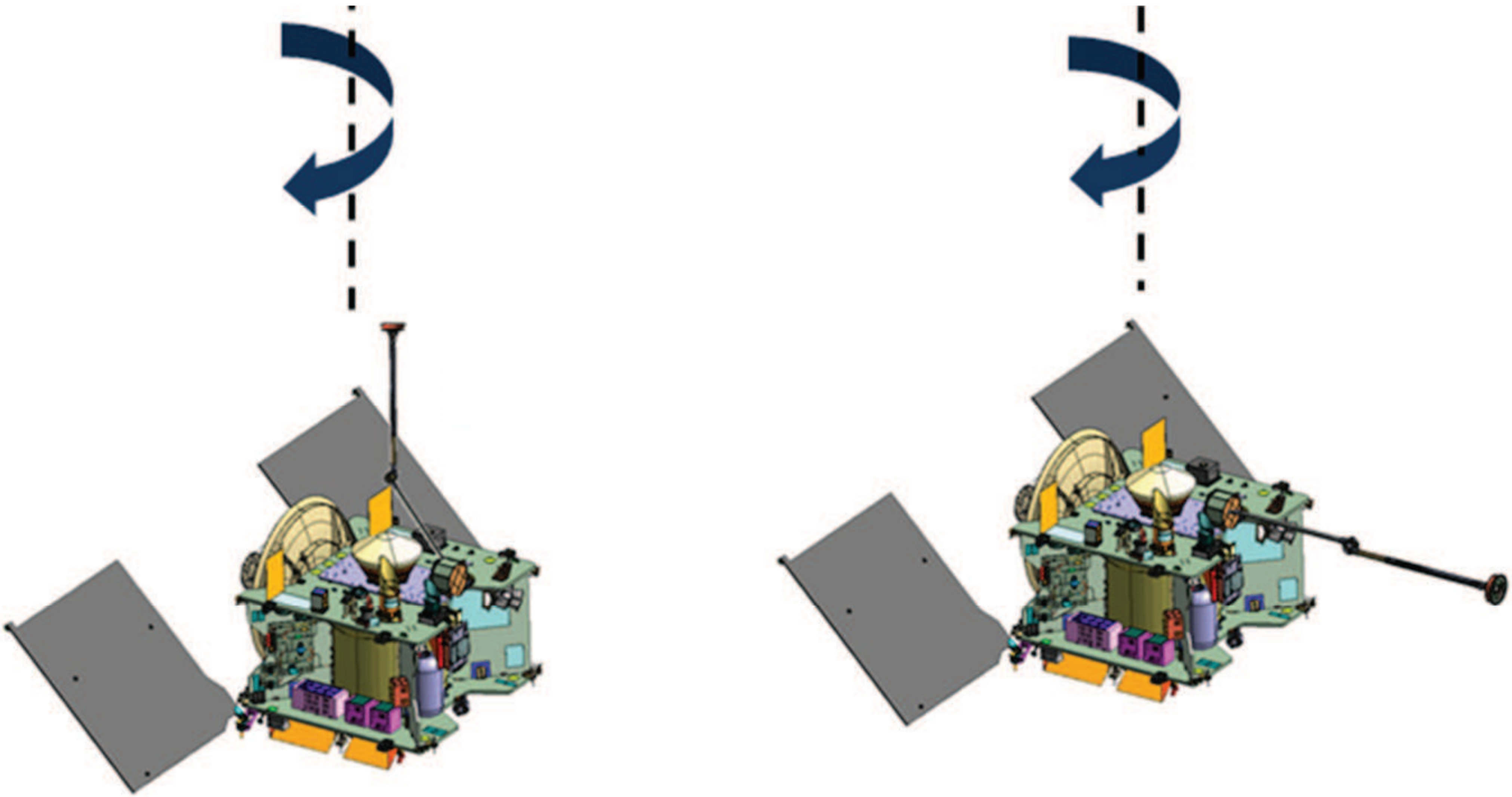

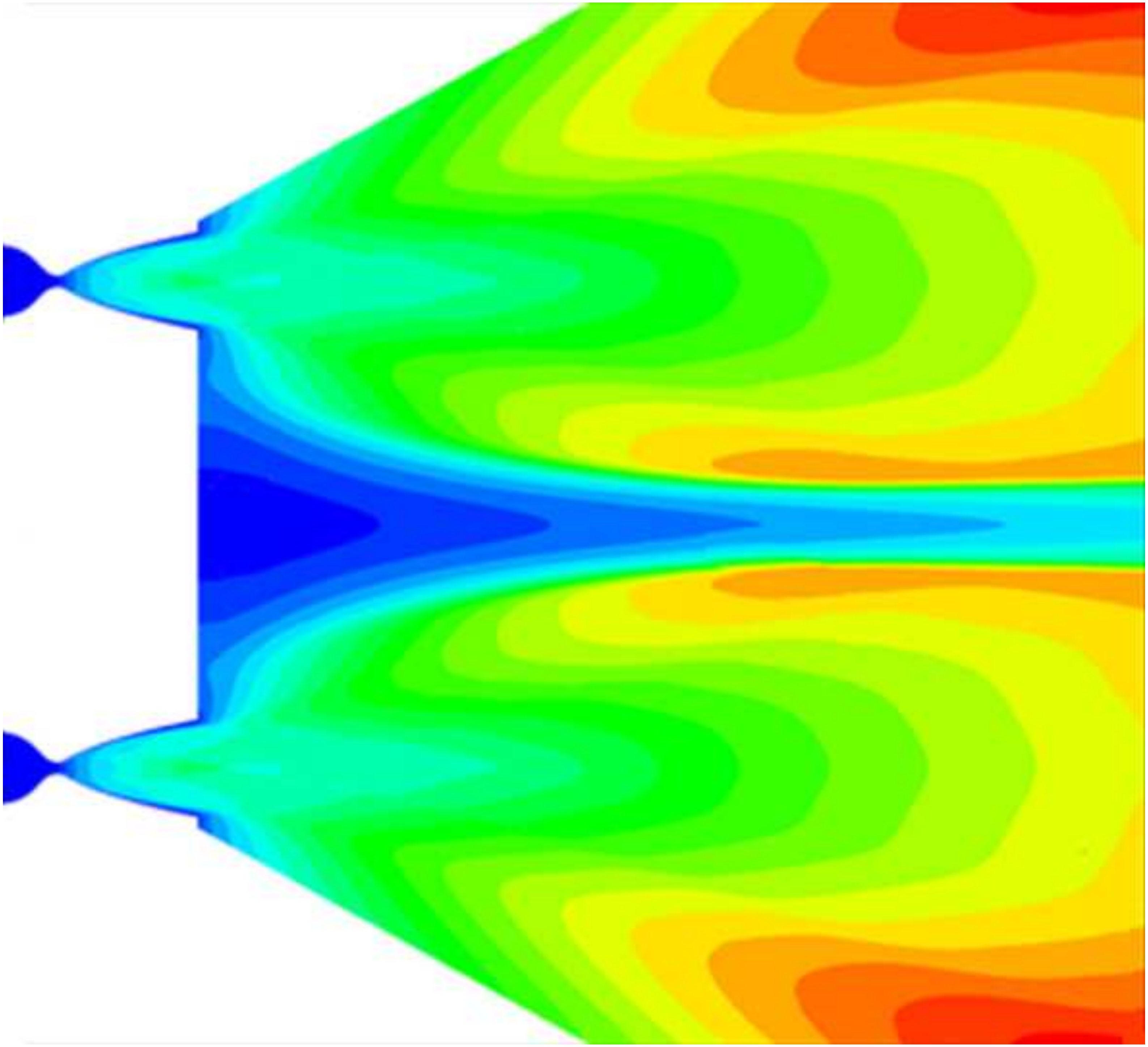

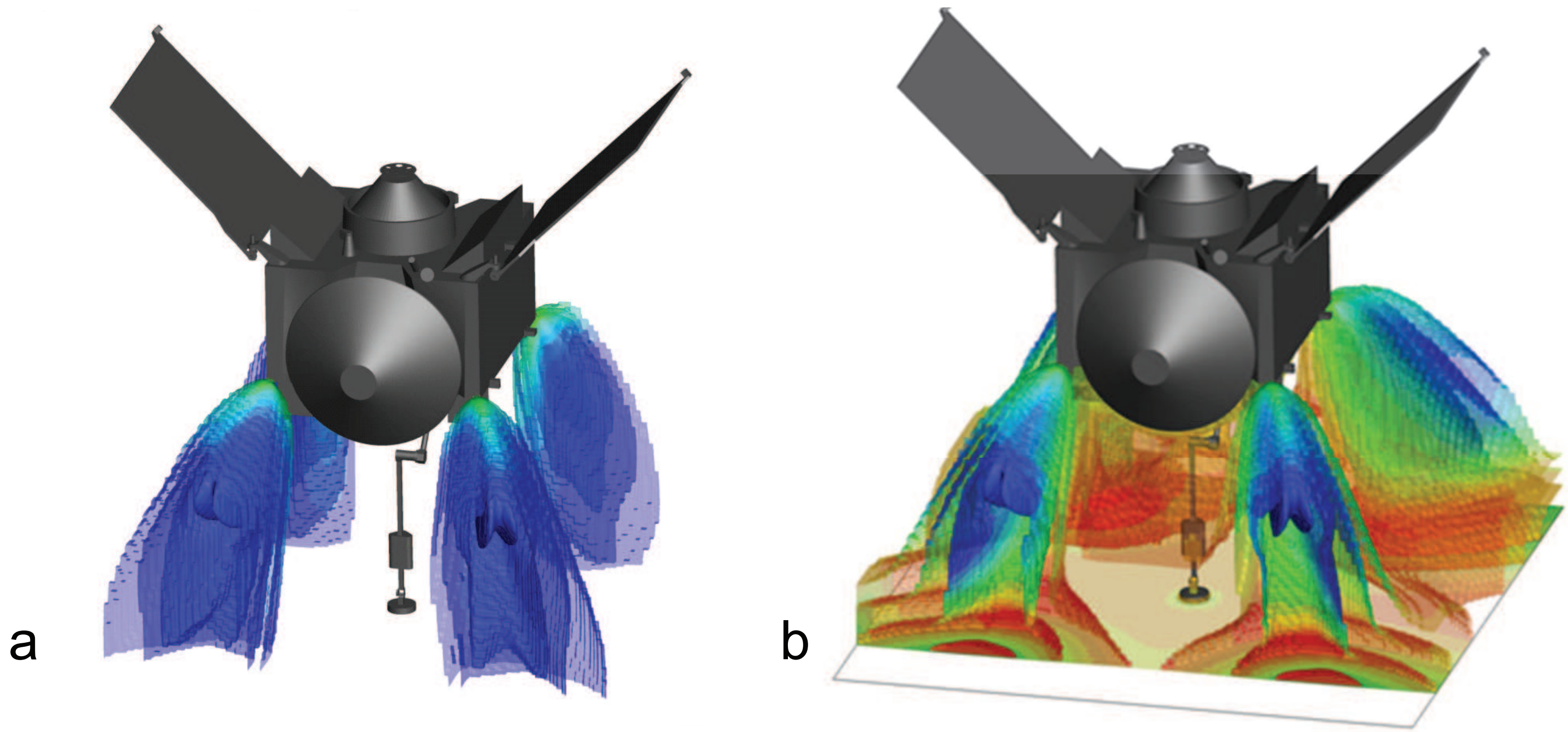
a
b

Response to reviewer

Click here to download figure Fig11.eps

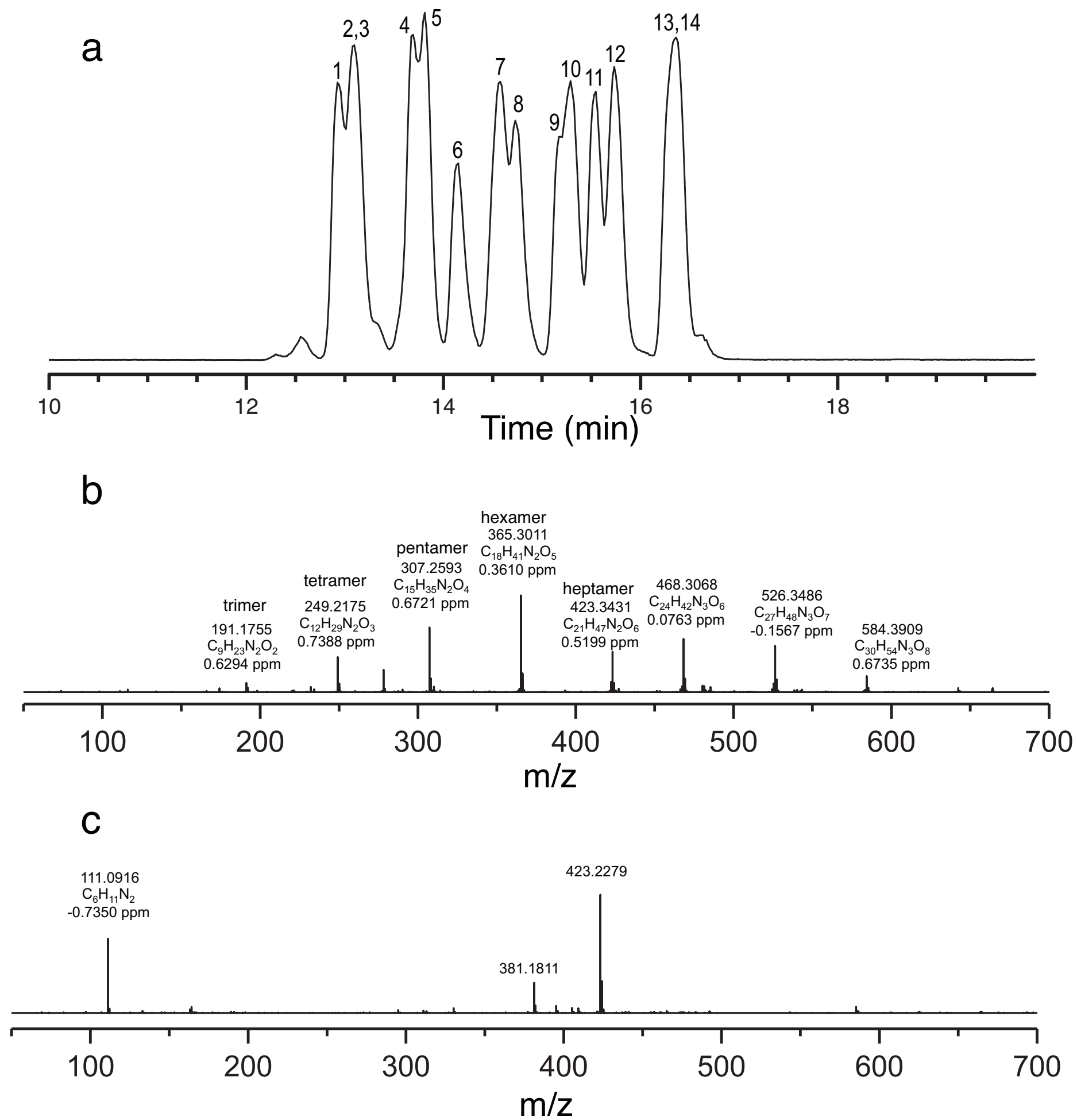

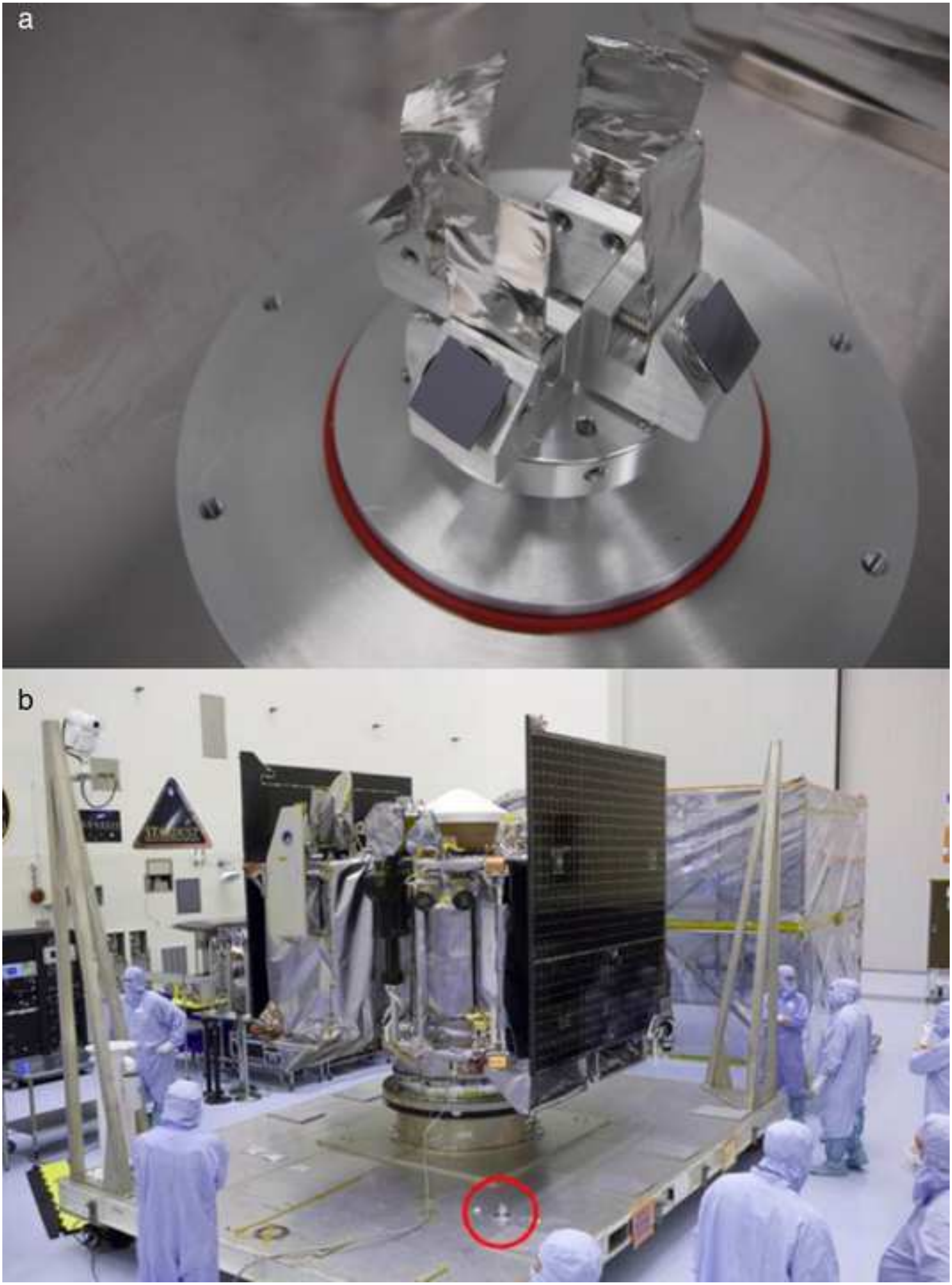
a
b

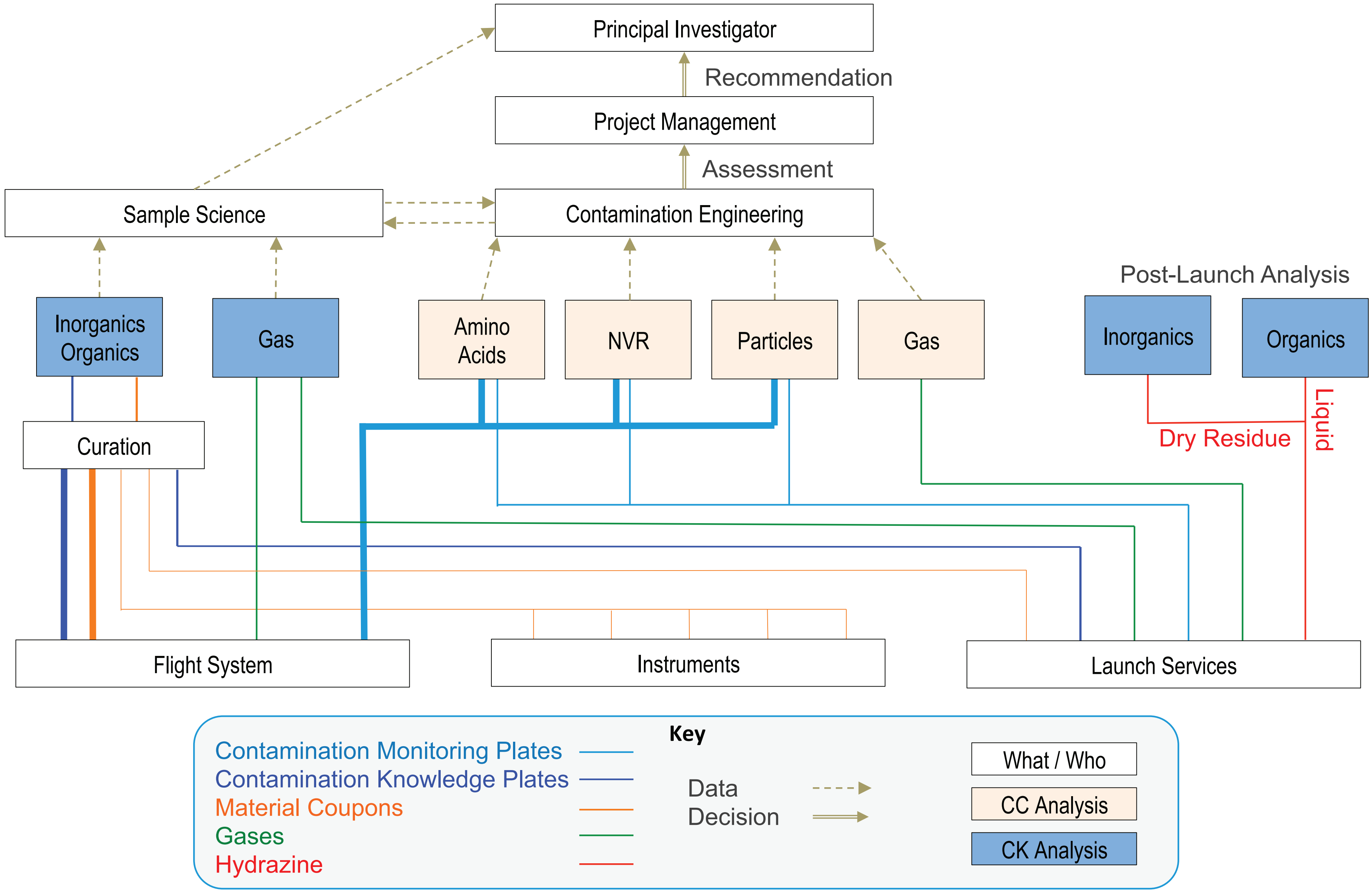
Principal Investigator
Recommendation
Project Management
Assessment
Sample Science
Contamination Engineering
Inorganics Organics
Gas
Amino Acids
NVR
Particles
Gas
Post-Launch Analysis
Inorganics
Organics
Curation
Dry Residue
Liquid
Flight System
Instruments
Launch Services
Key
Contamination Monitoring Plates
Contamination Knowledge Plates
Material Coupons
Gases
Hydrazine
Data
Decision
What / Who
CC Analysis
CK Analysis

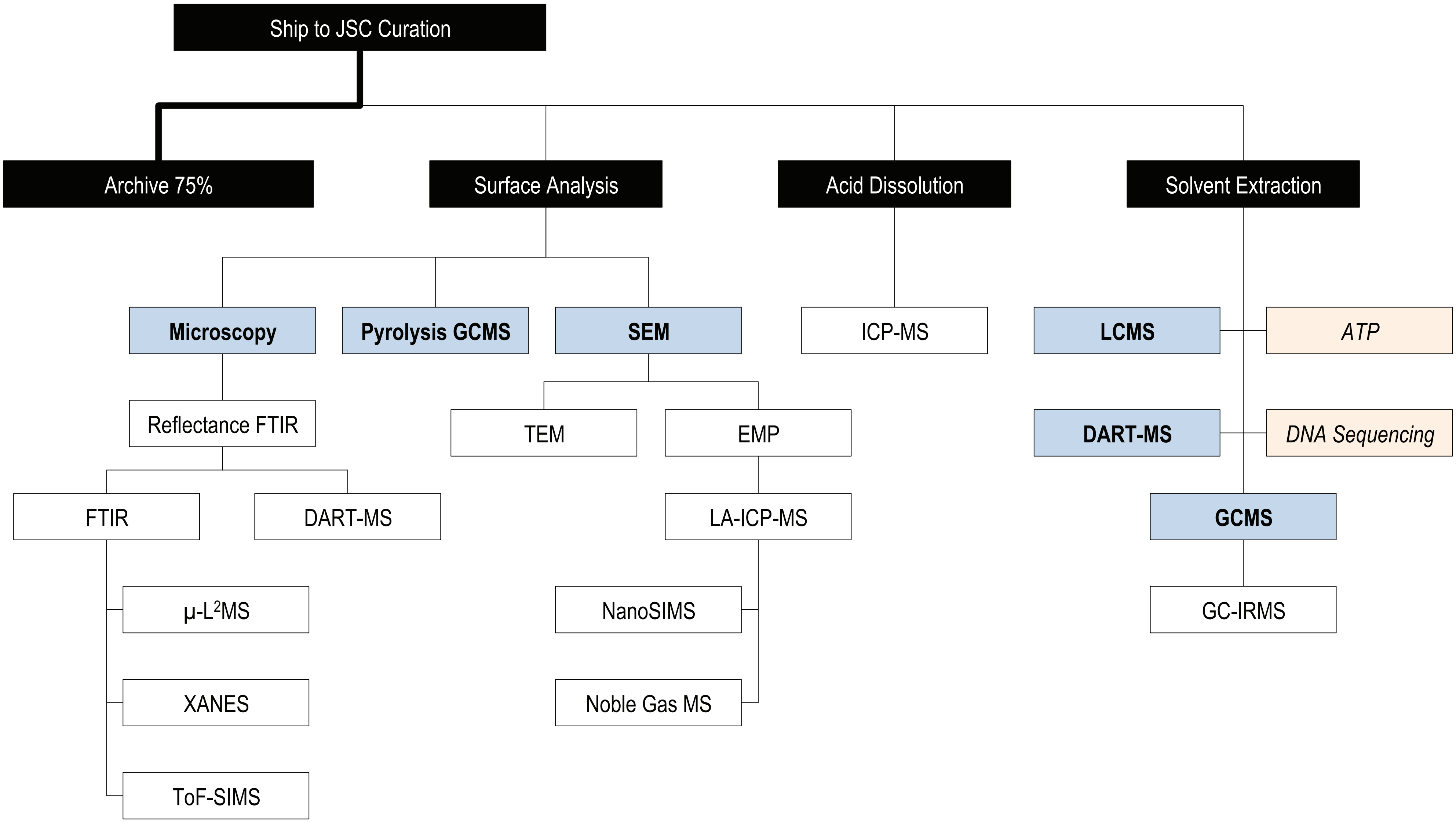
Ship to JSC Curation
Archive 75%
Surface Analysis
Acid Dissolution
Solvent Extraction
Microscopy
Pyrolysis GCMS
SEM
ICP-MS
LCMS
ATP
Reflectance FTIR
TEM
EMP
DART-MS
DNA Sequencing
FTIR
DART-MS
LA-ICP-MS
GCMS
μ-L²MS
NanoSIMS
GC-IRMS
XANES
Noble Gas MS
ToF-SIMS

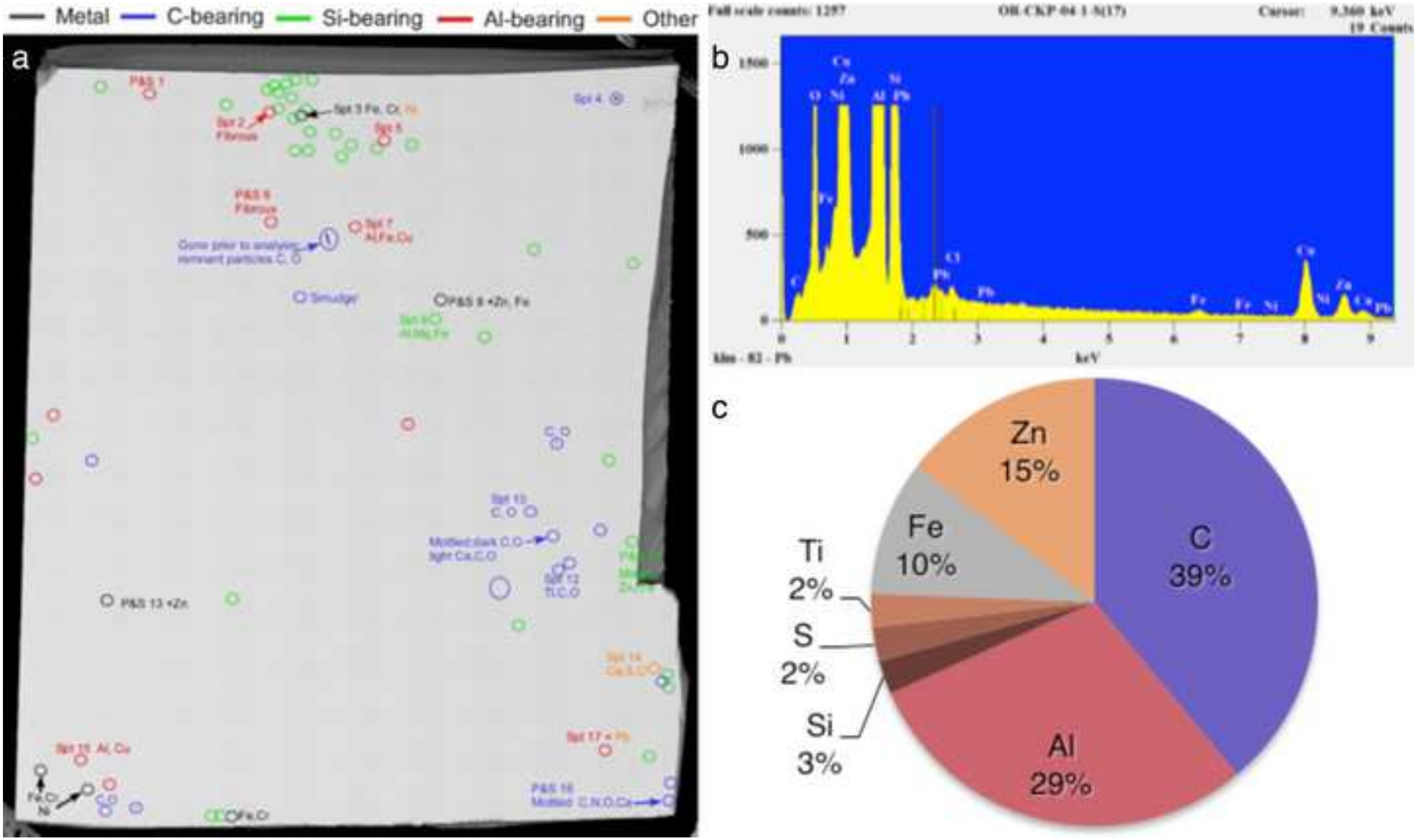

Metal
C-bearing
Si-bearing
Al-bearing
Other
a
b
c
keV
Zn
15%
Fe
10%
Ti
2%
S
2%
Si
3%
C
39%
Al
29%

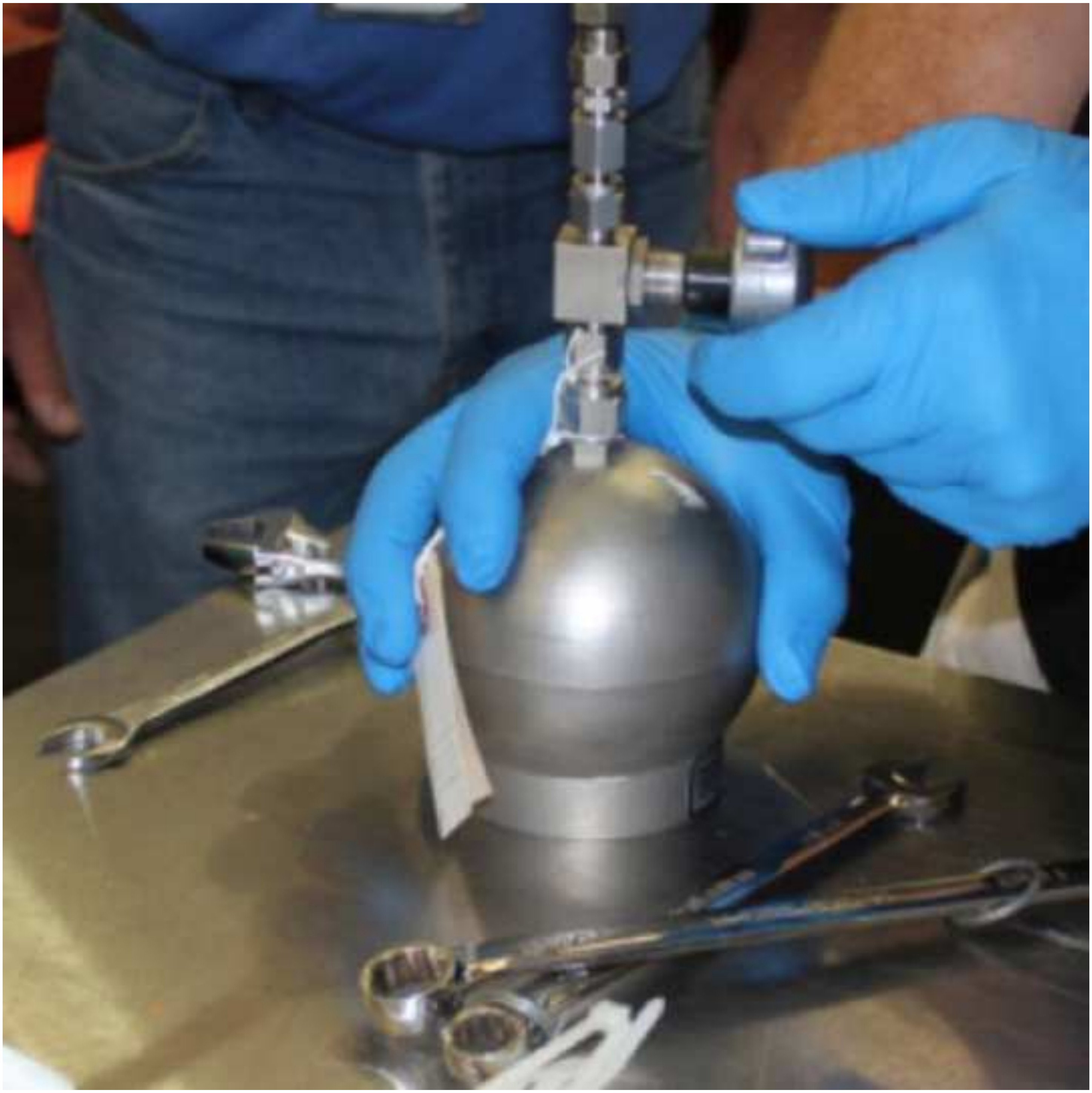

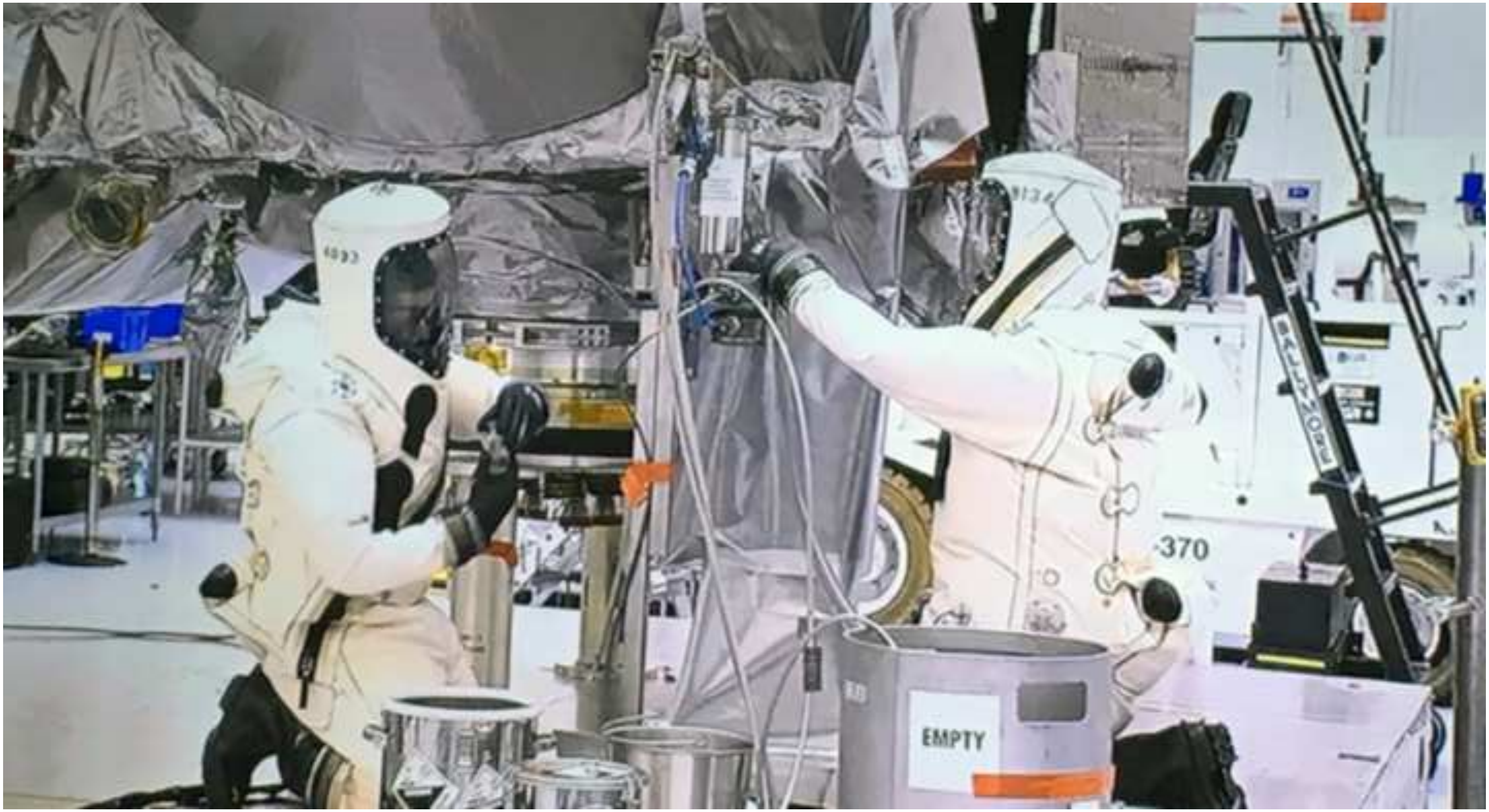
EMPTY
370
BALLYMORE

| Component | Aluminum | Epoxy | Foil | Honeycomb | Lubricant | Miscellaneous | NVR | Paint | Polymer | Sapphire | Steel | Titanium | Total |
|---|---|---|---|---|---|---|---|---|---|---|---|---|---|
| KSC | 2 | 4 | 1 | | 1 | 6 | 7 | | 4 | | | | 25 |
| OCAMS | | 3 | | | 2 | 2 | | 2 | 1 | | | | 10 |
| OLA | | 15 | 1 | | | | | | | | | | 16 |
| OTES | | | | | 1 | | | | | | | | 1 |
| OVIRS | | 2 | | | 1 | | | 1 | | | | | 4 |
| REXIS | | 1 | | | 1 | 2 | | | 1 | | | 1 | 6 |
| Spacecraft | | 4 | | 3 | | 8 | | | 1 | | | | 16 |
| SRC | 30 | 51 | 1 | 1 | 3 | 18 | 6 | | 3 | 3 | 45 | 1 | 162 |
| Support | 1 | | 1 | | 1 | 19 | 2 | | 11 | | | | 35 |
| TAGSAM | 28 | 11 | | | 1 | 9 | 4 | | | 6 | 57 | 4 | 120 |
| Total | 61 | 91 | 4 | 4 | 11 | 64 | 19 | 3 | 21 | 9 | 102 | 6 | 395 |

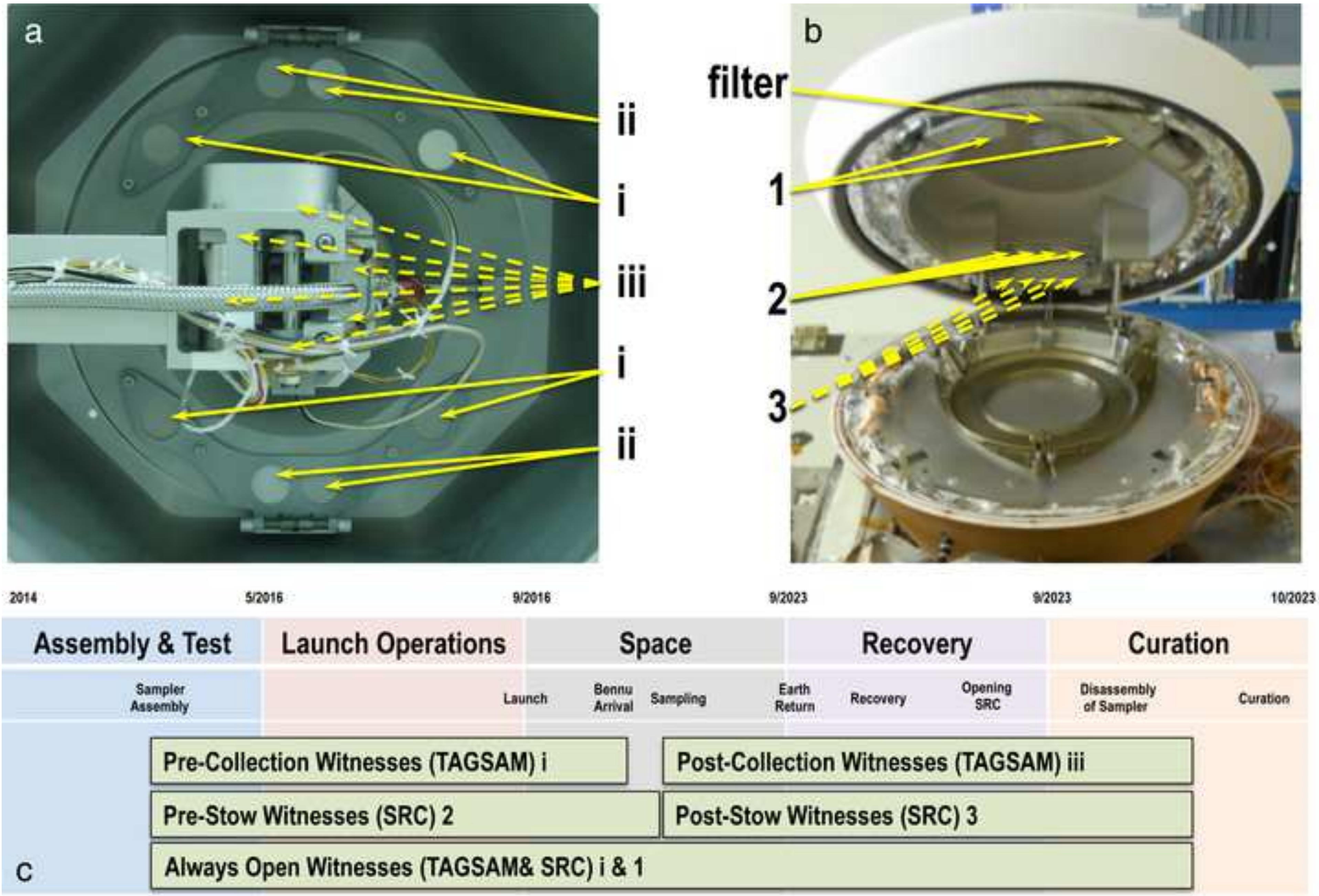
a
ii
i
iii
i
ii
b
filter
1
2
3
2014
5/2016
9/2016
9/2023
9/2023
10/2023 -
Assembly & Test
Launch Operations
Space
Recovery
Curation
Sampler Assembly
Launch
Bennu Arrival
Sampling
Earth Return
Recovery
Opening SRC
Disassembly of Sampler
Curation
Pre-Collection Witnesses (TAGSAM) i
Post-Collection Witnesses (TAGSAM) iii
Pre-Stow Witnesses (SRC) 2
Post-Stow Witnesses (SRC) 3
Always Open Witnesses (TAGSAM& SRC) i & 1
c

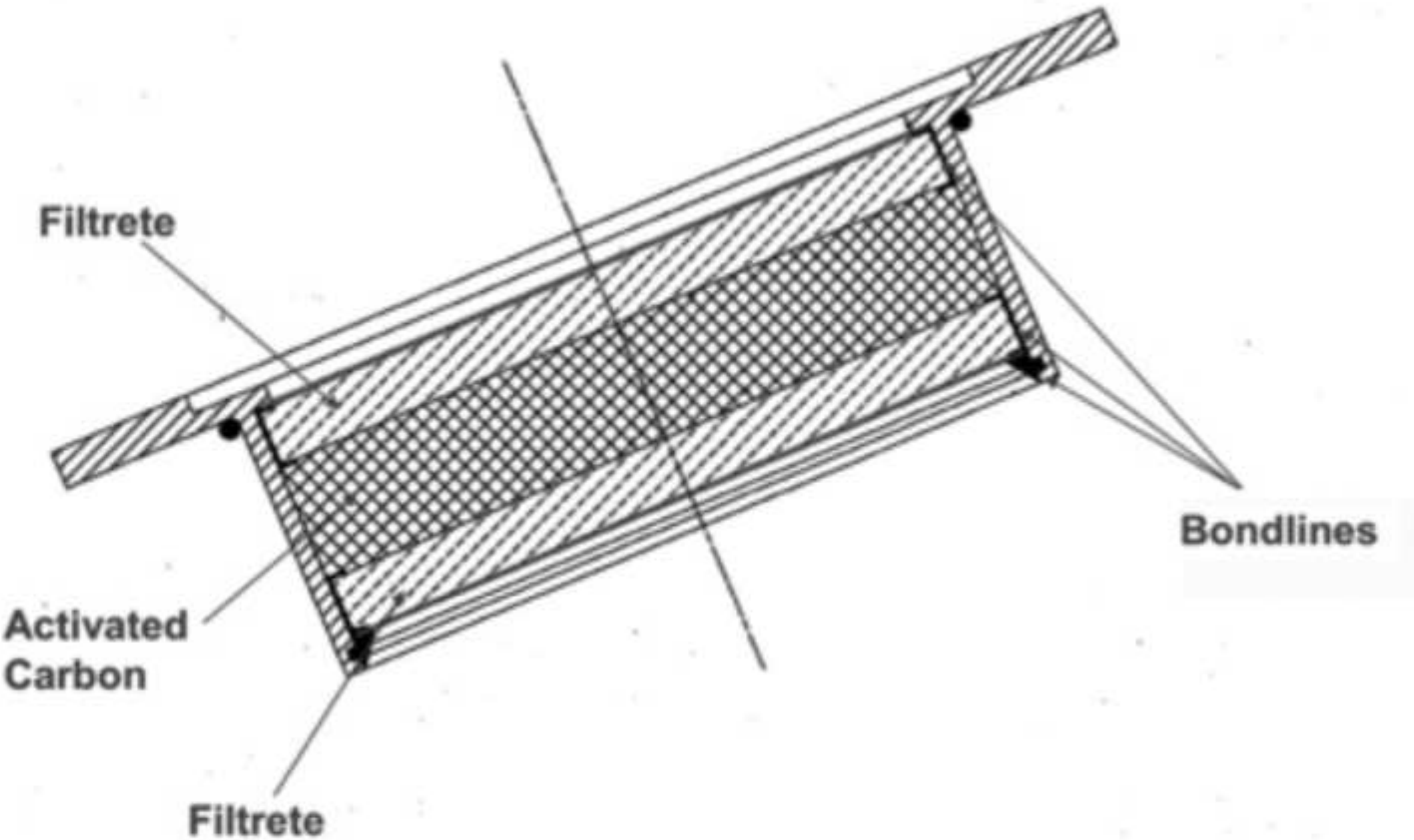
Filtrete
Bondlines
Activated Carbon
Filtrete

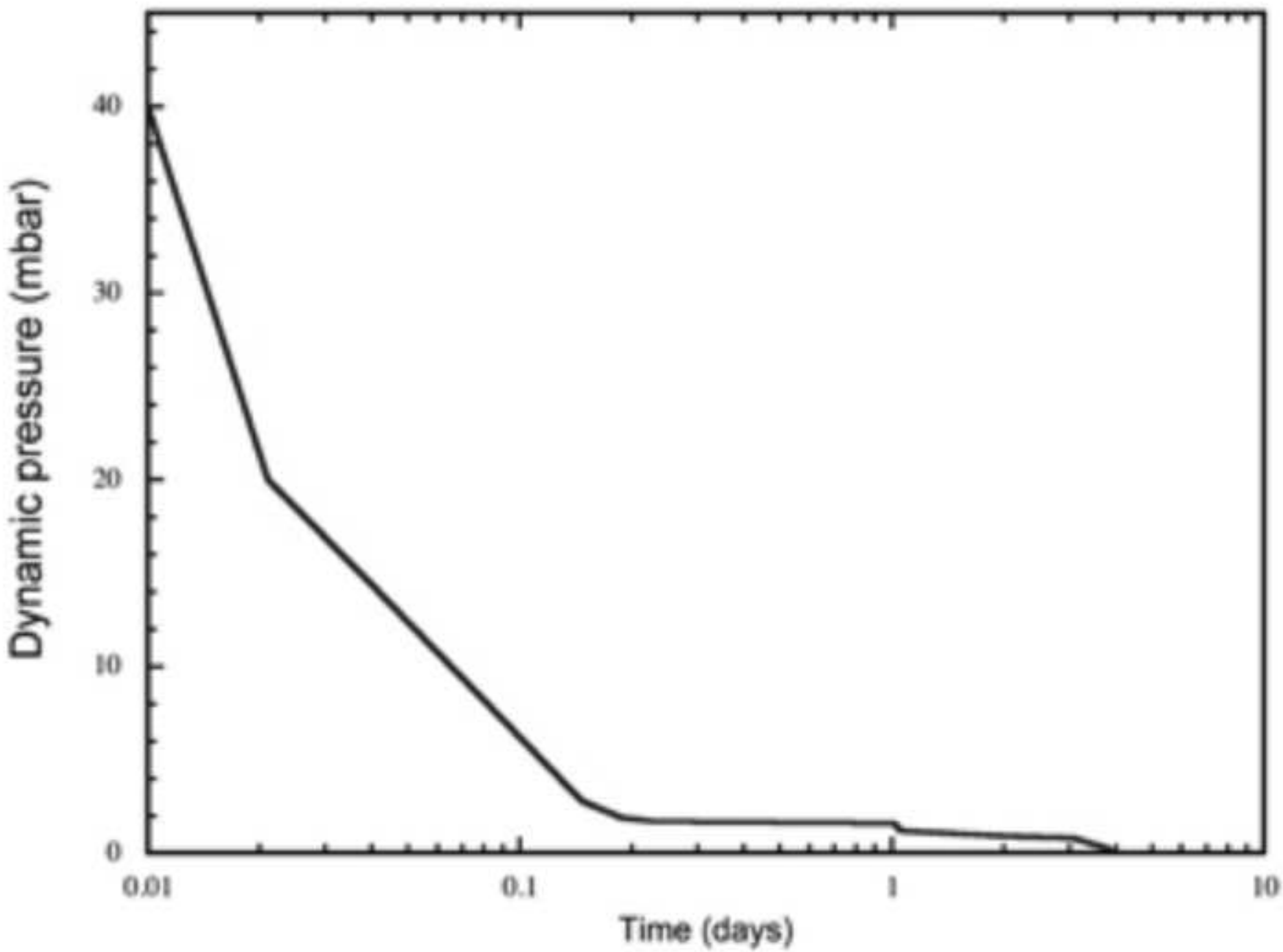
Dynamic pressure (mbar)
40
30
20
10
0
0.01
0.1
1
10
Time (days)

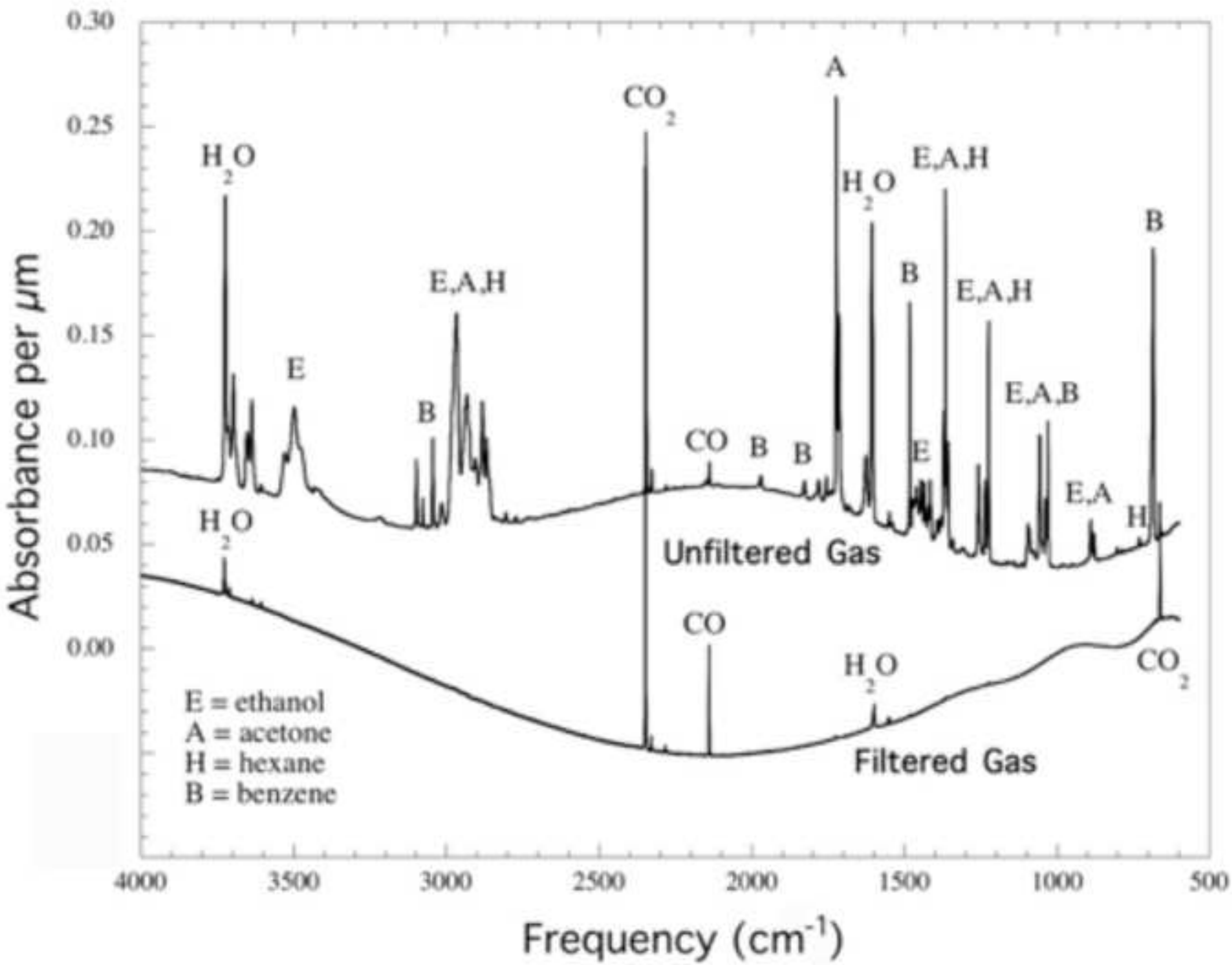
0.30
0.25
0.20
0.15
0.10
0.05
0.00
Absorbance per μm
H2O
E
B
E,A,H
CO2
CO
B
B
A
H2O
E,A,H
B
E
E,A,H
E,A,B
E,A
H
B
Unfiltered Gas
H2O
CO
H2O
CO2
Filtered Gas
E = ethanol
A = acetone
H = hexane
B = benzene
4000
3500
3000
2500
2000
1500
1000
500
Frequency (cm-1)

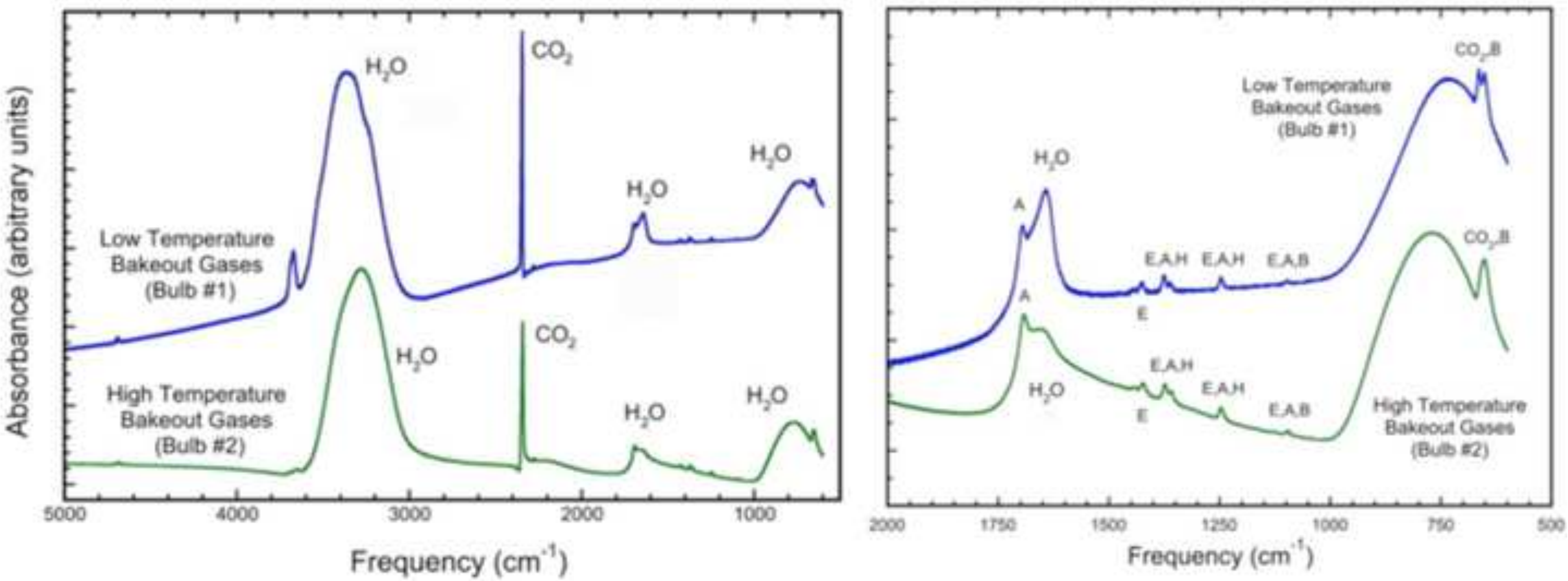
Absorbance (arbitrary units)
H2O
CO2
H2O
H2O
Low Temperature Bakeout Gases (Bulb #1)
High Temperature Bakeout Gases (Bulb #2)
H2O
CO2
H2O
H2O
5000
4000
3000
2000
1000
Frequency (cm-1)
CO2,B
Low Temperature Bakeout Gases (Bulb #1)
H2O
A
E,A,H
E,A,H
E,A,B
A
E
CO2,B
E,A,H
H2O
E,A,H
E
E,A,B
High Temperature Bakeout Gases (Bulb #2)
2000
1750
1500
1250
1000
750
500
Frequency (cm-1)

Response to reviewer

Click here to download figure Fig24.tif

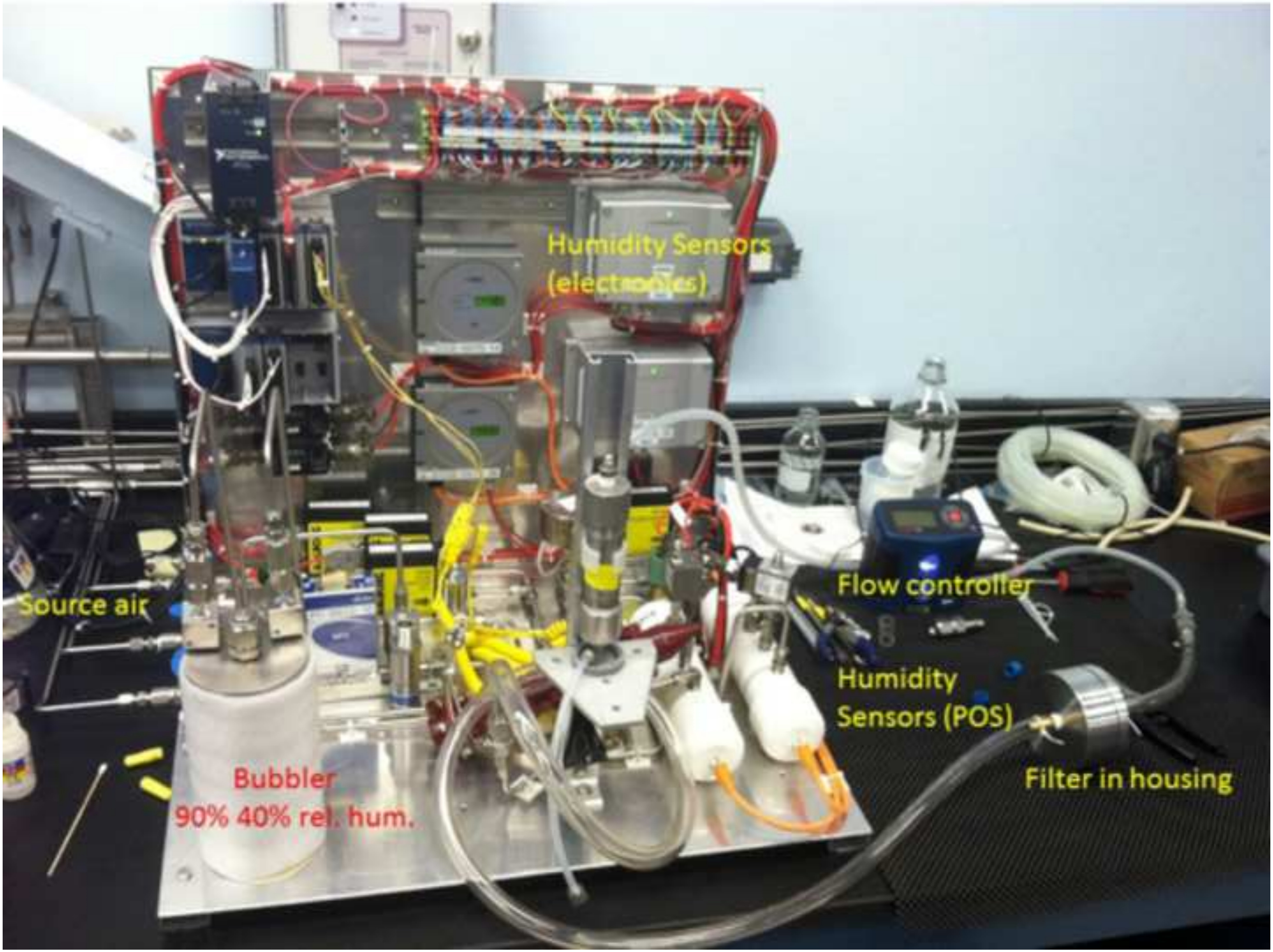

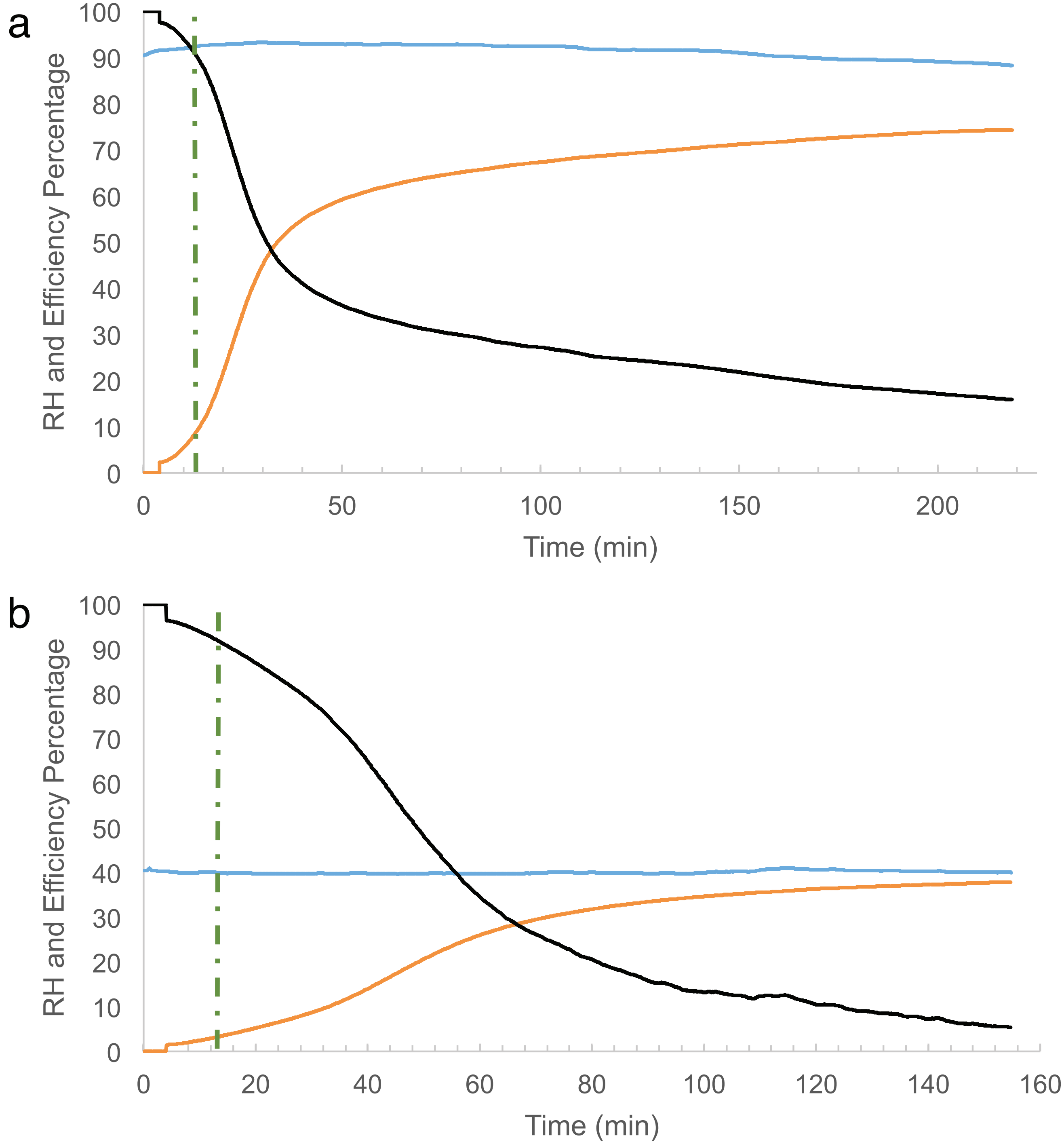
a
b
RH and Efficiency Percentage
Time (min)
0
10
20
30
40
50
60
70
80
90
100
50
150
200
120
140
160

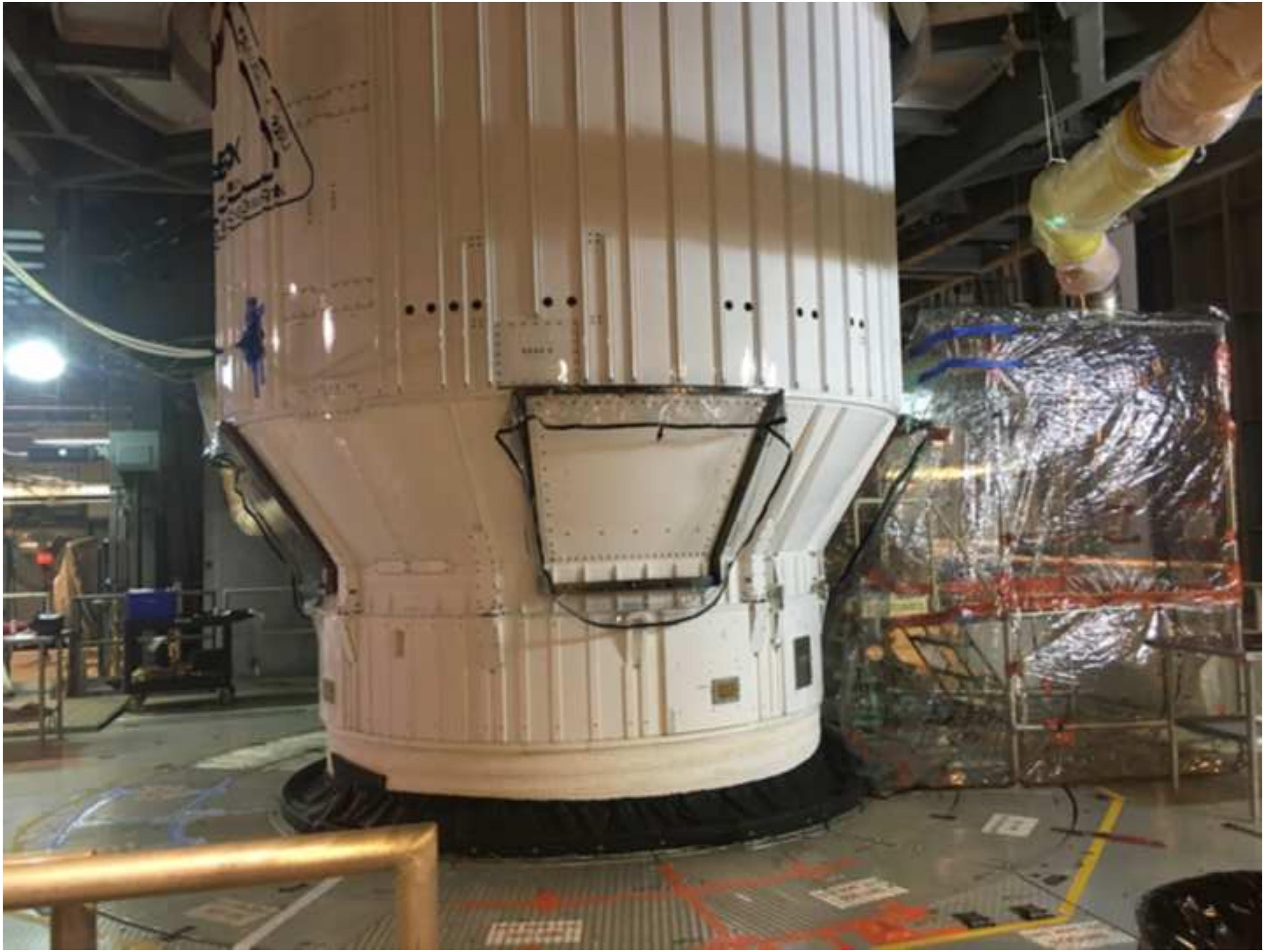

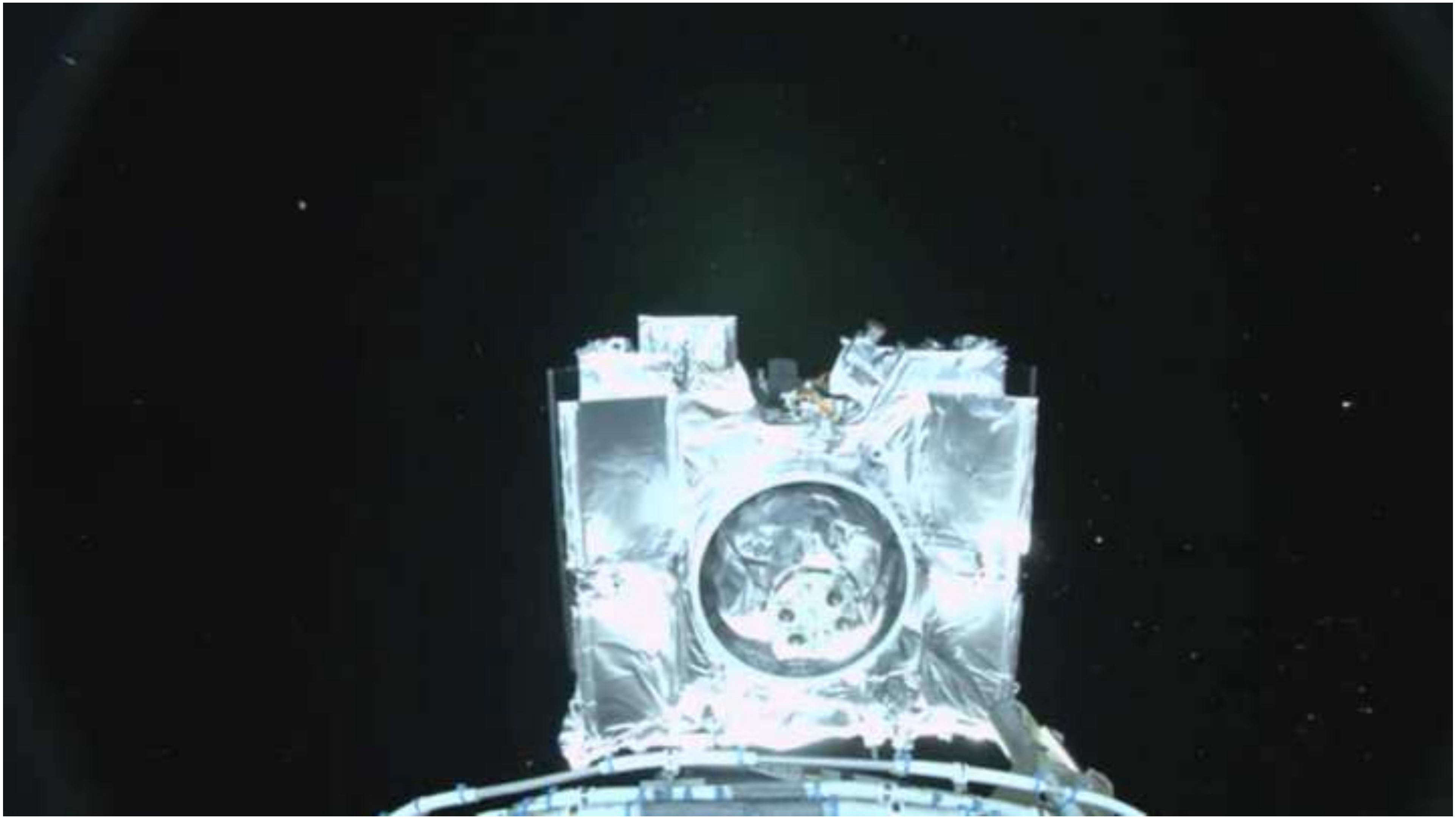

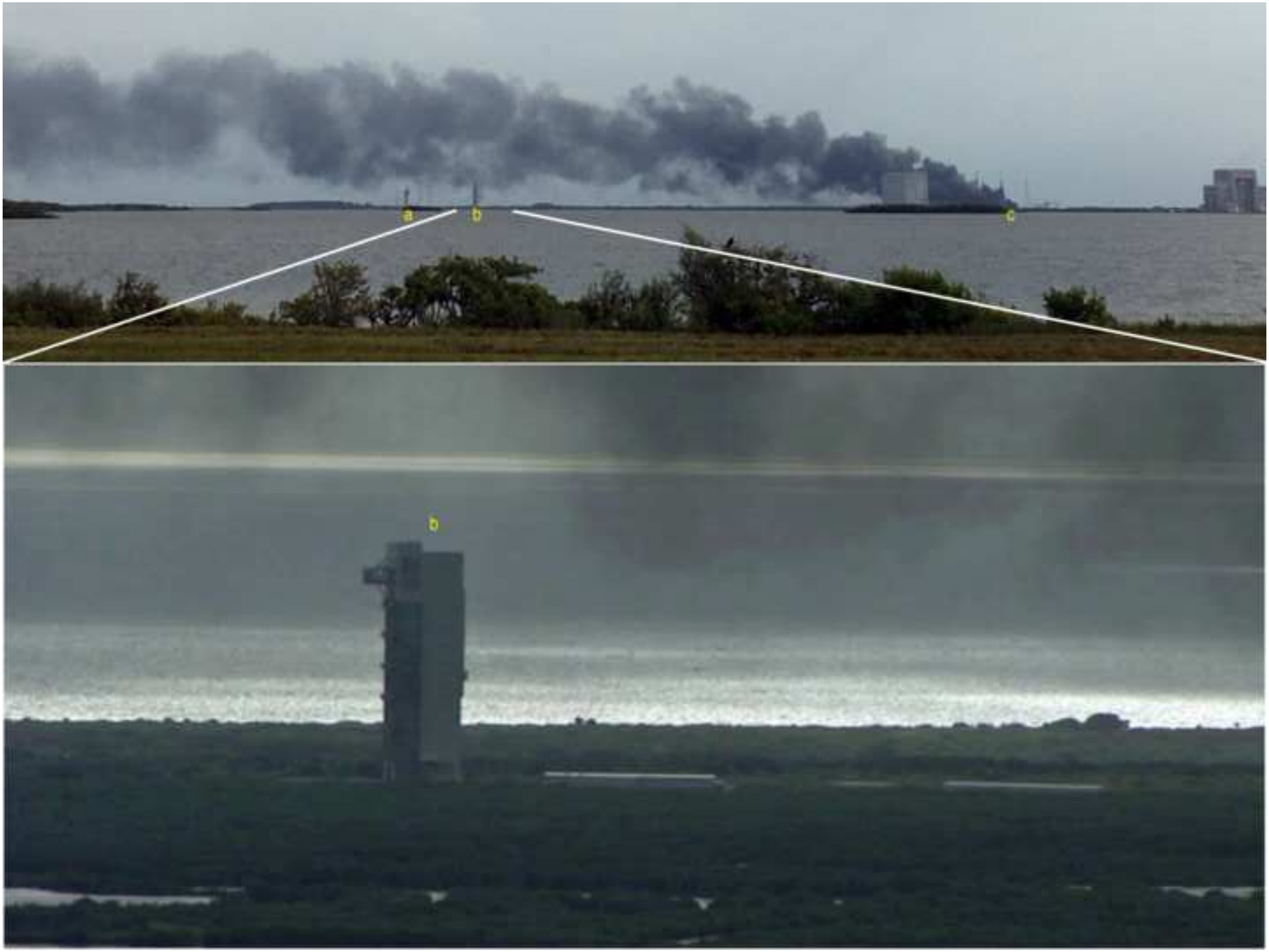
a
b
c
b

Click here to access/download
**Electronic Supplementary Materials**
OnlineSupplementalMaterials2.pdf